\documentclass[longauth]{aa} 

\usepackage{longtable}

\usepackage{xspace}
\usepackage{hyperref}
\usepackage{graphicx}
\usepackage{color}
\usepackage{lineno}
\usepackage{graphics}
\usepackage{subcaption}
\usepackage[section]{placeins}
\usepackage{longtable}
\usepackage{graphicx}
\usepackage[export]{adjustbox}
\usepackage{natbib}

\usepackage{txfonts}
\usepackage{tikz}

\begin{document}

   \title{The SOPHIE search for northern extrasolar planets-XIX. A system including a cold sub-Neptune potentially transiting a V = 6.5 star HD88986} 
   
   \author{N.~Heidari\inst{\ref{lam},\ref{iap},\thanks{CNES postdoctoral fellowship},\ref{OCA},\ref{iran}},
          I.~Boisse \inst{\ref{lam}},
          N.~C.~Hara \inst{\ref{geneva}},
          T.~G.~Wilson \inst{\ref{tom}},
          F.~Kiefer \inst{\ref{Lesia}}, 
          G.~H\'ebrard \inst{\ref{iap},\ref{OHP}},
          F.~Philipot \inst{\ref{Lesia}}, 
          S.~Hoyer \inst{\ref{lam}},
          K.~G.\ Stassun\inst{\ref{Stassun}},
          G.~W.~Henry \inst{\ref{henry}},
          N.~C.~Santos \inst{\ref{santos1},\ref{santos2}},
          L.~Acu\~{n}a \inst{\ref{lorena}},
          D.~Almasian \inst{\ref{iran}},
          L.~Arnold \inst{\ref{OHP},\ref{CFHT}},
          N.~Astudillo-Defru \inst{\ref{chile}},
          O.~Attia \inst{\ref{geneva}},
          X.~Bonfils \inst{\ref{grenoble}},
          F. Bouchy \inst{\ref{geneva}}, 
          V.~ Bourrier \inst{\ref{geneva}},
          B.~ Collet \inst{\ref{lam}},
          P.~Cort\'es-Zuleta \inst{\ref{lam}},
          A.~Carmona \inst{\ref{grenoble}},
          X.~Delfosse \inst{\ref{grenoble}},
          S.~Dalal \inst{\ref{dalal}},
          M. Deleuil \inst{\ref{lam}},
          O. D. S. Demangeon \inst{\ref{santos1},\ref{santos2}},
          R.~ F. D\'iaz \inst{\ref{diaz}},
          X.~Dumusque \inst{\ref{geneva}},
          D.~ Ehrenreich \inst{\ref{geneva}},
          T.~Forveille \inst{\ref{grenoble}},
          M.~J.~ Hobson \inst{\ref{lorena},\ref{melissa2}},
          J.~S.~ Jenkins\inst{\ref{jen},\ref{jen1}},
          J.~ M.~ Jenkins \inst{\ref{Jenkins}},
          A.~ M.~ Lagrange\inst{\ref{Lesia}}, 
          D.~ W. Latham \inst{\ref{jason}},
          P.~ Larue \inst{\ref{grenoble}},
          J.~Liu \inst{\ref{lam}},
          C.~Moutou \inst{\ref{Toulouse}},
          L.~ Mignon \inst{\ref{geneva}},
          H.~P.~Osborn \inst{\ref{bern},\ref{andrew}}, 
          F. ~Pepe\inst{\ref{geneva}},
          D.~ Rapetti\inst{\ref{Jenkins},\ref{spoc}},       J.~Rodrigues\inst{\ref{santos1},\ref{santos2},\ref{OFXB}},
          A.~Santerne \inst{\ref{lam}},
          D.~ Segransan \inst{\ref{geneva}},
          A. ~Shporer \inst{\ref{Avi-Shporer}},
          S.~Sulis \inst{\ref{lam}},
          G.~Torres \inst{\ref{Torres}},
          S.~ Udry \inst{\ref{geneva}},
          F.~ Vakili \inst{\ref{OCA}},
          A.~Vanderburg\inst{\ref{andrew}},
          O.~ Venot \inst{\ref{venot}},
          H.~G.~Vivien \inst{\ref{lam}},
      J.~I.~Vines \inst{\ref{vines}}
           }

    \institute{ Aix Marseille Univ, CNRS, CNES, LAM, Marseille, France.\label{lam} \email{heidari@iap.fr}
    \and
           Institut d'astrophysique de Paris, UMR 7095 CNRS université pierre et marie curie, 98 bis, boulevard Arago,  75014, Paris \label{iap}
    \and     
         Laboratoire J.-L. Lagrange, Observatoire de la C\^ote d’Azur (OCA), Universite de Nice-Sophia Antipolis (UNS), CNRS, Campus Valrose, 06108 Nice Cedex 2, France.\label{OCA}        
    \and   
         Department of Physics, Shahid Beheshti University, Tehran, Iran \label{iran}.     
    \and
       Observatoire de Gen\`eve,  Universit\'e de Gen\`eve, Chemin Pegasi, 51, 1290 Sauverny, Switzerland\label{geneva}
    \and     
        SUPA, School of Physics and Astronomy, University of St. Andrews, North Haugh, Fife KY16 9SS, UK. \label{tom}    
    \and
       LESIA, Observatoire de Paris, Université PSL, CNRS, Sorbonne Université, Université de Paris, 5 place Jules Janssen, 92195, Meudon, France \label{Lesia}  
    \and
       Observatoire de Haute-Provence, CNRS, Universit\'e d'Aix-Marseille, 04870 Saint-Michel-l'Observatoire, France \label{OHP}
    \and  
          Vanderbilt University, Department of Physics \& Astronomy, 6301 Stevenson Center Ln., Nashville, TN 37235, USA \label{Stassun}
      \and
      Center of Excellence in Information Systems, Tennessee State University,3500 John A. Merritt Blvd., P.O. Box 9501, Nashville, TN 37209, USA. \label{henry}  
     \and
         Instituto de Astrof\'isica e Ci\^encias do Espa\c{c}o, Universidade do Porto, CAUP, Rua das Estrelas, 4150-762 Porto, Portugal.\label{santos1}
    \and     
         Departamento de F\'isica e Astronomia, Faculdade de Ci\^encias, Universidade do Porto, Rua do Campo Alegre, 4169-007 Porto, Portugal.\label{santos2}
          \and
     Max-Planck-Institut für Astronomie, Heidelberg, Germany. \label{lorena}
     \and
        Canada France Hawaii Telescope Corporation (CFHT), 65-1238 Mamalahoa Hwy, Kamuela HI 96743  USA \label{CFHT}
        \and
         Departamento de Matem\'atica y F\'isica Aplicadas, Universidad Cat\'olica de la Sant\'isima Concepci\'on, Alonso de Rivera 2850, Concepci\'on, Chile\label{chile} 
          \and
       Univ. Grenoble Alpes, CNRS, IPAG, 38000 Grenoble, France\label{grenoble}  
    \and
    Astrophysics Group, University of Exeter, Exeter EX4 2QL, UK. \label{dalal}
    \and
    Telescope Science Institute, Baltimore, MD, USA. \label{juliet}
     \and
     Univ. de Toulouse, CNRS, IRAP, 14 avenue Belin, 31400 Toulouse, France \label{Toulouse}
     \and 
     Millennium Institute of Astrophysics (MAS), Nuncio Monseñor Sótero Sanz 100, Providencia, Santiago, Chile.\label{melissa2}
     \and
    International Center for Advanced Studies (ICAS) and ICIFI (CONICET), ECyT-UNSAM, Campus Miguelete, 25 de Mayo y Francia, (1650) Buenos Aires, Argentina.\label{diaz}
        \and 
       Center for Astrophysics \textbar \ Harvard \& Smithsonian, 60 Garden St, Cambridge, MA 02138, USA \label{jason}
      \and
       Physikalisches Institut, University of Bern, Gesellsschaftstrasse 6, 3012 Bern, Switzerland.\label{bern}
     \and
     Department of Physics and Kavli Institute for Astrophysics and Space Research, Massachusetts Institute of Technology, Cambridge, MA 02139, USA. \label{andrew}
     \and
     Instituto de Estudios Astrofísicos, Universidad Diego Portales,  Av. Ej\'ercito 441, Santiago, Chile\label{jen}
     \and
     NASA Ames Research Center, Moffett Field, CA, 94035, USA \label{Jenkins}
     \and
     Research Institute for Advanced Computer Science, Universities Space Research Association, Washington, DC 20024, USA \label{spoc}
     \and
     Observatoire François-Xavier Bagnoud -- OFXB, 3961 
     Saint-Luc, Switzerland\label{OFXB}
     \and
     Department of Physics and Kavli Institute for Astrophysics and Space Research, Massachusetts Institute of Technology, Cambridge, MA 02139, USA \label{Avi-Shporer}
     \and
     Center for Astrophysics $\vert$ Harvard \& Smithsonian, 60 Garden Street, Cambridge, MA 02138, USA \label{Torres}
     \and
     Université de Paris Cité and Univ Paris Est Creteil, CNRS, LISA, F-75013 Paris, France\label{venot}
     \and
     Centro de Astrof\'isica y Tecnolog\'ias Afines (CATA), Casilla 36-D, Santiago, Chile\label{jen1}
     \and
     Instituto de Astronom\'ia, Universidad Cat\'olica del Norte,
Angamos 0610, 1270709, Antofagasta, Chile\label{vines}
         }

  \date{Received XX, 2023; accepted XX, 2023}
 
  \abstract
   {Transiting planets with orbital periods longer than 40 d are extremely rare among the 5000+ planets discovered so far. The lack of discoveries of this population poses a challenge to research into planetary demographics, formation, and evolution. Here, we present the detection and characterization of HD88986\,b, a potentially transiting sub-Neptune, possessing the longest orbital period among known transiting small planets (< 4 R$_{\oplus}$) with a precise mass measurement ($\sigma_M/M$ > 25\%). Additionally, we identified the presence of a massive companion in a wider orbit around HD88986. To validate this discovery, we used a combination of more than 25 years of extensive radial velocity (RV) measurements (441 SOPHIE data points, 31 ELODIE data points, and 34 HIRES data points), Gaia DR3 data, 21 years of photometric observations with the automatic photoelectric telescope (APT), two sectors of TESS data, and a 7-day observation of CHEOPS. Our analysis reveals that HD88986\,b, based on two potential single transits on sector 21 and sector 48 which are both consistent with the predicted transit time from the RV model, is potentially transiting. The joint analysis of RV and photometric data show that HD88986\,b has a radius of 2.49$\pm$0.18 R$_{\oplus}$, a mass of 17.2$^{+4.0}_{-3.8}$ M$_{\oplus}$, and it orbits every 146.05$^{+0.43}_{-0.40}$ d around a subgiant HD88986 which is one of the closest and brightest exoplanet host stars (G2V type, R=1.543 $\pm$0.065 R$_{\odot}$, V=$6.47\pm 0.01$ mag, distance=33.37$\pm$0.04 pc). The nature of the outer, massive companion is still to be confirmed; a joint analysis of RVs, Hipparcos, and Gaia astrometric data shows that with a 3$\sigma$ confidence interval, its semi-major axis is between 16.7 and 38.8 au and its mass is between 68 and 284 M$_{Jup}$. HD88986\,b's wide orbit suggests the planet did not undergo significant mass loss due to extreme-ultraviolet radiation from its host star. Therefore, it probably maintained its primordial composition, allowing us to probe its formation scenario. Furthermore, the cold nature of HD88986\,b (460$\pm$8\, K), thanks to its long orbital period, will open up exciting opportunities for future studies of cold atmosphere composition characterization. Moreover, the existence of a massive companion alongside HD88986\,b makes this system an interesting case study for understanding planetary formation and evolution.}

 \keywords{planets and satellites: detection – techniques: photometric, radial velocities – stars: individual (HD88986, and TIC 1042868)}
\titlerunning{HD88986\,b: a long-period planet}
\authorrunning{N. Heidari et al.}
   \maketitle
%

\section{Introduction}

Among the 5000+\footnote{\url{https://exoplanetarchive.ipac.caltech.edu}} planets discovered so far (and many more candidates), transiting planets ($\sim$ 4000) have a considerable impact on our understanding of the formation and evolution of planetary systems. Such planets, when orbiting a bright host star that allows radial velocity (RV) follow-up, can be accurately characterized in terms of fundamental parameters such as mass and density, allowing for their internal structure to be modeled \cite[e.g.,][]{heidari2021hd207897,delrez2021transit}. Moreover, these objects provide us with a great opportunity to gather information about the composition and temperature of their atmospheres by transmission and/or emission spectroscopy \cite[e.g.,][]{tabernero2020horus}.

Among all the known transiting exoplanets with precise mass and radius measurements \citep{otegi2020revisited}, those with orbital periods exceeding 40 d are extremely rare, representing only $\sim$ 1\% of the total population (as of June 7, 2023). This scarcity poses a significant challenge to our understanding of planet demographics, formation, evolution, and the potential for habitability. The limited number of such long-period exoplanets has compelled many studies on exoplanet occurrence rates to focus primarily on planets with relatively short periods \citep[e.g.,][]{silburt2015statistical, petigura2013prevalence,fulton2017california}. Moreover, \cite{kopparapu2013habitable} showed that the inner boundary of the habitable zone, encircling main-sequence stars with spectral types earlier than approximately M4 ($T_{\rm eff}$ $>$ 2800 K), is longer than $\sim$ 11 d. This emphasizes the importance of exploring planetary systems with longer orbital periods in our pursuit of habitable planets. Additionally, new scientific exploration such as detecting exomoons has yet to be achieved. The importance of exomoon discovery is highlighted by our Moon's influence on Earth's spin dynamics \citep{li2014spin} as well as the prospective habitability of icy moons inside our own Solar System \citep{reynolds1983habitability}. The lack of successful exomoon discoveries could be linked to the dearth of long-period planets, as planets with longer orbital periods are more likely to harbor moons \citep{dobos2021survival}. These examples highlight the importance of detecting and studying this "missing population," thereby advancing our comprehension of exoplanet populations.

However, long-period transiting planets are particularly difficult to detect. The two primary methods used for detecting exoplanets, RV and transit photometry, each have their own challenges when it comes to detecting these elusive long-period planets. The RV method requires high-precision measurements over extended time spans to capture a full orbital period, while the transit method faces challenges due to both a lower transit probability and the limited baseline observations in most photometric surveys. These challenges are further compounded when dealing with smaller planets with a shallow transit depth and small RV semi-amplitude.

Here, we present the detection and characterization of a planetary system orbiting a V=6.5 mag G2-type star: HD88986 b, the longest-period transiting sub-Neptune among the known small planets (< 4 R$_{\oplus}$), accompanied by an outer massive companion. To accomplish this discovery, we employed a variety of observations, including photometric, spectroscopic, and astrometric techniques. The structure of this paper is organized as follows: Section \ref{spectroscopy} provides a description of the spectroscopic observations utilized in our analysis. In Section \ref{Stellar parameters}, the stellar properties of the host star are discussed. To identify the stellar rotational period, in Section \ref{stellar activity} we present an analysis of various stellar activity indicators. In Section \ref{rv-detection}, we analyze RV data that led to the discovery of the sub-Neptune planet and a long-term curvature. In Sections \ref{photometry_extrac} and \ref{joint2}, we present the detection of a single transit event in TESS data sector 21 and perform a joint analysis, respectively. In Section \ref{Additional_photometric_data}, the result of our investigations into detecting HD88986 b's second transit event in additional photometric data is presented. Finally, in Sections \ref{Astrometry} and \ref{Discution}, we discuss the origin of the long-term trend and conclude on this system, respectively. The paper also includes appendices presenting updated and new procedures for spectroscopic extractions on SOPHIE data.

\section{Spectroscopic observations}
\label{spectroscopy}
We have intensive spectroscopic observations of HD88986 spanning a remarkable duration of 25 years. These observations comprise a total of 506 data points obtained through the utilization of three high-resolution spectrographs: SOPHIE \citep{bouchy2013sophie+}, HIRES \citep{vogt1988hires}, and ELODIE \citep{baranne1996elodie}. The whole dataset is shown in Fig. \ref{RV}. It displays a clear long-term curvature on a time scale of at least 25 years, as well as other variations on shorter time scales.

\subsection{High-resolution spectroscopy with SOPHIE}

\begin{figure}
    \centering
    \includegraphics[width=0.499\textwidth]{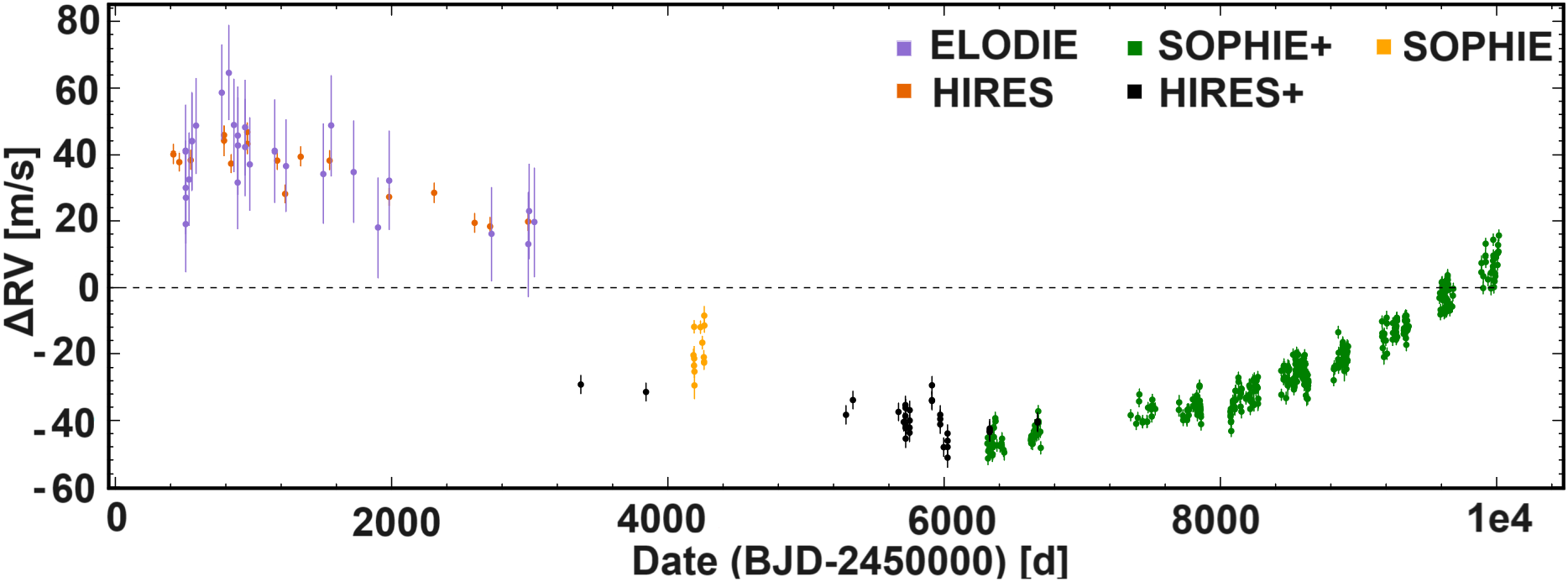} 
    \caption{Radial velocity measurements of HD88986 from ELODIE, HIRES, HIRES+, SOPHIE and SOPHIE+.}
    \label{RV}
\end{figure}

HD88986 has been monitored by the SOPHIE high-precision spectrograph mounted on the 1.93-m telescope at the Haute-Provence Observatory (OHP, France). The observations were carried out as part of Recherche de Planètes Extrasolaires (RPE) program 1, also known as SP1, which is a high-precision program to search for Neptunes and Super-Earths orbiting bright stars in the solar neighborhood \citep{courcol2015sophie,hara2020,heidari2021hd207897}. Over a period of more than 15 years, spanning from 2007 December 7$^{th}$ to 2023 March 12$^{th}$, 441 high-resolution spectra were collected for this star (see Fig. \ref{RV}). The observations were conducted in SOPHIE high resolution (HR) mode (resolving power of $\lambda / \Delta \lambda \approx 75 000$ at 550 nm), with simultaneous Thorium-Argon (Th-Ar) or Fabry-Perot (FP) calibration light measurements. The latter allows us to track instrumental drift, ensuring precise and accurate RV measurements.

In June 2011, hexagonal fibers were installed in the SOPHIE spectrograph. This led to achieving an RV precision of 1-2 m s$^{-1}$ but also about 50 m s$^{-1}$ jump in the measured RVs of the standard stars \citep{bouchy2013sophie+}. Hence, we separated the data before June 2011 (12 data points with the name SOPHIE) from the data after (429 data points with the name SOPHIE+). The exposure time was set for both data sets at 600-900 s to average the stellar oscillations, achieving a median signal-to-noise ratio (S/N) of 158 per pixel at 550~nm.

The RV was derived using the SOPHIE data reduction system \citep[DRS,][]{bouchy2009sophie}. The DRS encompasses several crucial steps, including spectrum extraction, removal of telluric lines, correction for CCD charge transfer inefficiency (CTI), computation of the cross-correlation function (CCF) between the spectra and a binary mask, barycentric RV correction, and ultimately fitting Gaussian profiles to the CCFs to extract the RVs \citep{baranne1996elodie,pepe2002coralie}. We note that to extract HD88986 RVs, we used a G2 mask. Additionally, prior to RV extraction, we corrected the spectra for the atmospheric dispersion effect (see Appendix \ref{Atmospheric}).

Once the RVs were extracted, we performed a correction for the nightly instrumental drift, which was measured through simultaneous calibration light observations. For this purpose, SOPHIE fiber A was dedicated to star observations, while fiber B was used to monitor a calibration lamp. In this configuration, sky background observations were not possible. Thus, it is crucial to identify and flag spectra contaminated by moonlight to ensure the accuracy of the RV analysis. By considering the phase and position of the Moon at the time of observation (see Appendix \ref{moon} for more details), we identified 31 spectra contaminated by the Moon. Consequently, we conservatively excluded these spectra from our analysis. We note that the inclusion or exclusion of these data points has no significant effect on our final results. Furthermore, we removed 1 data point that was identified as a 3$\sigma$ RV outlier, along with an additional 17 measurements due to low signal-to-noise ratio (S/N $<$ 50) and invalid calibration lamp flux. Consequently, a total of 50 data points were excluded from the SOPHIE+ measurements, and our final analysis incorporated 379 SOPHIE+ spectra, ensuring the reliability of our dataset for subsequent analysis.

SOPHIE experiences a long-term instrumental variation \citep{courcol2015sophie}, which is tracked by observing so-called "constant stars" every night. To account for this variation, following \cite{courcol2015sophie}, we constructed a master time series using the RVs of these constant stars, which we then subtracted from the HD88986 data. A detailed description of our update on the construction of the RV master constant time series can be found in Appendix \ref{master_cons}. The mean uncertainty of our final RV SOPHIE+ data is 1.2 m/s, with a root mean square (RMS) of 15.30 m/s. The final SOPHIE dataset is provided in Table \ref{tab:rvs sophie}.

\begin{table}
\caption{\label{stellar parameters} Stellar properties of HD88986}
\centering
\resizebox{\columnwidth}{!}{%
\begin{tabular}{lll}
\hline
\hline
Identifiers &  & \\
 & TIC 1042868 & \\
 & HD 88986 & \\
 & HIP 50316 & \\
 & Gaia DR3 741184091114529792  & \\
 & 2MASS J10162809+2840571    &  \\
 \hline
\hline
Parameter & HD88986 & References\\
\hline

 & Astrometric properties &\\
  & &\\
Parallax (mas) & $29.9864 \pm 0.0205$ & Gaia DR3\\
AEN $\varepsilon$ (mas)& 0.135& Gaia EDR3\\
Significance of $\varepsilon$ &20.6& Gaia EDR3\\
Distance (pc) & $33.37\pm0.04$ & Gaia DR3\\
$\alpha$ (h m s) & $10:16:28.08$ & Gaia DR3\\ 
$\delta$ (d m s) & $28:40:56.94$ & Gaia DR3 \\ 
  & &\\
 & Photometric properties &\\
   & &\\
B-V & $0.635\pm0.006$ & HIP\\
V(mag) & $6.47\pm 0.01$  & HIP \\
Gaia(mag) & $6.315\pm 0.003$ & Gaia DR3\\
Gaia$_{BP}$ (mag)&  $6.628\pm0.003$& Gaia DR3 \\
Gaia$_{RP}$ (mag)& $5.822\pm 0.004$& Gaia DR3 \\
TESS(mag) & $5.8706\pm 0.0061$ & TESS \\
J(mag) & $5.247\pm 0.024$ & 2MASS \\
H(mag) & $4.946\pm 0.023$ & 2MASS \\
K$_{s}$(mag) & $4.884\pm 0.020$ & 2MASS \\
W$_{1}$ (mag) & $4.895\pm 0.239$ & WISE \\
W$_{2}$(mag) & $4.762\pm 0.085$ & WISE \\
W$_{3}$(mag) & $4.933\pm 0.014$ & WISE \\
W$_{4}$(mag) & $4.873\pm 0.029$ & WISE \\
  & &\\
 & Spectroscopic properties &\\
   & &\\
Spectral type & G2V & MK classification\\
$\log~g$ (cm s$^{-2}$)& $4.16\pm0.03$& Sect \ref{Stellar parameters}\\
$\xi_{t}$ (km s$^{-1})$  &$1.11\pm0.02$& Sect \ref{Stellar parameters}\\
$\log (R'_{HK})$ & $-5.04\pm0.10$ & Sect \ref{stellar activity}\\
v$\sin i$ (km s$^{-1}$)& $3.3\pm1.0$& SOPHIE DRS\\
$[Fe/H]$ & $0.06\pm0.02$& Sect \ref{Stellar parameters}\\
T$_{\rm eff}$ (K) &$5861\pm17$& Sect \ref{Stellar parameters} \\
  & &\\ 
  & Bulk properties &\\
  & &\\
Mass (M$_{\odot}$) &$1.25\pm0.05^{*}$& SED Sect \ref{Stellar parameters}\\
Radius(R$_{\odot}$) & $1.543\pm0.010^{*}$ & SED Sect \ref{Stellar parameters}\\
$P_{rot}$(d) & $25^{+8}_{-6}$ & Sect \ref{stellar activity}\\
Age (Gyr)& $7.9\pm1.3^{*}$& Sect \ref{Stellar parameters}\\
\hline
\end{tabular}
}
\tablefoot{$^{*}$ adopted the systematic uncertainty floor as suggested by \cite{tayar2022guide} throughout this study.}
\end{table}

\begin{figure}
\centering
\includegraphics[width=0.45\textwidth]{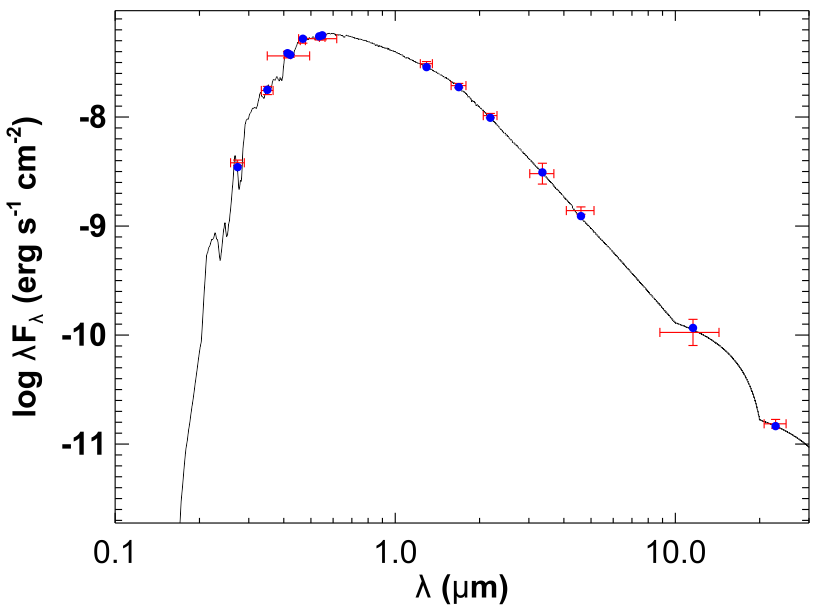}
\caption{Spectral energy distribution of HD88986. Red symbols represent the observed photometric measurements, whereas the horizontal bars represent the effective width of the passband. Blue symbols are the model fluxes from the best-fit Kurucz atmosphere model (black).}
\label{fig:sed}
\end{figure}

\subsection{High-resolution spectroscopy with HIRES}

The star was observed using the HIRES spectrograph from 1996 December 2$^{nd}$ to 2014 January 19$^{th}$, spanning 17 years during which 51 high-resolution spectra were obtained (see Fig. \ref{RV}). Detailed information regarding data reduction and observation can be found in \cite{butler2017lces}. HIRES data experience a small jump of $1.5 \pm 0.1$ m/s resulting from a CCD change in August 2004, as well as a long-term drift ($\lesssim 1$ m/s) and a small intra-night drift, as identified by \cite{tal2019correcting}. To account for these effects, we utilized the systematically corrected HIRES data obtained from the Vizier catalog access tool \footnote{\url{https://vizier.u-strasbg.fr/viz-bin/VizieR}} following the methodology outlined in \cite{tal2019correcting}. The mean uncertainty of the HIRES data is 1.2 m/s, with a RMS of 11.1 m/s. To address any residual offsets from the CCD change in the HIRES data, we fit an offset between the data obtained before (HIRES) and after (HIRES+) the CCD change. The HIRES RVs are presented in Table \ref{tab:rvs hires}. We note that our moon contamination criteria (see Appendix \ref{moon}) were not applied to HIRES data since these data are only utilized to investigate the origin of the long-term curvature and are not intended for high-precision detection of the sub-Neptune.

\subsection{High-resolution spectroscopy with ELODIE}

ELODIE was a high-resolution spectrograph, mounted on the 1.93-m telescope at OHP, which was in particular used to discover the first exoplanet in 1995 \citep{mayor1995jupiter}. HD88986 was observed by ELODIE from 1997 February 28 to 2004 January 29, gathering 31 high-resolution spectra (see Fig. \ref{RV}). The K0 numerical mask is used to extract the RVs \citep{baranne1996elodie}. The exposure time varied from 600 to 900 s, resulting in a mean uncertainty of 9.0 m s$^{-1}$ and RMS of 13.0 m s$^{-1}$, which is close to the intrinsic stability of ELODIE. We note that 3 data points were removed due to their low S/N (< 50). They are listed in Table \ref{tab:rvs elodie}. Similar to the HIRES data, the moon contamination criteria were not applied, as this data was not utilized for the detection of the sub-Neptune but to investigate the origin of the long-term curvature.

\section{Stellar parameters}
\label{Stellar parameters}
HD88986 is a G2V star with a G magnitude of 6.3. To obtain the stellar atmospheric parameters, we co-added all SOPHIE+ spectra (428) after correcting for the RV variation of the star, barycentric Earth radial velocity, and background pollution due to the calibration lamp \citep{heidari:tel-04043297, hobson2019exoplanet}. This results in a high S/N per pixel spectrum of 3174 at 550 nm. Then we calculated the effective temperature ($T_{\rm eff}$), metallicity ([Fe/H]), and surface gravity ($\log g$), using the procedure described in \cite{santos2013sweet} and \cite{sousa2018sweet}. The resulting $T_{\rm eff}$, [Fe/H], and $\log g$ together with other stellar parameters of HD88986, are presented in Table \ref{stellar parameters}.

We performed an analysis of the broadband spectral energy distribution (SED) of the star together with the {\it Gaia\/} EDR3 parallax \citep[with no systematic offset applied; see, e.g.,][]{stassun2021parallax}, in order to determine an empirical measurement of the stellar radius, following the procedures described in \citet{stassun2016eclipsing,stassun2017accurate,stassun2018evidence}. We obtained the B$_T$ V$_T$ magnitudes from {\it Tycho-2} \citep{hog2001tycho}, the JHK$_S$ magnitudes from {\it 2MASS} \citep{cutri20032mass}, the W1--W4 magnitudes from {\it WISE} \citep{wright2010wide}, the $uvby$ Str\"omgren magnitudes from \citet{paunzen2015new}, and the G G$_{\rm BP}$ G$_{\rm RP}$ magnitudes from {\it Gaia} \citep{Gaia2016, Gaia2021}. We also used the UV measurement at 274~nm from the {\it TD1} UV satellite \citep{thompson1978catalogue}. Together, the available photometry spans the full stellar SED over the wavelength range 0.2--22~$\mu$m (see Figure~\ref{fig:sed}). Then, we performed a fit using Kurucz's stellar atmosphere models with the $T_{\rm eff}$, $\log g$, and [Fe/H] adopted from the spectroscopic analysis. The remaining free parameter is the extinction $A_V$, which we fixed at zero due to the proximity of the system to Earth.

The resulting fit (Figure~\ref{fig:sed}) has a reduced $\chi^2$ of 1.4. Integrating the model SED gives the bolometric flux at Earth, $F_{\rm bol} = 7.28 \pm 0.17 \times 10^{-8}$ erg~s$^{-1}$~cm$^{-2}$. Taking the $F_{\rm bol}$ and $T_{\rm eff}$ together with the {\it Gaia\/} parallax, gives the stellar radius, $R_\star = 1.543 \pm 0.010$~R$_\odot$, placing the star within the subgiant range \citep[1.5-3~R$\odot$;][]{huber2017asteroseismology,berger2018revised}. In addition, we can estimate the stellar mass from the empirical relations of \citet{torres2010accurate}, giving $M_\star = 1.19 \pm 0.07$~M$_\odot$, which is consistent with the value of $1.25 \pm 0.05$~M$_\odot$ determined empirically via $R_\star$ and $\log g$. We acknowledge the possibility that our formal error budget for radius, mass, and stellar age could be underestimated, as suggested by \cite{tayar2022guide}. For stars such as HD88986, the systematic uncertainty floor could rather be up to $\approx$ 4.2\% for radius, $\approx$ 5\% for mass, and $\approx$ 20\% for age. Hence, we conservatively adopt these relative uncertainties, reported in brackets in Table \ref{stellar parameters}, to obtain more realistic stellar parameter errors throughout this study.

 Finally, we can use the observed  $\log R'_{\rm HK}$ (see Sect. \ref{stellar activity}) to estimate the stellar age via empirical activity-age relations. We obtain an age of $\tau_\star = 7.9 \pm 1.3$~Gyr via the empirical relations of \citet{mamajek2008improved}.

\section{Stellar rotation and activity}
\label{stellar activity}

To study the host star's activity and rotational period, we used $\log R'_{\rm HK}$, S-index, Na index, and CCF bisector measurements from the SOPHIE+ spectrograph, along with the S-index values from the HIRES and HIRES+ spectrographs. The bisector spans are derived from the SOPHIE DRS using the method described by \cite{queloz2001no}. We extracted the Na index as introduced in \cite{da2011long}. The HIRES and HIRES+ S-index values were acquired from \cite{butler2017lces}. For the extraction of $\log R'_{\rm HK}$ and S-index from the SOPHIE+ spectra, we followed the procedure outlined by \cite{noyes1984rotation} and \cite{boisse2010sophie}, respectively. The key step before deriving the $\log R'_{\rm HK}$ and S-index is subtracting the background contamination light from the Th-Ar or FP calibration lamp from the stellar spectra. To correct this, we used the direct measurement method (see Appendix \ref{activity_back}).

To estimate the rotational period of the star, we summed 176 HD88986 SOPHIE+ spectra which fulfilled two criteria. First, the spectra with S/N $>$ 50 in the first (bluest, $\lambda$ $\sim$ 3955 A$^{^\circ}$) order of spectra where CaII H\&K lines are located. Second, the spectra with minimal contamination due to the background light. This led to the value of $\log R'_{\rm HK} =-5.04\pm 0.10$. Furthermore, we investigated potential changes in magnetic activity over an approximately 11-year SOPHIE+ observation span. For this purpose, the dataset was divided into three distinct observational seasons spanning from 2012 to 2023: 2012-2016, 2016-2019, and 2019-2023. We observed a gradual increase in the $\log R'_{\rm HK}$ parameter, with values transitioning from -5.12 $\pm$0.10 during the initial subset to -5.05$\pm$0.10 in the second subset, and ultimately stabilizing at -5.00$\pm$0.10 in the final subset. All these values are in good agreement with the value of $\log R'_{\rm HK} =-5.22$ and $\log R'_{\rm HK} =-5.07$ reported by \cite{radick2018patterns} and \cite{hall2007activity}, respectively. Finally, we estimated a rotational period of 25$^{+8}_{-6}$ d following \cite{noyes1984rotation}, a value consistent with the \citet{mamajek2008improved} recipe, which yields 26.3$\pm$ 3.1 d. 

\begin{figure}
\centering
\includegraphics[width=0.45\textwidth]{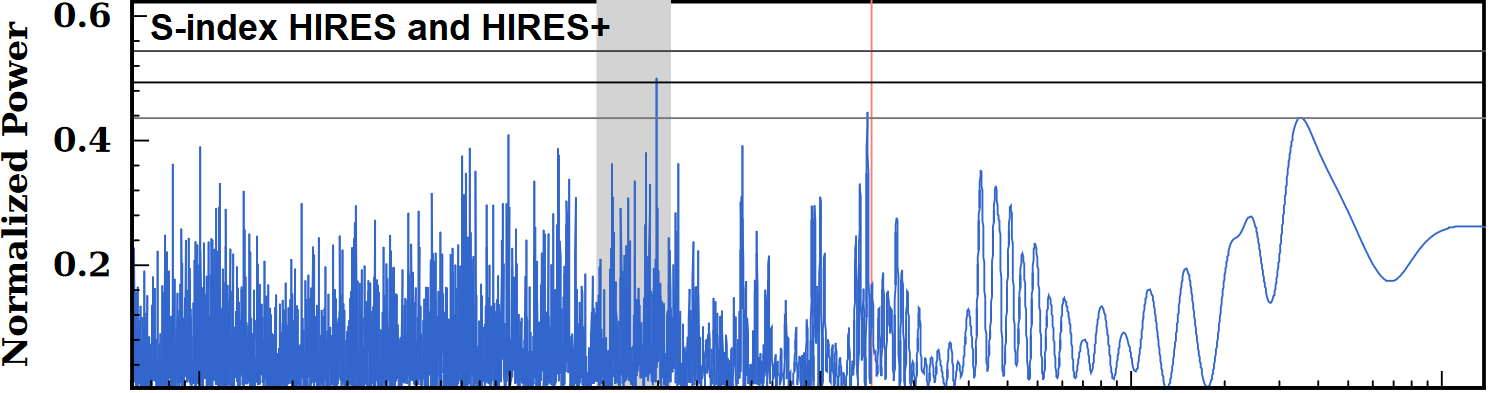}

\includegraphics[width=0.45\textwidth]{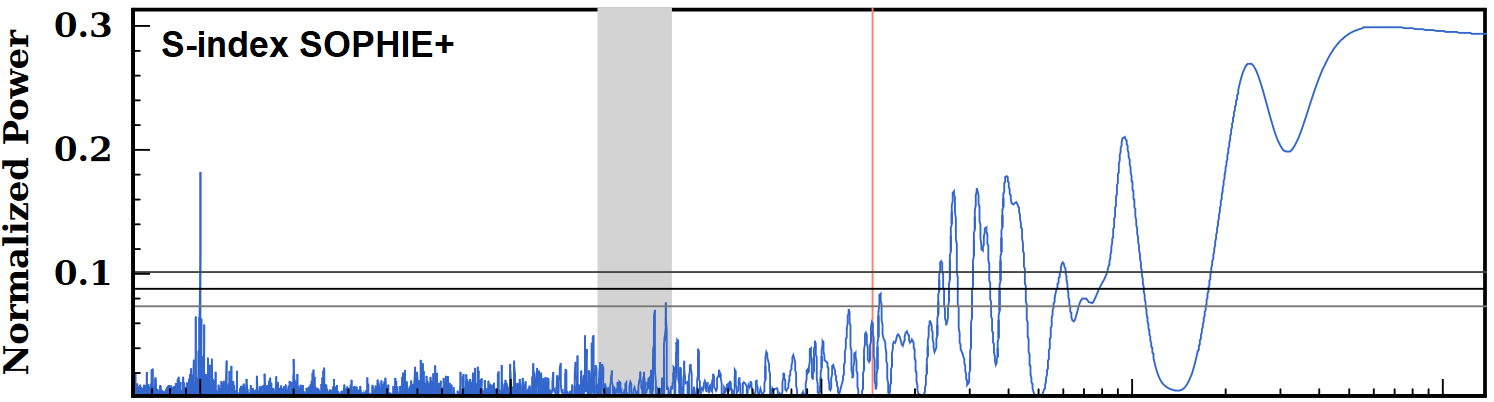}

\includegraphics[width=0.45\textwidth]{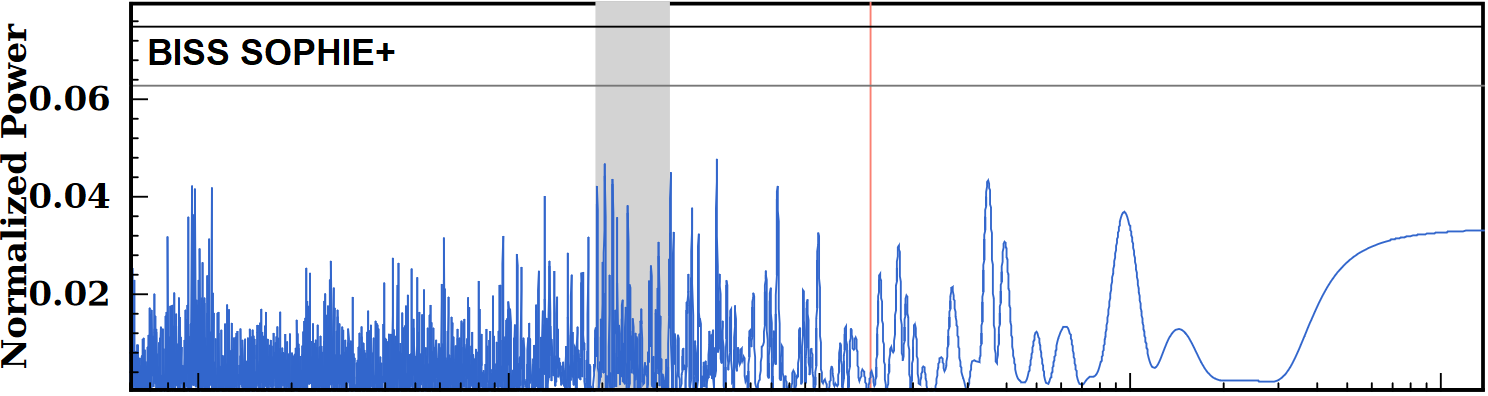}

\includegraphics[width=0.45\textwidth]{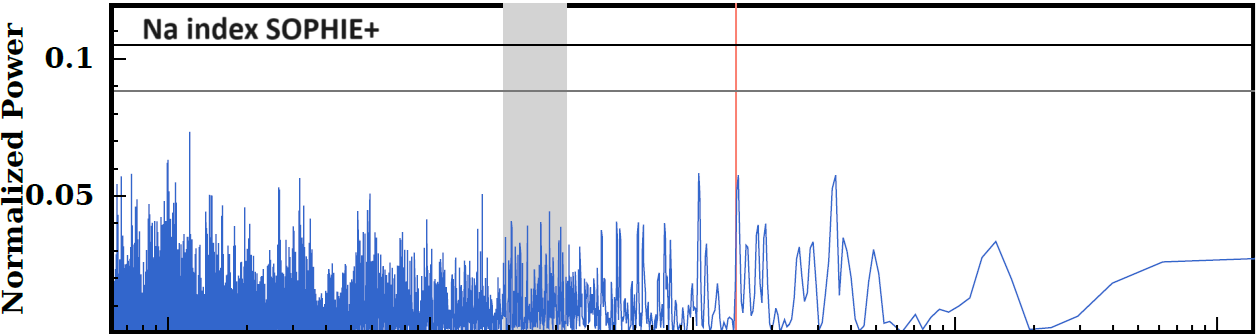}

\includegraphics[width=0.45\textwidth]{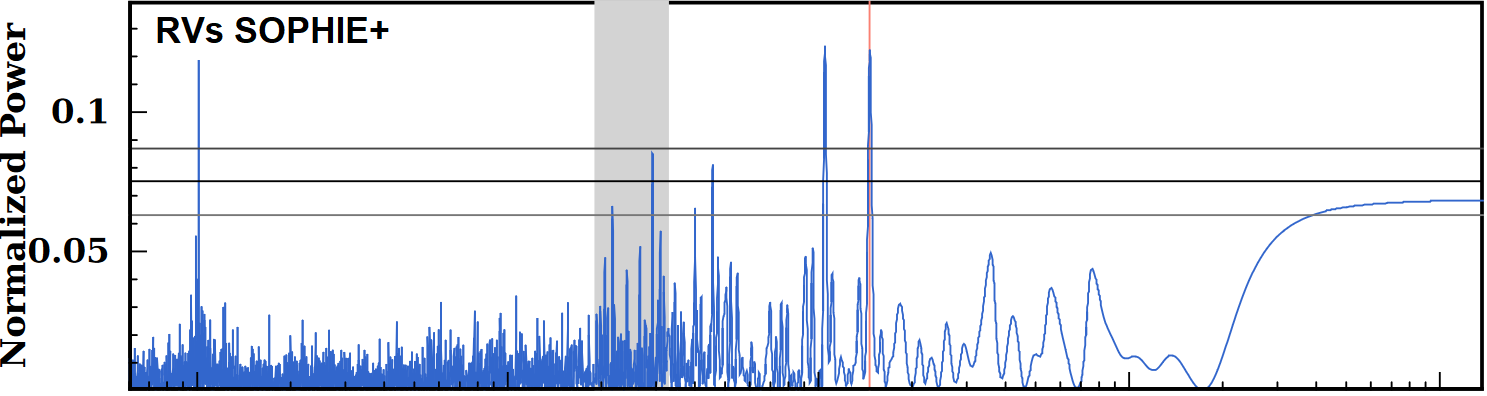}

\includegraphics[width=0.45\textwidth]{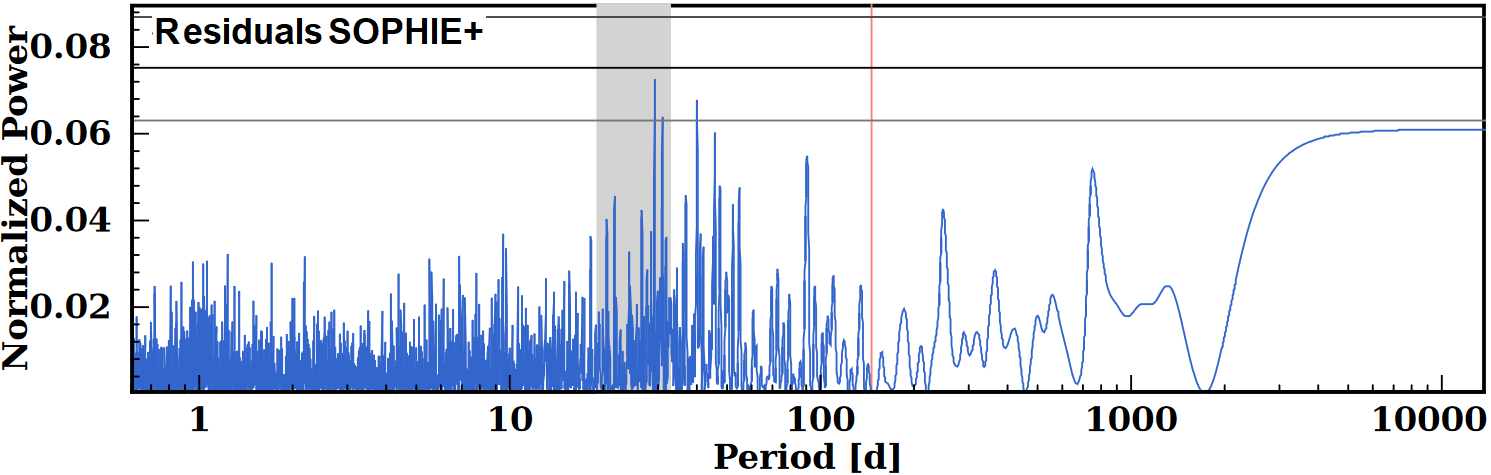}
\caption{Periodogram of RVs and activity indicators of HD88986. From \emph{top} to \emph{bottom}: HIRES and HIRES+ S-index, SOPHIE+ S-index, bisector, RVs, and residuals of RVs after Keplerian fit on the 146.1 d. The vertical red line illustrates the planet candidates on 146.1 d, which have no corresponding peak in activity indicators. The vertical gray strip marks the estimated rotational period of the star. Also, the horizontal lines show the FAP level of 10\%, 1\%, and 0.1 \%, respectively \citep{2008MNRAS.385.1279B}}.
\label{fig:periodogram}
\end{figure} 

We searched for rotational modulation in the Simple Aperture Photometric \citep[SAP, see][]{twicken:PA2010SPIE,morris:PA2020KDPH} TESS data (see Sect. \ref{photometry_extrac} and \ref{tess48} for details of the observations) using the Lomb-Scargle periodogram \citep{lomb1976least,scargle1982studies,vanderplas2018understanding}. No convincing signal was found. This was expected, given the star's quiet nature and also the limited $\sim$ 27 d observation window of TESS, which made the clear visibility of a 25 d signal difficult. We also note that additional photometric data from the T8 automatic photoelectric telescope (APT) did not show any photometric variability related to the stellar rotational period (see Sect. \ref{apt}).

To constrain the stellar rotation, we conducted an analysis of the activity indicator periodogram using the Data and Analysis Center for Exoplanets \citep[DACE,][]{delisle2016analytical}\footnote{Available at \url{https://dace.unige.ch}}. For this analysis, we utilized the SOPHIE+ S-index to ensure comparability with the S-index measurements obtained from HIRES and HIRES+ instruments. We excluded a total of 68 S-index data points due to their dependency on S/N ($<$ 30 in SOPHIE+ order 1, $\lambda$ $\sim$ 3955 A$^{^\circ}$) and significant contamination caused by the calibration lamp. Additionally, one data point was excluded due to its identification as a 5$\sigma$ outlier. In a similar vein, 102 data points from the Na index were omitted due to their reliance on S/N ($<$70 in SOPHIE+ order 30, $\lambda$ $\sim$ 5931 A$^{^\circ}$) and their susceptibility to contamination by the telluric lines. Fig. \ref{fig:periodogram} displays the periodogram of activity indicators. The HIRES and HIRES+ S-index measurements, along with the SOPHIE+ S-index, reveal periodic signals at 29.6 d \citep[false alarm probability (FAP) < 1\%,][]{2008MNRAS.385.1279B} and 31.5 d (FAP < 10\%), respectively. These results are consistent with the estimated rotational period of the star at 25$^{+8}_{-6}$ d. Furthermore, there is an activity signal at 141.2 d in the periodogram of the HIRES and HIRES+ S-index, which will be further discussed in the following section.

\section{Detection of the sub-Neptune HD88986 b in the SOPHIE+ RVs}
\label{rv-detection}
\subsection{RV periodogram}

To perform our RV periodogram analysis, we only used SOPHIE+ RV data because it contains an extensive number of data points (378), a long baseline, a higher RV accuracy, and superior sampling compared to other instruments. The analysis employed the default noise model in DACE, assuming an additional Gaussian white noise with nominal error bars of 1.5 m/s on the SOPHIE+ RVs. After removing the long-term curvature using a second-order polynomial model, the periodogram of SOPHIE+ RVs showed significant peaks at 104.8 and 146.1 d, falling below the analytical FAP \citep{2008MNRAS.385.1279B} of 0.1\% (see Fig. \ref{fig:periodogram}, fourth panel). These two signals are yearly aliases of one another.

\begin{figure}[h!]
\centering
\includegraphics[width=0.45\textwidth]{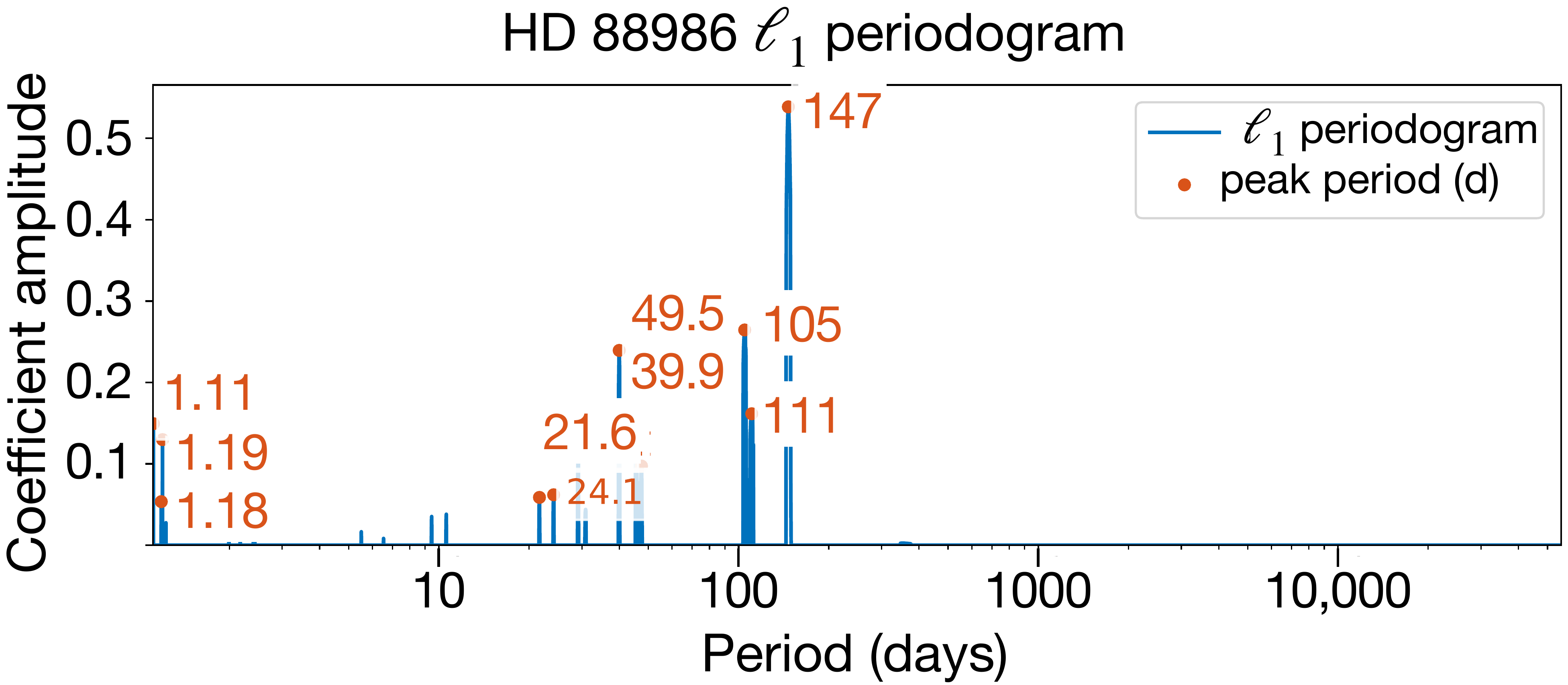}
\caption{$\ell_1$ periodogram of SOPHIE+ data. The identified periods are shown in red. }
\label{l1}
\end{figure}

To determine the favored alias between two signals, we compute the $\ell_1$ periodogram. This one takes in a frequency grid and an assumed covariance matrix of the noise as input. It aims to find a representation of the RV time series as a sum of a small number of sinusoids whose frequencies are in the input grid. It outputs a figure that has a similar aspect as a regular periodogram but with fewer peaks due to aliasing. The peaks can be assigned a FAP, whose interpretation is close to the FAP of a regular periodogram peak. The signals found to be statistically significant might vary from one noise model to another. To explore this aspect, as in \cite{hara2020}, we considered several candidate noise models based on the periodicities found in the ancillary indicators. We tried 1200 noise models, all Gaussian, such that the covariance is the sum of a white noise term of amplitude $\sigma_W$, a red noise term with Gaussian decay of amplitude  $\sigma_R$ and timescale $\tau_R$, and a quasi-periodic component \citep{haywood2014planets} with amplitude $\sigma_{QP}$, timescale $\tau_{QP}$ and period $P^\star$ and harmonic complexity equal to 1. We tried all combinations with $\sigma_W, \sigma_R, \sigma_{QP}=0,0.3,0.6,0.9,1.2,1.5$ m/s, $\tau_R=0,3,6$ day, $P^\star=29$ d and $\tau_{QP}=20,40,60$ d or $P^\star=40$ d and $\tau_{QP}=20,50,80$ d. We ranked the models with cross-validation as well as Laplace approximation of the Bayesian evidence. We find that for the 20\% highest ranked models, a peak with a period between 145 and 149 d consistently appears. The model with the highest Laplace approximation of the evidence is obtained with $\sigma_W=\sigma_{QP} = 1.2$ m/s, $\sigma_R = 1.5$ m/s, $\tau_R = 0$ d, $\tau_{QP}=20$ d. The corresponding $\ell_1$ periodogram is shown in Fig. \ref{l1}. The highest peak appears at 146.5 d, which is compatible with the 146.1 d signal given the frequency resolution set by the timespan of observations. This signal presents a FAP of 2.45$\times$ 10$^{-6}$, which is clearly statistically significant. We, therefore, conclude that the true signal is at 146 d, while the signal at 104.8 d represents its yearly aliases. Other signals appearing on the $\ell_1$ periodogram are not statistically significant.

Subsequently, we performed a circular Keplerian fit (see Sect. \ref{rv_several} and \ref{Detection}) to remove the signal at 146.1 d and investigated the resulting RV residuals. The periodogram displayed two signals at 29.2 d and 40.0 d, with FAP values below 10\% (see Fig. \ref{fig:periodogram}, bottom panel).

\begin{figure*}
        \noindent
        \centering
        \begin{tikzpicture}
        \path (0,0) node[above right]{\includegraphics[scale=0.46]{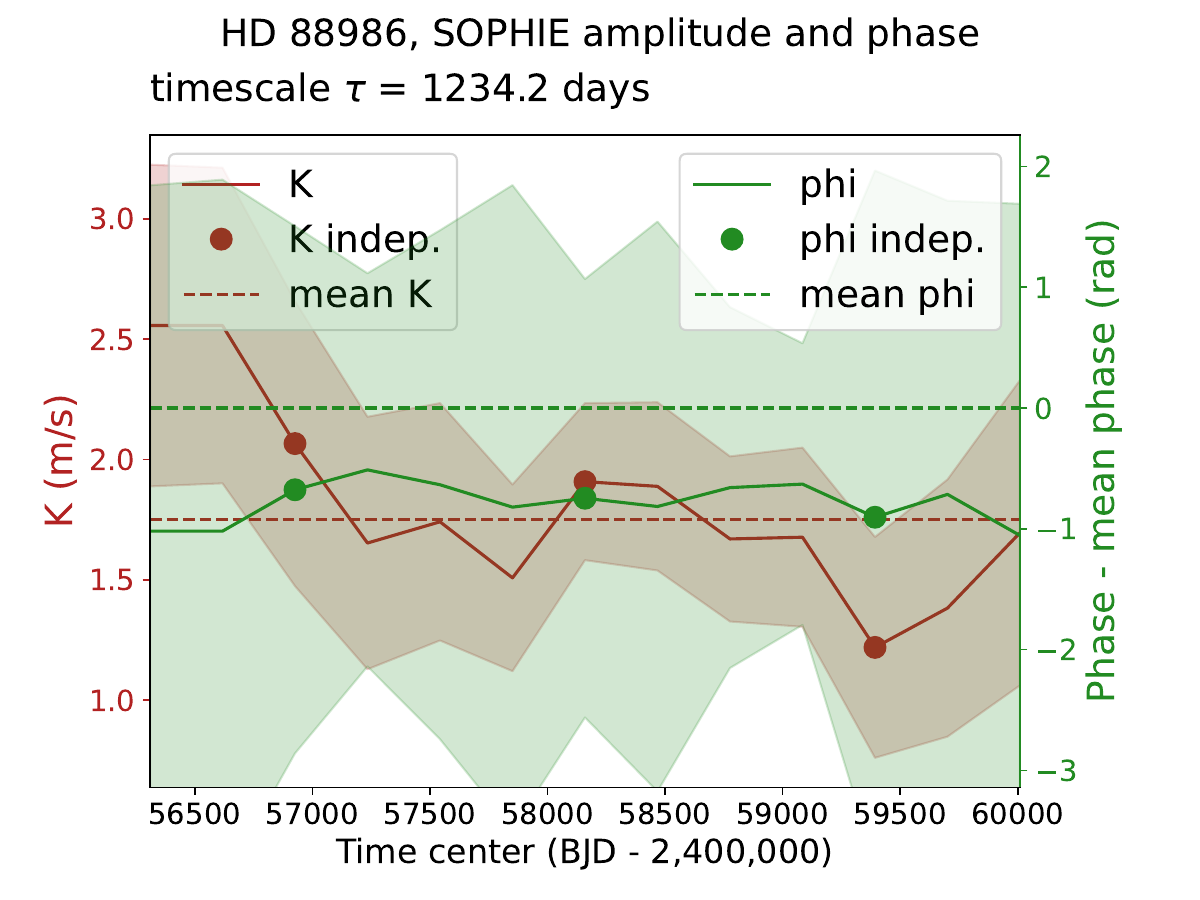}};
        \path (1.2,6.6) node[above right]{\large(a)};
        \path (9.5,0) node[above right]{\includegraphics[scale=0.46]{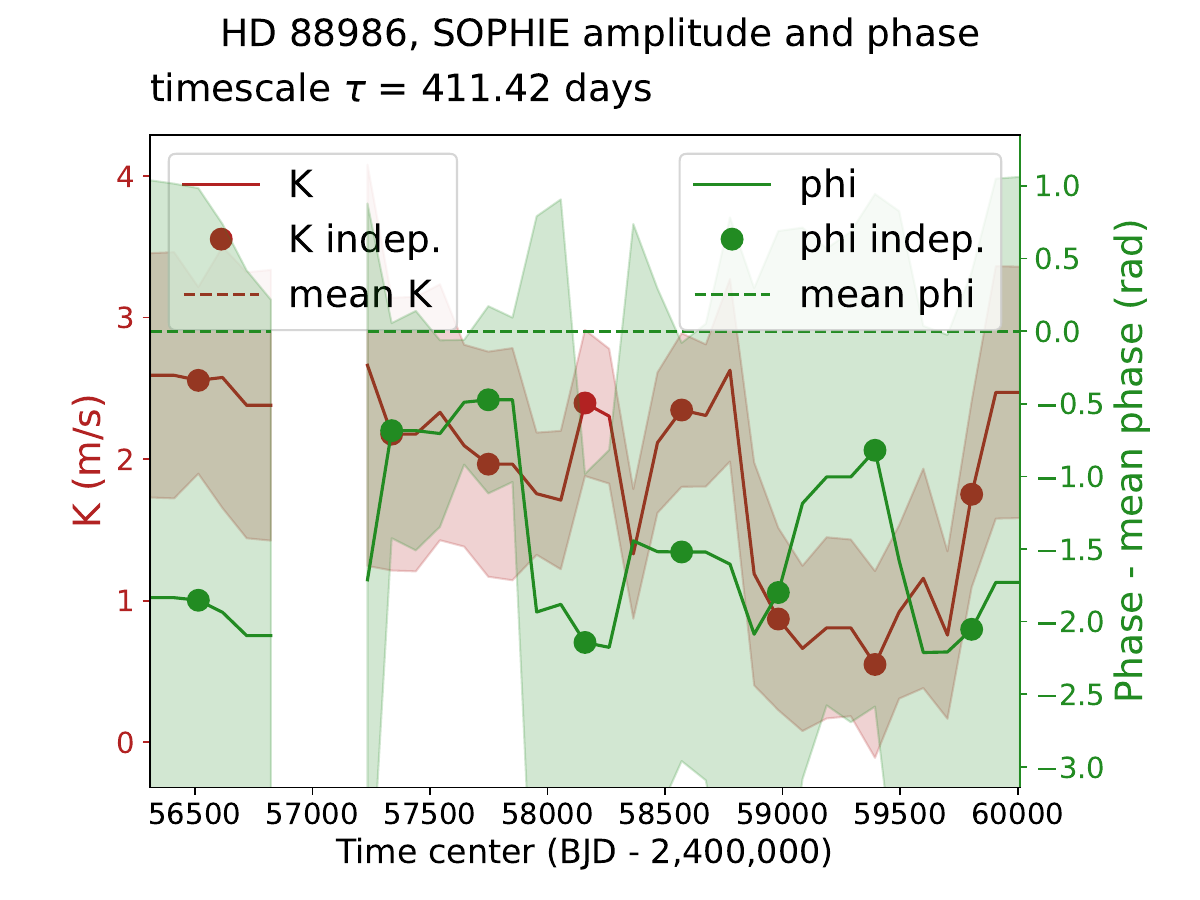}};
\path (10,6.6) node[above right]{\large(b)};
        \end{tikzpicture}
 \caption{Amplitude (red) and phase (green) of a 146.1 d signal as a function of time for different sizes of time windows. Solid lines correspond to estimate and shaded areas to $\pm $ 1 $\sigma$ uncertainties.  Denoting by $T_\mathrm{obs}$ the total time span of observations, a) and b) are obtained with windows of size $T_\mathrm{obs}/3 = $ 1234.2 d and $T_\mathrm{obs}/9$ = 411.42 d, respectively.}
 \label{fig:amplitude_phase}
\end{figure*}

To conclude our analysis, we checked that the 146 d signal has a constant phase and amplitude following the methodology of \cite{Hara2022_periodicity}. It simply consists of computing g the phase and amplitude of a signal at a given period with a moving time window. To perform this analysis, it is crucial to have realistic error bars. As visible in Fig. \ref{fig:periodogram}, the SOPHIE+ S-index exhibits low-frequency variations that are most likely due to a magnetic cycle. We expect a higher dispersion of RVs when S-index values are high \citep{borgniet2015, meunier2021, haradelisle2023}. Following \cite{diaz2016}, we added a white noise jitter term, as well as a jitter scaled with the value of the $\log R'_{\mathrm{HK}}$, and fit those along with a polynomial line and the 146 d signal. We used those values to compute the amplitude and phase as a function of time, shown in Fig. \ref{fig:amplitude_phase}. We here consider windows of size $T_\mathrm{obs}/3 = $ 1211 days and $T_\mathrm{obs}/9 = 411$ days where $T_\mathrm{obs}$ is the total time span of observations. The most notable feature is the drop in amplitude in Fig. \ref{fig:amplitude_phase}.b at BJD 2459000 - 2460000. We performed the quantitative analysis presented in \cite{Hara2022_periodicity} and determined that the phase and amplitudes are consistent with being constant. The drop in amplitude is likely due to the fact that the epoch BJD 2459000 - 2460000 corresponds to an activity maximum, and it is possible that the activity pattern changes and masks the planetary signal in the corresponding observational seasons.

To summarize, both periodograms exhibit a strong periodicity of about 146 d in RVs, which falls within a different period than the estimated star's rotation period (25$^{+8}_{-6}$ d; see Sect. \ref{stellar activity}). Notably, the periodogram of the HIRES and HIRES+ S-index reveals an activity signal at 141.2 d, which differs from the RV signal by five days and has a relatively low strength (FAP $\sim$10\%). None of the SOPHIE+ activity indicators showed a periodicity at 146.1 d. Moreover, no correlation was found between RV residuals after removing the trend and both S-index (Pearson's coefficient R= 0.03) and bisector (Pearson's coefficient R= 0.13). Furthermore, our analysis above demonstrates consistent phase and amplitude of the 146.1 d signal. Consequently, it is unlikely that RV periodicity at 146.1 d has an activity origin. Additionally, it is noteworthy that a survey encompassing all SOPHIE constant stars, as well as other stars observed by SOPHIE, did not reveal the same RV periodicity, confirming that the observed periodicity is not due to instrumental artifacts. Therefore, the RV signal at 146.1 d is likely to have a planetary origin. Throughout the rest of the paper, we attribute this periodicity to the planet HD88986 b.

Furthermore, given the two periodic signals of 31.5 d in the SOPHIE+ S-index and 29.6 d in the HIRES and HIRES+ S-index, along with the estimated star rotation period (25$^{+8}_{-6}$ d), the RV residuals signal at 29.2 d is likely the result of stellar rotational modulation. We take into account this signal in our analysis for the rest of the paper, testing three different methods to model it (see below in Sect. \ref{rv_several}). Finally, since the RV residuals signal at 40.0 d
is statistically insignificant, further observations are required to
determine whether this signal has an astrophysical origin or is
simply noise.

 We note that among the RVs presented in this paper, only those obtained with SOPHIE+ allow the low-amplitude signal of planet HD88986 b to be detected. This is due both to the large number of measurements and to their high accuracy. ELODIE, SOPHIE, HIRES, and HIRES+ RVs do not assist here in the discovery of that planet. Therefore, to avoid any offsets or potential systematics among instruments, we only use the SOPHIE+ data in the fits of HD88986 b presented below in the continuation of this section and Sect. \ref{joint2}. The other RVs datasets are only included in section \ref{RV_astrometry} for constraining the outer companion.

\subsection{RV models}
\label{rv_several}
To adequately describe the data, we took into account several SOPHIE+ RV-only models and performed model comparisons. The RV analysis was carried out by \texttt{juliet} \citep{espinoza2019juliet}, which employs \emph{radvel} \citep{fulton2018radvel} to model RVs and \emph{george} \citep{ambikasaran2015fast} and \emph{celerite} \citep{foreman2017fast} to model potential activity effects on the data through Gaussian process regression (GPs). For each tested model, \texttt{juliet} computes the Bayesian log evidence (ln Z). If the Bayesian log-evidence difference ($\Delta$lnZ) between a model and another exceeds two, it is moderately favored over the latter; a difference greater than five indicates strong favorability \citep{trotta2008bayes}.  The models are indistinguishable when the difference in Bayesian log evidence is $\Delta$lnZ $\leq$ 2. In this case, the model with the fewest free parameters would be chosen. 

We defined our RV model as follows:

\begin{equation}
\label{eq1}
M(t) = K(t) + \epsilon (t) + \mu+ Q (t^{2})+ A(t),
\end{equation}

where K(t) is the Keplerian model and the $\epsilon$(t) $\sim$ $\mathcal{N} (0,\sigma(t)^2 + \sigma_{w}^2),$ is a normal distribution ($\mathcal{N}$) of white-gaussian noise where $\sigma$ (t) is the uncertainty of each RV point at time t, and $\sigma_{w}$ is a jitter term. Additionally, $\mu$ is a systematic RV offset of the instrument, and Q and A are defined as quadratic and linear terms, respectively, to model the long-term curvature. The model is tested for both scenarios of the eccentricity-free and circular orbit (Co). To explore the possible effect of stellar activity on the planet's parameters, we model the stellar activity in three different ways:

\begin{itemize}

\item a sinusoidal orbit.

\item an exponential GP kernel (EXP-GP) with the form of $k_{i,j}= \sigma^{2}_{GP} exp(-|t_{i}-t_{j}|/ T_{GP}) $, where $\sigma_{GP}$ is the amplitude of GP modulation, and  $T_{GP}$ presents the characteristic timescale \citep{ambikasaran2015fast}.

\item a quasi-periodic GP kernel \citep[QP-GP,][]{foreman2017fast} with the form of $\kappa_{i,j} =B_{GP}/(2+C_{GP})~ e^{(-|t_{i}-t_{j}|/ L_{GP})}(cos (2 \pi |t_{i}-t_{j}|/P_{rot; GP})+(1+C_{GP}))$, where $B_{GP}$ amplifies the kernel, $C_{GP}$ is a constant scaling term, $L_{GP}$ is a correlation time-scale component, and finally $P_{rot; GP}$ is the rotational modulation.

\begin{table}[h!]
\centering
\resizebox{\columnwidth}{!}{%
\begin{tabular}{lccccc}
\hline
Model & $\Delta$ lnZ & period (d) & K (m/s)& T$_{c}$ (BJD-2400000 d) \\
\hline
No planet&-97.0&----&----&----\\
No planet+EXP-GP&10.2&----&----&----\\
No planet+QP-GP&17.5&----&----&----\\
1Co& 0 & 146.0$\pm$0.5 & 1.7$\pm$0.2& 58897$\pm$3\\
1eccentricity-free& 1.6 & 145.8$^{+0.4}_{-0.3}$& 1.9$\pm$0.3 &58884$^{+8}_{-7}$\\
1Co+sinusoidal& 5.3& 146.2$\pm$0.4 & 1.6$\pm$ 0.2 & 58899$\pm$3\\
1Co+EXP-GP& 20.0&146.3$\pm$ 0.6&1.6$\pm$0.4 & 58897$^{+5}_{-6}$\\
1Co+QP-GP&22.3&146.3$\pm$0.6&1.7$\pm$0.4 &58897$\pm$6\\
\hline
\end{tabular}%
}
\caption {Different tested models on the SOPHIE+ RV-only data along with model comparisons with \texttt{juliet}. K refers to the RV semi-amplitude and T$_{c}$ is the time of mid-transit from RV fit. Additionally, in the model names, “Co” represents a circular orbit. The final choice model is indicated in bold.}
\label{rv-model}
\end{table}

\end{itemize}
In addition to the models above, we also executed three no-planet models wherein we assumed the absence of any planetary signal in the RVs (i.e., in equation \ref{eq1}: K(t)=0). Table \ref{rv-model} provides a summary of the results obtained from testing various models. Furthermore, Table \ref{prior_rv-only} presents the priors employed in the analysis, along with detailed descriptions of all parameters. Throughout this paper, we conducted \texttt{juliet} runs for each model using a consistent setup, employing a configuration with 3$\times$ number of free parameters as the number of walkers, executed 10000 steps per walker, and discarded the initial 3000 steps as burn-in.

\subsection{Detection of the sub-Neptune HD88986 b}
\label{Detection}
We set a uniform prior to the planetary period between 135 d and 155 d. For the mid-transit time, we applied a uniform prior defined by a time window of 146 d which is consistent with the planetary period duration. For the other parameters, we used fairly broad uniform priors. The results of the no-planet model as well as the circular and eccentric fits are shown in the Table \ref{rv-model}. We note that this table includes T$_{c}$, the derived center time of the inferior conjunction, as a transit of this planet is reported below in Sect. \ref{photometry_extrac}. The results of the circular and eccentric models are consistent and both are statistically significant compared to the no-planet model ($\Delta$lnZ $\geq$ 97) which confirms a clear detection of the planetary signal. The eccentricity-free model, however, exhibits bimodality on the periastron argument $\omega$. Given that the two models are statistically indistinguishable ($\Delta$lnZ $\leq$ 2), and their posterior distributions are consistent, we continued to model the HD88986 b planet with a circular orbit as it has fewer free parameters.

The last three lines of Table \ref{rv-model} present the results of the fits adopting the three different models for the stellar activity (see Sect. \ref{rv_several}). In the sinusoidal model, we employed a uniform prior for the activity period ranging from 25 to 35 d, as we detected a potential stellar signal at 29 d (see Sect. \ref{stellar activity}). Regarding the QP-GP hyperparameter, we imposed some constraints. On $P_{rot; GP}$, we used a Gaussian prior centered at 29 d with a standard deviation of 3 d. On the $C_{GP}$, we initially tried a wide range of Jeffreys priors from 10$^{-20}$ to 100. The posterior distribution of $C_{GP}$ reached the prior boundary at 10$^{-20}$, indicating that this parameter converged to zero. This result motivates us to set the $C_{GP}$ to 10$^{-20}$, which is consistent with zero. We note that the models with fixed/unfixed $C_{GP}$ are statistically indistinguishable ($\Delta$lnZ $\leq$ 2). Therefore, we fixed $C_{GP}$ and continued using the QP-GP model with three free parameters. A broad uniform prior is taken into account for all remaining parameters of different models, as presented in Table \ref{prior_rv-only}.

The results of the five different planet models reported in Table \ref{rv-model} are consistent. This strongly argues in favor of the detection of the planet HD88986 b with those parameters. Notably, all planet models accounting for the potential activity signal are strongly favored statistically ($\Delta$ ln Z$\geq$ 5) when compared to models that do not consider the stellar activity. Among these models, the planet models with simultaneous GP kernels (1Co+EXP-GP and 1Co+QP-GP) are strongly favored over the sinusoidal model (1Co+ sinusoidal). Additionally, the model incorporating the simultaneous QP-GP kernel displays moderate statistical favorability compared to the simultaneous EXP-GP kernel. We note that for the 1Co+sinusoidal model, we have a bimodality in the posterior distribution of the stellar rotation period at 29 d and a much smaller peak at about 31 d, which might be explained by differential stellar rotation. 

In addition to the different planetary models and one no-planet model explored above, we extended our analysis to encompass two GP-only (EXP-GP and QP-GP) models (see Table \ref{rv-model}). These models exhibit a higher statistical strength than those when considering the planet alone (1Co and 1eccentricity-free), suggesting that the presence of RV variabilities is primarily driven by dominant stellar activity signals rather than a planetary influence. However, a detailed examination outlined in Table \ref{rv-model} reveals a strong preference: models incorporating both the planet candida with the GP components are strongly favored ($\Delta$lnZ $\geq$ $\sim$ 5) over GP-only models. This robust preference, the consistency of planet parameters across various models, and the detection of a strong periodicity in our periodogram analysis (refer to Sect. \ref{rv-detection}), coupled with the consistent phase and amplitude of the signal (refer to Sect. \ref{rv-detection}), ensures the credibility of our detection.

Finally, for the sake of completeness, we incorporated all RV data presented in this paper, with the exception of ELODIE, owing to its significantly larger error bars (approximately 8 times larger) in comparison to other datasets. This additional model aimed to assess the consistency of the remaining RV data with the results obtained from SOPHIE+ RVs. To account for the long-term curvature in the data, we applied a two-Keplerian model. Our analysis yielded consistent results (Period= 146.9$\pm$0.3 d, K= 1.7$\pm$ 0.2 m/s, T${c}$= 58903 $\pm$ 3) when compared with other models listed in Table \ref{rv-model}. After removing the second Keplerian orbit, the RV RMS values were 5.5 m/s, 5.6 m/s, 6.2 m/s, and 3.2 m/s for HIRES, HIRES+, SOPHIE, and SOPHIE+ respectively. These RMS values, coupled with the limited number of available data points and sparse sampling of data from instruments other than SOPHIE+, indicate the challenges faced in detecting signals of such shallow amplitude in the other datasets beyond SOPHIE+. Additionally, this reaffirms the right selection of only SOPHIE+ data for presenting the detection and characterization of this low-mass planet, underscoring the accuracy of our results and ensuring the absence of additional instrumental offsets among the instruments. It is pertinent to highlight that satisfactory convergence was not achieved for certain parameters of the second Keplerian orbit, such as eccentricity and $\omega$, as the orbit of this massive companion remains incomplete (see Sect. \ref{Astrometry}).

To conclude, the results of all investigated models agree within the error bars. Accordingly, we chose the 1Co+ QP-GP model on SOPHIE+ data as our final choice because it is strongly favored statistically over the model without GP and has moderate favorability compared to the Co+ EXP-GP model.


\section{First transit detected in TESS sector 21 in February 2020}
\label{photometry_extrac}
 
\begin{figure*}
\includegraphics[width=\textwidth]{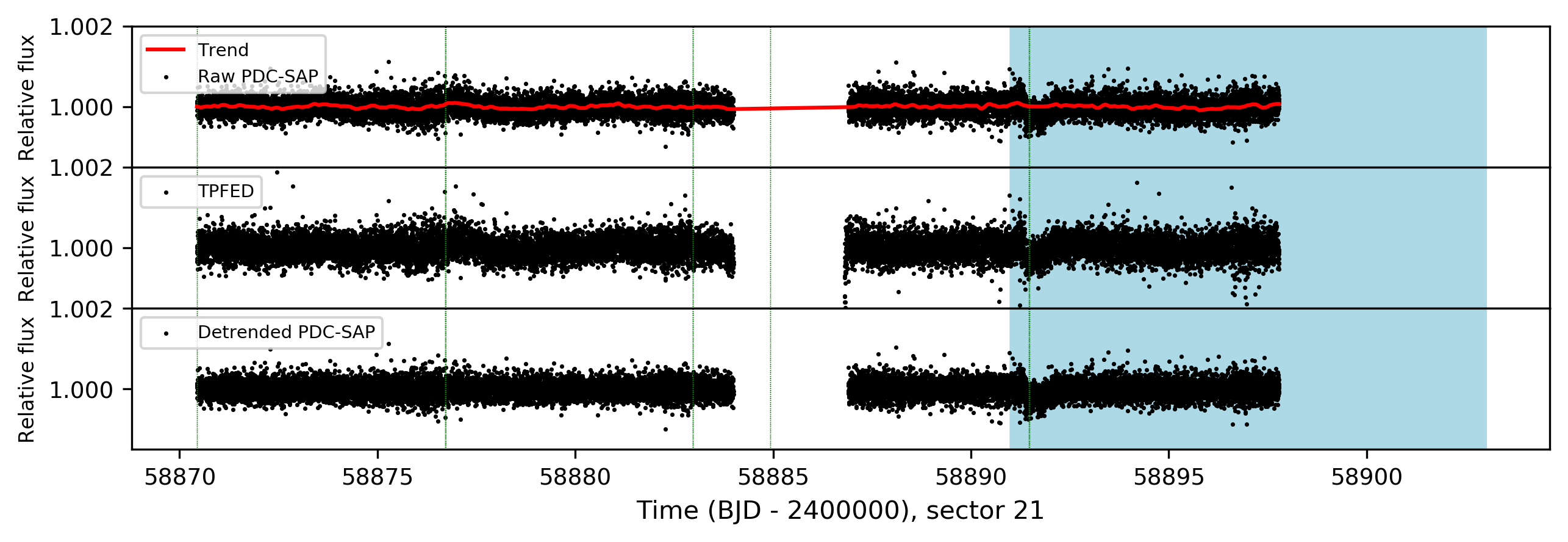}
\caption{TESS observation of HD88986 in sector 21 in 2020 February. \emph{Top}: The normalized TESS PDC-SAP light curve of sector 21 (black dots) along with the best-fit trend (red curve) to the data. The green vertical lines represent the times of the spacecraft's momentum dumps. \emph{Middle}: The normalized re-extracted light curve of sector 21 (black dots). See the text for more information. \emph{Bottom}: The final detrended light curve. The expected HD88986 b transit event from SOPHIE+ RVs (Sect. \ref{Detection}), with 1 sigma uncertainties, is highlighted in blue, and a single transit event is found in the TESS photometric data within} this region.
\label{lc-21}
\end{figure*}

HD88986 was observed in TESS sector 21\footnote{\url{https://heasarc.gsfc.nasa.gov/cgi-bin/tess/webtess/wtv.py}} with camera 1 in a 2-minute cadence from 2020 January 21 to 2020 February 18. The photometric data were produced by the Presearch Data Conditioning-Simple Aperture Photometry (PDC-SAP) pipeline \citep{Stumpe2012,smith2012kepler, stumpe2014multiscale}, provided by the Science Processing Operations Center \citep[SPOC;][]{jenkins2016tess} at NASA Ames Research Center. The normalized raw TESS photometric data are plotted in the top panel of Fig. \ref{lc-21}.

TESS data from sector 21 revealed a potential single transit candidate with T$_{c}$=2458891.6 (corresponding to 2020 February 12), a duration of about 16 h, and a depth of $\sim$ 220 ppm. Remarkably, this T$_{c}$ is in agreement, within uncertainty, with all the T$_{c}$ values predicted above from the RV fits of HD88986 b (Sect. \ref{Detection}, Table \ref{rv-model}). We note that, neither the TESS SPOC nor Quick Look pipelines (QLP) detected this transit signature, as they require at least two transits to generate a Threshold Crossing Event (TCE) that would be vetted by the TESS Science Office.

To investigate whether that potential single, shallow transit in sector 21 is a false positive scenario, we performed a test by calculating the mean in-transit and mean out-of-transit flux, along with the difference between them (see Fig. \ref{test_spoc}). This approach allows us to examine the offset between the different image positions and the actual position of the target star, providing valuable insights into false positive scenarios. While interpreting the difference images from saturated stars like HD88986 is particularly challenging, we observed that most of the energy in the transit feature is associated with the upper end of the bleed of the saturated pixels in the core of the stellar image. Therefore, it is likely that the transit feature is indeed associated with the host star.

Additionally, as Fig. \ref{lc-21} shows, a telescope reaction wheel momentum dump occurred during the transiting event \citep{fausnaugh2020tess}. The sector 21 light curve seems to have been only minimally impacted by the other momentum dumps occurring within this sector. This suggests the robustness of the applied momentum dump correction. However, to ensure that the signal is not the cause of a pipeline's imperfect momentum correction and produces an ingress-like feature, we re-extracted the light curve.

To do so, we use the TPFED/FFIED tool (hereafter TPFED) recently developed by \cite{Wilson2023} to conduct custom extractions of the TESS sector 21 data using the calibrated target pixel files (TPFs) with the default quality bitmask. In brief, we extracted target fluxes for a range of custom aperture masks created with radii of two to four pixels in steps of 0.1 pixels centered on the target. It should be noted that as the target does not fall in the exact center of a pixel increasing the aperture mask radius by 0.1 pixels can result in unique noncircular masks. All produced light curves were background-corrected after determining the sky level using custom background masks. We then detrended the data using two methods. Firstly, we conducted Principal Component Analyses (PCA) on the pixel values within our custom background masks to determine the scattered-light flux contribution to the light curves and then removed these systematics by using the five prime principal components as basis vectors in a linear model. Secondly, we corrected flux modulation due to spacecraft jitter by retrieving the co-trending basis vectors (CBVs), and two-second cadence engineering quaternion measurements for the cameras that observed HD88986. Following the method used in \citet{Delrez2021}, we computed the means and averages of the quaternions over the scientific observational cadences and subsequently used these vectors along with the CBVs to remove any flux trends. The final light curve is presented in the middle panel of Fig. \ref{lc-21}, and the potential single transit is clearly seen within the dataset.

In a quest to further explore possible sources of instrumental noise that
could impact the detection or shape of the potential, shallow single
transit, we explored alternative methodologies following Rapetti et al.~(in preparation). We used the adaptation of the technique Pixel Level Decorrelation~\citep[hereafter PLD;][]{Deming_2009,Luger_2016, Luger_2018} implemented in the \texttt{PLDCorrector} class of the community Python package \textsf{Lightkurve}\footnote{\url{https://docs.lightkurve.org}}. This method employs (i) a spline polynomial fit to describe stellar variability, (ii) PCA eigenmodes to model the background light, and (iii) the PLD technique to account for pointing and mechanical effects. As an additional approach, we employed a version of PDC as adapted in the \texttt{CBVCorrector} class of \textsf{Lightkurve}, utilizing the CBV technique that the PDC method of the SPOC pipeline employs (hereafter we refer to this corrector as CBV).

Before applying the PLD, we added the background flux and errors estimated by the TESS SPOC pipeline back onto the SAP light curve. Flux level, fraction, and crowding adjustments are applied to the corrected light curves. To automatically optimize the selection of parameter values for the correctors, we evaluate the resulting light curve using the Savitsky-Golay Combined Differential Photometric Precision (sgCDPP) proxy algorithm \citep{Gilliland_2011,Van_Cleve_2016} implemented in \textsf{Lightkurve}, for durations of 30, 60, 120, 160, and 200 minutes (see the legend of Fig. \ref{pld_rcq}). For a grid of corrector parameter values (for further details on the parameters and the grid, see Rapetti et al., in prep.), we calculated the harmonic mean (HM) of these quantities and selected the corrected light curve that minimizes the HM. In addition, the final sgCDPP metrics can be compared to those obtained for the SPOC PDC-SAP corrected light curve (see Fig. \ref{pld_rcq}). For this comparison, we also calculated the over-fitting metric implemented in \textsf{Lightkurve} (see Fig. \ref{pld_rcq}) to measure the broad-band power spectrum via a Lomb-Scargle periodogram and assess the level of introduced noise in the corrected light curves.

In addition to the methods detailed above for extracting data from the 2-minute TPFs, we extracted the full frame image (FFI) light curve from the TESS image using a strategy similar to \cite{2016ApJS..222...14V}. In particular, we created 20 different photometric apertures, 10 circular apertures, and 10 shaped like the TESS point spread function at the star location on the detector. We then calculated light curves from each aperture and corrected for systematics by performing a linear least-squares fit modeling on the light curve with time series the mean, standard deviation, kurtosis, and skew of the spacecraft quaternion measurements within each exposure \citep[e.g.,][]{2019ApJ...881L..19V}, the SPOC PDC CBVs, the background flux time series, and a basis spline to model slow variability. We performed the least-squares linear fit iteratively, removing outliers until convergence. Once the light curves were corrected for systematics, we corrected for dilution from other nearby stars and identified the one with the best photometric precision, which we used for our FFI analysis. The final resulting light curve is plotted in Fig. \ref{ffi}.

The TESS light curves of sector 21, produced using different approaches, all find a feature where the potential single, shallow transit was identified (see Fig.\ref{lc-21}, \ref{pld_rcq} and \ref{ffi}). Additionally, its mid-time always agrees with the predicted T$_{c}$ reported in Sect. \ref{rv-detection} for HD88986 b, whatever the chosen RV model is (see Table \ref{rv-model}). Furthermore, the feature exhibits fair similarity and robustness across various tested approaches for correcting the instrumental variation in the TESS data. The consistent detection of this feature through various methodologies, combined with its robust nature, supports the fact that this feature is unlikely to be attributed to instrumental effects. However, the precise shape of the transit is not clearly discernible in the SAP data (as seen in the first panel of Fig. \ref{pld_rcq}). This lack of clarity is expected due to a momentum dump occurring at the time of the transit, in particular for such a shallow transit. Similar to other instances of momentum dumps observed in the SAP data, this event introduces instrumental variations. Additionally, the SAP data is not fully corrected for the scattered light that might affect the exact shape of the potential single transit. Given the robustness of the feature, as well as the fact that the observed T$_{c}$ of the potential single transit agrees, within uncertainties, with the T$_{c}$ obtained from all the RV-only models (see Sect. \ref{rv-detection}), it is likely that this feature corresponds to a single transit attributed to HD88986 b.

We chose PDC-SAP data for our final analysis as all the newly extracted light curves are fairly consistent with the PDC-SAP light curve. We detrended the light curve using GP with an approximate Matern kernel introduced in \cite{foreman2017fast}. The reason for this choice is that there is no evidence of existing quasi-periodic oscillations in the TESS light curves. Before applying this method to more accurately measure the planet's radii, we masked out the in-transit and immediately surrounding data points. The final detrended light curve is shown in Fig. \ref{lc-21} bottom panel, which we use for the rest of this work. We note that no additional signal was found by performing a Transit Least Squares \citep{hippke2019optimized} algorithm on this data.

Additionally, we searched for any potential light contamination caused by neighboring stars on the light curve of HD88986. Within the aperture set by PDC-SAP, there is only one neighboring star. Because this star is one of the $\sim$ 1 million new Gaia DR3 sources, its light contamination is not corrected by SPOC. This star has a magnitude of G=12.3 ($\Delta$G$_{mag}$= 6 compared to HD88986) and is located $1.4"$ west of HD88986. There are no values for RP magnitude and renormalized unit weight error (RUWE), which is a measurement of the goodness of the star's astrometric solution. The poor behavior of this star could be due to blending with HD88986. Because the Gaia RP bandpass is comparable to the TESS bandpass, one can assess the level of contamination using the Gaia RP fluxes of these stars. In our case, however, the lack of RP flux data for the neighbor star prevents us from estimating the contamination effect.

\section{Joint analysis of SOPHIE+ and TESS sector 21}
\label{joint2}

We performed a joint modeling of TESS photometric data of sector 21 and SOPHIE+ RVs using \texttt{juliet}. \texttt{juliet}, in addition to the packages mentioned above for RV modeling (see Sect. \ref{rv_several}), employs batman \citep{batman} for transit fitting. Following the RV-only study in Sect. \ref{rv-detection}, we used the planet model with a simultaneous QP-GP model for modeling RVs in our joint modeling. Here, we tested both eccentricity free and zero models, as combining RV and transit data provides more constraints to fitting the orbital parameters including eccentricity and the argument of periastron. We employed the same priors for RV-related parameters, as detailed in Section \ref{rv-detection}, with the exception of T${c}$. For T${c}$, we set a Gaussian prior centered at 58891.6 with a standard deviation of 5 d. To parameterize the limb darkening coefficient for TESS photometry, we applied a linear law through a parameter of q. This choice is motivated by the limited number of informative in-transit data points when modeling single transit events \citep{2019MNRAS.489.3149S}. We note that applying a quadratic limb-darkening law had no effect on our final results. Additionally, the results of the spectral analysis provided in Sect. \ref{Stellar parameters} were used to set a Gaussian prior to the star density $\rho_{*}$.  Finally, we added a jitter term of $\sigma$ to TESS photometric data. Table \ref{prior_joint} shows all of the employed priors and the description of the parameters. 
\begin{table}
\centering
\caption {Median values and 68\% confidence interval of parameters for HD88986 b based on the joint analysis of the photometric and RV data by \texttt{juliet} (see Sect. \ref{joint2} and Fig. \ref{sec_48}).
}
\label{joint}
\resizebox{\columnwidth}{!}{%
\begin{tabular}{lc}
\hline
Parameter (unit)& Posterior HD88986 b\\
\hline
Stellar parameters:&\\
$\rho_{*}$(kg/m$^3$)&472$^{+38}_{-35}$\\
&\\
Planet parameters&\\
P$^{*}$ (d)& 146.05$^{+0.43}_{-0.40}$\\
T$_{c}$ (BJD-2400000)& 58891.690$\pm$0.003
\\
K (m/s)& 1.85$\pm$ 0.34
\\
e&0.24$\pm$0.05\\
$\omega$ ($^\circ$)&306$^{+10}_{-11}$ \\
R$_{p}$/R$_{*}$&0.0148$\pm$ 0.0004\\
b &0.21$^{+0.17}_{-0.14}$        
\\ 
a/R$_{*}$ &81.1$\pm$2.1\\ 
&\\
TESS instrumental parameters:&\\
M$_{TESS}$ (ppm)&-0.0000001$\pm$0.0000014\\
$\sigma_{\omega, TESS}$ (ppm)&143.4$\pm$ 1.3\\
q1& 0.27$^{+0.17}_{-0.14}$\\
&\\
SOPHIE+ instrumental parameters:&\\
$\sigma_{SOPHIE+}$ (m/s)&2.09$^{+0.16}_{-0.17}$\\
mu$_{SOPHIE+}$ (m/s)& 29090.6$\pm$ 0.4\\
&\\
Drift on SOPHIE+:&\\
A (m/s) &0.0165$\pm$0.0003\\ 
Q (m/s) &0.0000039$\pm$0.0000002\\
&\\ 
QP-GP on SOPHIE+:&\\
B$_{GP}$ (m/s)& 4.0$^{+1.1}_{-0.8}$\\
C$_{GP}$ (m/s)& 10$^{-20}$ (fixed)\\ 
P$_{rot}$ (d)& 30.0$\pm$2.0\\
L$_{GP}$ (d)&29$^{+37}_{-18}$\\
\hline
Derived planet parameters:&\\
a (au)&0.58$\pm$0.04\\
i ($^{\circ}$)&89.9$\pm$0.1
\\ 
R$_{p}$ (R$_{\oplus}$)&2.49$\pm$0.18\\ 
M$_{p}$ (M$_{\oplus}$)&17.2$^{+4.0}_{-3.8}$\\
$\rho_{p}$ (g cm$^{-3}$)&6.1$^{+3.3}_{-2.3}$\\
T$_{eq}$(k)&460$\pm$8\\
\hline
\end{tabular}%
}
\tablefoot{$^{*}$ See the refined period in Table \ref{refined}.}
\end{table}

The joint modeling of TESS sector 21 and SOPHIE+ RVs with eccentricity-free and circular orbits yields consistent results. However, the eccentricity-free model exhibits strong statistical favorability ($\Delta$lnZ $>$ 9). Therefore, we present the resulting parameters of this model in Table \ref{joint} and depict its best-fit on combined photometry and RVs in Fig. \ref{full}. A corner plot of all parameters is included in Appendix \ref{fig:corner}. We note that the dilution factor was not taken into account in our analysis. This is due to the fact that when a wide uniform prior range of 0 to 1 is used, the dilution value tends to converge toward the lower edge of the prior 0. This outcome is unacceptable because there is only one faint nearby star in the TESS aperture, and it is approximately 6 magnitudes fainter than the primary star in the G band (refer to equation 6 of \cite{espinoza2019juliet} for more details). Such behavior can be expected in transits with low S/Ns \citep{espinoza2019juliet} and it is particularly relevant in our case as we are modeling only one transit. Additionally, due to the unavailability of GAIA RP magnitude for the neighbor star, we were unable to estimate and constrain the dilution factor accurately. However, considering that the neighbor star is faint, we expect any contamination effect from it to be negligible.

Based on our final parameters, the transiting planet HD88986\,b is a sub-Neptune with a period of 146.05$^{+0.43}_{-0.40}$~d. It exhibits an eccentricity of 0.23$\pm$0.06, alongside a radius of 2.49$\pm$0.18 R$_{\oplus}$ and a mass of 17.2$^{+4.0}_{-3.8}$ M$_{\oplus}$, corresponding to a high mean density of 6.1$^{+3.3}_{-2.3}$ ~gr\,cm$^{-3}$. Additionally, the planet has an equilibrium temperature of 460$\pm$8\, K \citep[$T_{eq} = T_{*} \sqrt{(R_{*}/2a)}$,][]{mendez2017equilibrium}, making it a relatively cool planet.

\section{Search for a second transit with additional photometric data}
\label{Additional_photometric_data}

The analysis presented above in Sect. \ref{joint2} predicts another transit of HD88986\,b should have occurred in 2022 February. We used CHEOPS and TESS sector 48 to attempt the detection of the second transit.

 \begin{figure*}
 \centering
\includegraphics[width=2\columnwidth]{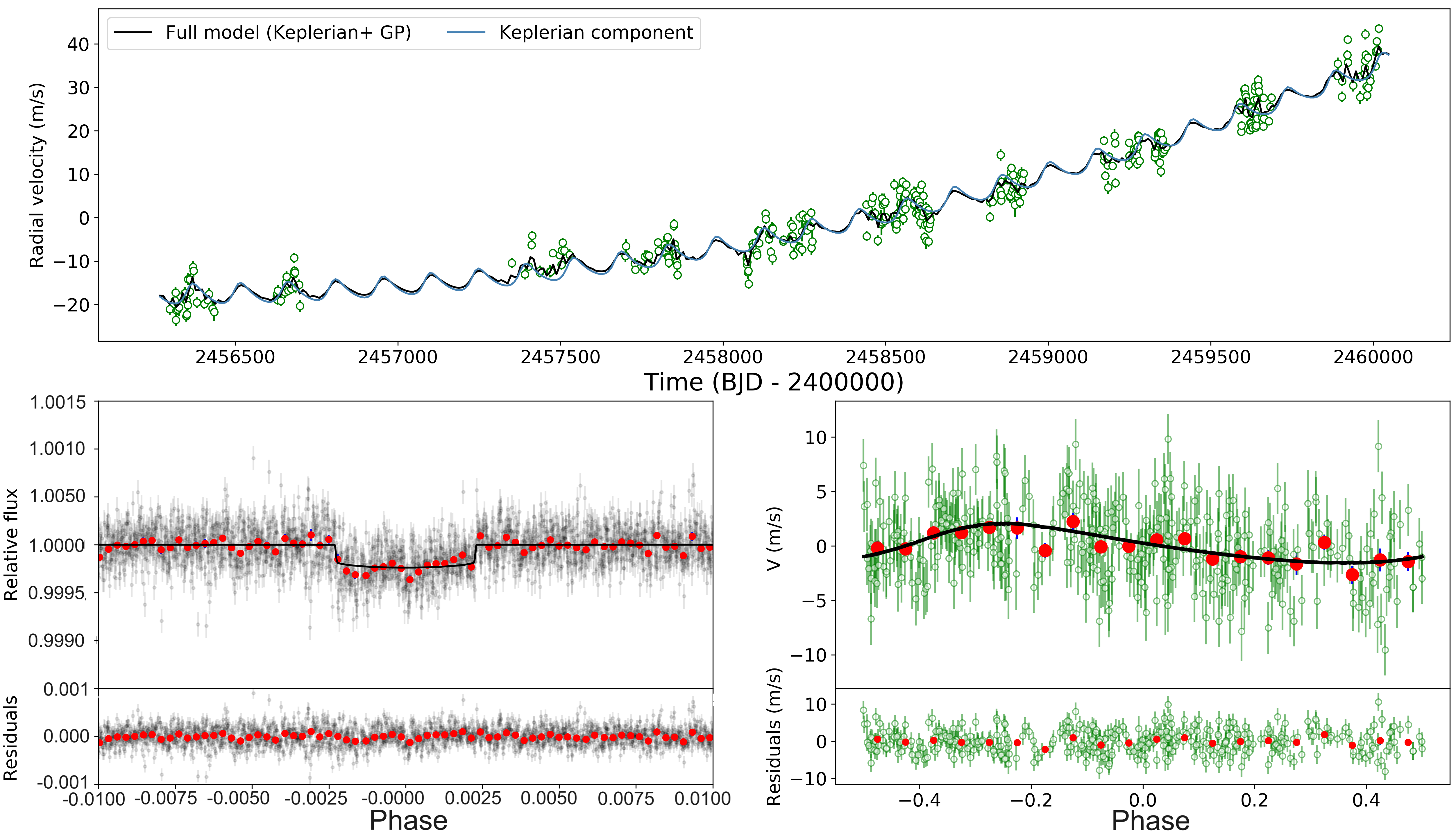}
\caption{Joint analysis of SOPHIE+ and TESS sector 21 observations of HD88986\,b. \emph{Top}: SOPHIE+ data overplotted by the best-fit orbit model. \emph{Bottom-left:}  Phase-folded TESS PDC-SAP photometric data of sector 21. The data are binned (red points) in 1 hour. The black line shows the best-fit transit model. \emph{Bottom-right}: Phase-folded SOPHIE+ RVs of HD88986\,b at the period of 146.05 d. The red points depict the binned data, utilizing a bin size of 0.05 in orbital phase units. The black line represents the best-fit orbit model.}
\label{full}
\end{figure*}

\subsection{CHEOPS photometry}

The {\it CHEOPS} spacecraft is a 30\, cm ESA space telescope \citep{Benz2021} that conducts ultra-high-precision photometry to characterize planets \citep{Bonfanti2021,Delrez2021,Lacedelli2022} and their atmospheres \citep{Lendl2020,Hooton2022}, but it has also been used to aid in the discovery of new planets \citep{Leleu2021,Osborn2022,Serrano2022,Wilson2022}. 

As derived in Sect. \ref{joint2}, the period of HD88986 b is 146.05$^{+0.43}_{-0.40}$~d. We predicted that the next transiting event would fall within the region of the TESS sector 48 which includes the gap between orbits. Therefore, to search for a second transit event, we obtained one visit of CHEOPS observation (PI: N. Heidari) spanning 167.4\, hr between 2022-02-08 and 2022-02-15 with an exposure time of 3.4\,s. This allowed us to cover the transit period's uncertainty from joint analysis of RVs and sector 21 by $\sim$ 2$\sigma$ and the TESS gap to be covered.

The data were processed with the latest version of the {\it CHEOPS} Data Reduction Pipeline (DRP v13; \citealt{Hoyer2020}) that conducts frame calibration, instrumental and environmental correction, and aperture photometry using predefined radii ($R$ = 22.5\arcsec, 25.0\arcsec, and 30.0\arcsec) as well as a noise-optimized radius. The DRP-produced flux contamination was subtracted from the light curves. We retrieved the data and corresponding instrumental basis vectors and assessed the quality using the {\sc pycheops} Python package \citep{Maxted2022} and found that the DEFAULT aperture minimized the root mean square (RMS) noise. Therefore, we used this aperture for our further analysis. These are plotted in the upper panel of Fig. \ref{sec_48}, on which the expected time of the transit, derived from the SOPHIE+ RVs and TESS sector 21 joint analysis (Sect. \ref{joint2}), is indicated in blue.

In previous studies, it has been noted that environmental effects (i.e., spacecraft temperature and illumination) and the presence of nearby contaminants can induce flux modulation in light curves \citep{Morris2021,Maxted2022,Wilson2022}. In order to correct for these effects and search for the smallest transit signals in our transit search analysis, we conduct a principal component analysis on the auto-correlation function of the {\it CHEOPS} frames using the methodology detailed in \citet{Wilson2022}. The process has been shown to monitor PSF shape changes, and so any effects that alter the {\it CHEOPS} PSF, such as environmental and contamination effects, are measured by this tool and can be removed by using the produced principal components as the basis vectors in a linear model detrending. Further examples of applications of this tool can be seen in \cite{Hoyer2022,Ehrenreich2023,Hawthorn2023}.

To assess the existence of a transiting body with the {\it CHEOPS} observations, we conduct a statistical analysis using a newly developed tool \citep{Wilson2023}. In brief, we use the PSF-based PCA components produced following \citep{Wilson2022} above in combination with the instrumental basis vectors to construct a linear noise model that is fit simultaneously with either a 0 or 1 planet transit model that allows us to compute the True and False Inclusion Probabilities (TIP and FIP; \citealt{Hara2022}) for the presence of transit in the data. These are calculated using the Bayes Evidences and posterior distributions of the 0 and 1 planet fits. For this study, we conduct this analysis twice: one with a period prior constrained by the transit model from the RV data and the second time with no period prior. For both cases and for all transit T$_0$ values within the {\it CHEOPS} dataset, we find FIP$\sim$1, which statistically means that there is no transit in the lightcurve.

This non-detection by {\it CHEOPS} leads to two potential conclusions: (1) the rejection of the presumed transit event in TESS sector 21 as a spurious feature, or (2) the transit might be occurring at a time that deviates more than $\sim$ 2$\sigma$ away from our predicted ephemeris. We note that although the {\it CHEOPS} data did not reveal any transit features, this observation was valuable in covering the gap in TESS sector 48 data (see Sect. \ref{tess48} and Fig. \ref{sec_48}), substantially contributing to the refinement of the planet's period.

\subsection{TESS photometry sector 48}
\label{tess48}

\begin{figure*}
\centering
\includegraphics[width=2\columnwidth]{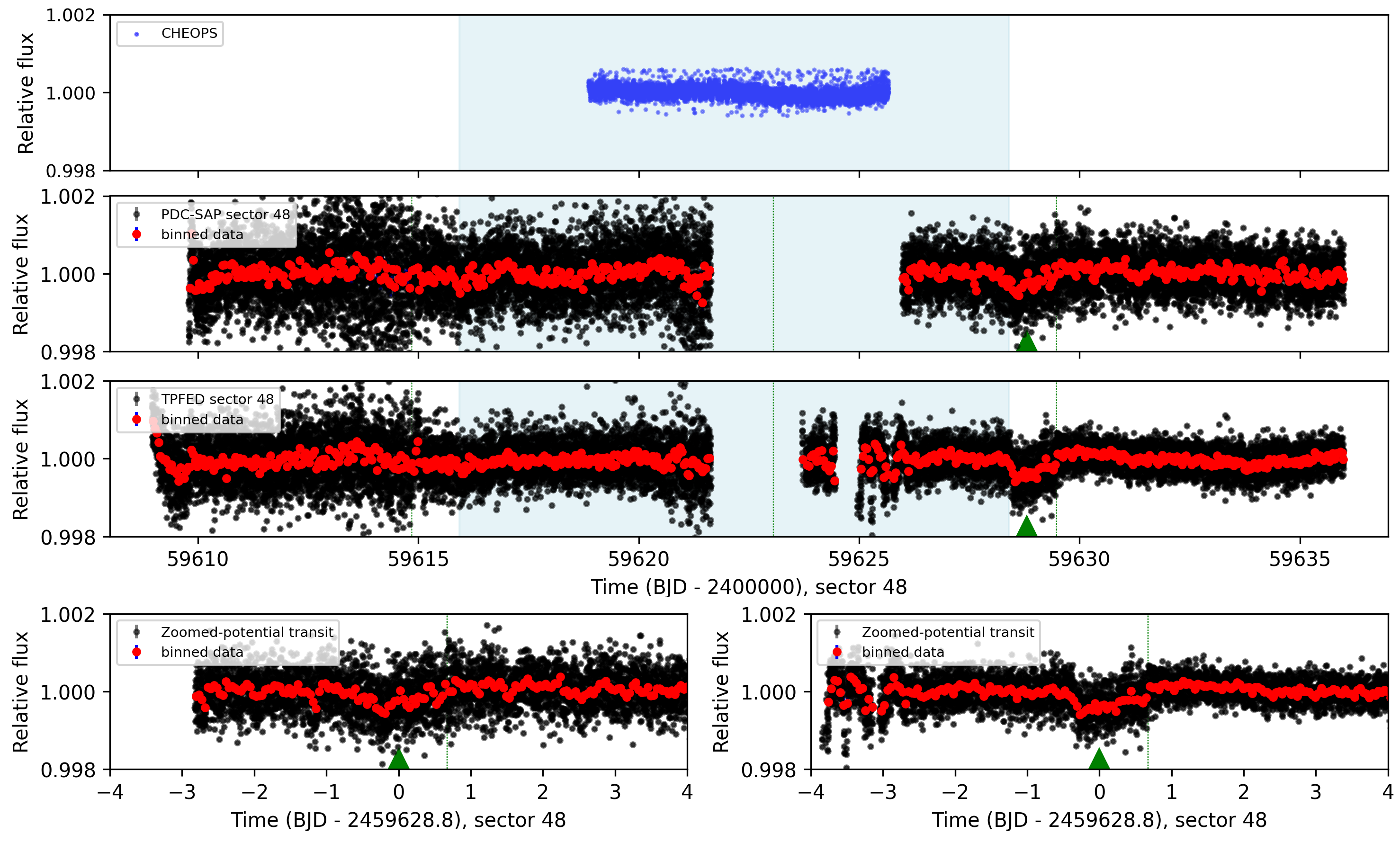}
\caption{CHEOPS and TESS observations of HD88986 in 2022 February. The predicted time of HD88986\,b's second transit event, based on the best-fit model of combining RVs and the photometric light curve of sector 21 (see Sect. \ref{joint2}), is highlighted in blue. \emph{Top}: The CHEOPS photometric data. \emph{Second:} TESS (black dots) PDC-SAP light curve of sector 48. The TESS data are binned (red points) in 1-hour increments. The potential transit-like feature is marked by a green triangle. The vertical dashed, green lines are the spacecraft momentum dumps. \emph{Third}: The re-extracted TESS light curve (see the text for more explanation). \emph{Bottom:} Zoomed on the potential transit event of HD88986\,b on PDC-SAP (\emph{left}) and re-extracted (\emph{right}) data. As the plot indicates, the two light curves are noticeably different. As a result, we did not include those data in our joint analysis presented in
Sect. \ref{joint2}.}
\label{sec_48}
\end{figure*}

TESS conducted a second observation of this star with a cadence of 2 minutes, spanning from 2022 January 28 to February 26. The PDC-SAP data provided by SPOC from this observation is depicted in Figure \ref{sec_48} (second panel). Remarkably proximate to the 3$\sigma$ anticipated transit region, as determined through the joint modeling of RVs and TESS sector 21 data, a second potential transit-like feature with a T$_{c}$ of about $\sim$ 59628.8 and a period of 147.4 d from the first transit is observed in the figure. We note that it was particularly lucky to cover two transits of that long-period planet with TESS, as only two sectors of TESS covered that star.

Similar to the first potential transit in sector 21, we tested the mean in-transit and mean out-of-transit flux, along with the difference between them (see Fig. \ref{test_spoc}). This test confirmed that the feature in sector 48 is likely related to the host star. However, in contrast to sector 21 with a standard deviation of $\sigma$=204.4 ppm, TESS sector 48 displays a substantial dispersion with $\sigma$=635.8 ppm. This noticeable difference might be linked to the presence of residual systematics that may have persisted even after the SPOC correction.

Therefore, following the methodology outlined in section \ref{photometry_extrac} using the TPFED tool, we conducted a customized extraction of TESS sector 48. The photometric results obtained from this approach are presented in Fig. \ref{sec_48} (third panel), with the transit-like feature zoomed in for better visualization (fourth panel). While the resulting custom extraction of the TESS light curve for sector 21 was fairly similar to the SPOC light curve, they are noticeably different for sector 48 (fourth panels). 

To further explore the potential source of instrumental noises, similar to sector 21, we also extracted the light curve using PLD and CBV approaches from 2-minute TPFs cadence, along with the FFI data (see sect. \ref{photometry_extrac} for more information about the methods). The resulting extracted data (see Figs. \ref{pld_rcq} and \ref{ffi}) following these methods also confirmed the noticeable difference between PDC-SAP data and the independently extracted light curves. Moreover, the variations within the transit further complicate our understanding of the TESS photometric data in sector 48, leading to its exclusion from our analysis. Additionally, we note that by the inclusion of PDC-SAP data of sector 48 in our joint modeling, the final results remain consistent compared to our model with only RVs and sector 21 in Sect. \ref{joint2}.

Fig. \ref{2_min_compatib} shows the phase-folded TESS PDC-SAP and FFI data for sectors 21 and 48 corresponding to the 147.4 d period. We note that the FFI light curve is detrended by the spline approach using the \texttt{Wotan} package \citep{2019AJ....158..143H}. While the consistency between the two potential transits in the PDC-SAP data remains uncertain, the two transits exhibit good consistency within the FFI data, particularly concerning transit depth and duration. 

In Table \ref{refined}, we summarize all the possible period solutions for HD88986\,b, including the periods obtained through the final choice of the RV-only model (see Sect. \ref{rv_several}), combined RVs with TESS sector 21 (see Sect. \ref{joint2}), and combined RVs with both TESS sector 21 and 48. The orbital period derived from RV data combined with two potential transits agrees (at 3$\sigma$) with the period calculated using RVs and single transit in sector 21, and also agrees (at 2$\sigma$) with the RV-only period. One could expect an even better agreement; possible persistent instrumental effects not perfectly taken into account in our models might be the cause. This agreement between all period solutions, arguing here
for an actual detection of a transit of HD88986\,b. Still, the noise in sector 48 light curve and the differences between reduction methods keep us prudent about the transit detection in sector 48, which we chose not to include in our final fit (Sect. \ref{joint2}). Conducting follow-up photometric observations for this system, with the goal of identifying HD88986\,b's second transit event, would strongly confirm that the planet is transiting while also providing a much better constraint on the planet's period. 

\begin{table}[h!]
\centering
\caption {Possible solution for HD88986\,b's period.}
\begin{tabular}{lccccc}
\hline
Models& HD88986\,b's period\\
\hline
RVs-only&146.3$\pm$0.6\\
RVs+ sector 21&146.1$\pm$0.4\\
RVs+ sector 21+ sector 48& 147.4$\pm$0.1\\
\hline
\end{tabular}
%
\label{refined}
\end{table}

\section{Constraining a long-term companion}
\label{Astrometry}

In this section, we examined different scenarios to determine the origin of long-term curvature seen in the RVs in addition to HD88986\,b. Stellar activity or a wide-orbit companion are two possibilities. To explore them, we used long-term photometric data obtained with the T8 APT, the combined RVs from ELODIE, HIRES, and SOPHIE instruments, as well as astrometric data from GAIA and Hipparcos.

\subsection{APT photometric observations}
\label{apt}
To characterize the origin of long-term curvature observed in RVs (see Fig. \ref{RV}), we used 1335 photometric observations of HD88986 covering 21 observing 
seasons from 1995-1996 to 2019-2020, except the four observing seasons 2015-16 through 2018-19, during which the star was not observed. The observations were 
acquired with the T8 0.80~m APT at Fairborn Observatory in southern Arizona. The T8 APT is equipped with a two-channel photometer that uses two EMI 9124QB bi-alkali photomultiplier tubes 
to measure the stellar brightness simultaneously in the Str\"omgren $b$ and 
$y$ passbands.  

The observations are made differentially with respect to three nearby
comparison stars. We measured the difference in brightness between our 
program star HD88986 (star~d) and the comparison stars (stars a: HD89557 (G= 7.3 mag, G8 III), b: HD87667 (G=7.3, F5), and c: HD88476 (G=6.6, G8 III))
and created differential magnitudes in the following six combinations:  d-a, 
d-b, d-c, c-a, c-b, and b-a.  Intercomparison of these six light curves
shows that the comparison star~a (HD89557) is the only one that appears to be 
constant to the limit of our precision, so we present our results as 
differential magnitudes in the sense star~d minus star~a, which we designate 
as d-a.

To improve the photometric precision of the individual nightly observations, 
we combined the differential $b$ and $y$ magnitudes into a single $(b+y)/2$ passband. The precision of a single differential observation with T8, as 
measured from pairs of constant comparison stars, typically ranges between 
0.001~mag and 0.0015~mag on good nights. The T8 APT is described in 
\citet{henry1999techniques}, where further details of the telescope, precision photometer, 
and observing and data reduction procedures can be found.

\begin{figure}
\centering
\includegraphics[width=0.35\textwidth, angle=270]{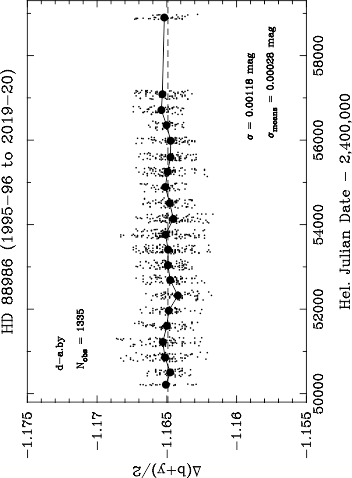}

\caption{Nightly Str\"omgren $(b+y)/2$ band photometry of 
HD88986 from 21 observing seasons from 1995-96 to 2020-21 (small filled 
circles) scatter about their mean (dashed line) with a standard deviation of 
0.00118~mag.  Seasonal means from the 21 seasons (large filled circles) 
scatter about their mean with a standard deviation of 0.00028~mag. No significant variations nor periodicities are detected.}
\label{henry1}
\end{figure}

Figure~\ref{henry1} plots the 1335 nightly observations in the $(b+y)/2$ passband photometry from d-a obtained across the 21 observing seasons as small filled circles. The mean of all the nightly observations,
-1.16492~mag, is plotted as the dashed line in the figure.  The standard 
deviation of the nightly observations from their mean is 0.00118~mag, 
consistent with the precision of the measurements. The 21 seasonal means 
of these data are plotted as large filled circles. The 
standard deviations of the individual seasonal means are roughly the size of 
the plot symbols. The standard deviation of the 21 seasonal means from the 
mean of the seasonal means is 0.00028~mag, indicating that there is no 
long-term variability in HD88986 to the limit of our photometric precision.

Table \ref{aptdata} summarizes observations in the $(b+y)/2$ passband photometry from d-a. The standard 
deviations of the nightly observations for each observing season indicate little or no 
short-term variability within each observing season. Frequency analysis of each 
individual observing season using the method of \citet{vanivcek1971further} confirms the lack of any periodic variability. \citet{henry2022nine} show extensive examples of this method of period analysis.

In a study conducted by \cite{lovis2011harps}, the activity cycles of 311 FGK stars were analyzed. It shows that RV semi-amplitudes can be induced up to approximately 25 m/s by the stellar long-period activity. However, in the case of HD88986, the long-period RV semi-amplitude is at least about 40 m. s$^{-1}$ (see the SOPHIE+ RVs in Fig. \ref{full}), and also as stated in this section, there is no evidence of long-term photometric variability up to the limits of our precision. This suggests that the origin of this long-term curvature is likely unrelated to stellar activity and instead points toward the presence of a third body in the system.

\subsection{Combining RV and Hipparcos/Gaia astrometry data}
\label{RV_astrometry}

In order to improve the characterization of the outer massive companion, we modeled simultaneously the available RV and absolute astrometry data. The use of an MCMC algorithm enables us to explore the different solutions for each orbital parameter and for the companion mass compatible with the data. This algorithm, introduced by \cite{Philipot2023}, is used to fit Keplerian orbits based on the emcee 3.0 \citep{Foreman_Mackey_2013} and HTOF \citep{Brandt_2021_htof} packages. The likelihood computation is similar to that of the ORVARA code \citep{Brandt_2021_orvara}.

\begin{table}
\centering
\caption {Median values and 68\% confidence interval for parameters of the outer companion of HD88986 based on the joint analysis of the Hipparcos/Gaia astrometric and RV data.
}
\resizebox{\columnwidth}{!}{%
\begin{tabular}{lc}
\hline
Parameter (unit)& The outer companion posteriors \\
\hline
Stellar parameters&\\
$M_{*}$ ($M_{\odot}$) & 1.20$_{-0.06}^{0.07}$\\ [0.1cm]
Parallax (mas) & 30.025$ \pm $0.023\\ [0.1cm]
$\sigma_{*}$ (m/s) & 3.3$_{-0.1}^{+0.2}$\\ [0.1cm]
&\\
The outer companion parameters&\\
P (year)& 116$_{-34}^{+40}$\\ [0.1cm]
T$_{c}$ (BJD-2400000 d)& 65000$\pm$2000 \\ [0.1cm] 
$\textit{a}$ (au)& 26.2$_{-5.5}^{+6.4}$ \\ [0.1cm]
$\sqrt{e}cos\omega$& 0.47$^{+0.16}_{-0.24}$\\ [0.1cm] 
$\sqrt{e}sin\omega$& -0.47$^{+0.13}_{-0.10}$\\ [0.1cm]
e&0.46$ \pm $ 0.13\\ [0.1cm]
$\omega$ ($^{\circ}$)& 314.9$_{-21.4}^{+14.4}$\\ [0.1cm]
$\textit{I}$ ($^{\circ}$)& 54.5$^{+21.3}_{-18.8}$ or 135.5$^{+18.4}_{-19.4}$\\ [0.1cm]
$\Omega$ ($^{\circ}$)& 29.1$^{+8.5}_{-3.6}$\\ [0.1cm]
$M_{c}$ ($M_{Jup}$)& $145_{-48}^{+73}$\\ [0.1cm]
&\\
Instrumental parameters:&\\
mu$_{ELODIE}$ (m/s)& 29220$_{-110}^{+180}$\\ [0.1cm]
mu$_{SOPHIE}$ (m/s)& 29320$_{-110}^{+180}$\\ [0.1cm]
mu$_{SOPHIE+}$ (m/s)& 29320$_{-110}^{+180}$\\ [0.1cm]
mu$_{HIRES}$ (m/s)& 240$_{-110}^{+180}$\\ [0.1cm]
mu$_{HIRES+}$ (m/s)& 240$_{-110}^{+180}$\\ [0.1cm]
\hline
\end{tabular}%
}
\label{joint_planet_C}
\end{table}

We considered the ELODIE, SOPHIE, SOPHIE+, HIRES, and HIRES+ RV data, presented previously, coupled with the proper motion and position values calculated in the Hipparcos-Gaia Catalog of Accelerations \citep{Brandt_HGCA_2021} from Hipparcos (\cite{Perryman1997} and \cite{Leeuwen2007}) and Gaia data release 3 (DR3; \cite{Gaia2016}, \cite{Gaia2021}) measurements. For the fit, we considered a Gaussian prior for the stellar mass and parallax, based on the values published by \cite{Kervella2022}, and a sin(\textit{I}) prior for the orbital inclination. For the semi-major axis (\textit{a}), the companion mass, the eccentricity, the longitude of the ascending node, the argument of periastron, the periastron passage time, and the jitter, we set uniform priors. In addition, as we use RV data from different instruments, we added an instrument offset for each dataset, also with uniform priors.

\begin{figure}[]
 \centering
\includegraphics[width=0.45\textwidth]{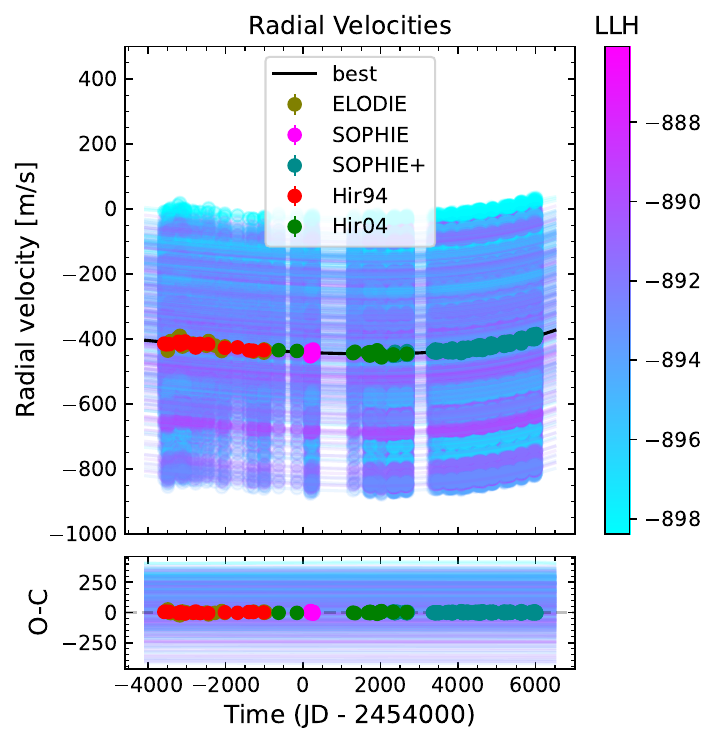}

\includegraphics[width=0.5\textwidth]{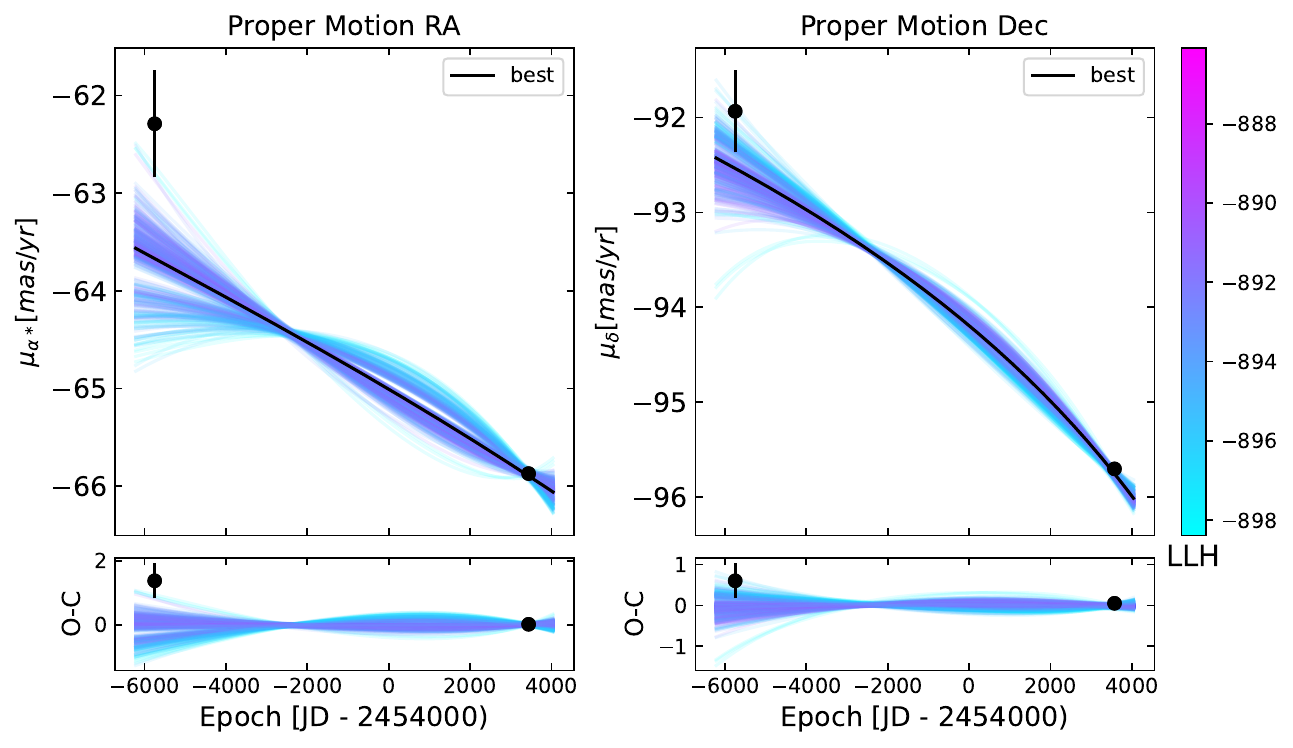}
\caption{Orbital fits for HD88986 outer massive companion. \textit{Top}: Fit of the HD88986 RV data points. \textit{Bottom}: Fit of the HD88986 proper motion measurements in right ascension (left) and declination (right). The black points correspond to the Hipparcos and Gaia EDR3 data points. In each plot, the black curve corresponds to the best fit. The color bar indicates the log-likelihood corresponding to the different fits plotted. 
\label{RV_PM}} 
\end{figure}

As the mass of HD88986\,b is low and its orbital period much smaller than the Hipparcos and Gaia DR3 observation windows (1227 and 1038 d, respectively), the proper motion variation of HD88986 induced by the planet HD88986\,b is negligible. We have therefore only fitted the orbit of the outer massive companion (Figure \ref{RV_PM}). However, as the RV data covers only a small part of the RV variation due to the outer companion, the star's RV remains poorly constrained and a wide range of solutions is compatible with the data, with similar likelihood values. We thus obtain an interval, with a confidence index of 3$\sigma$, between 16.7 and 38.8 au for the semi-major axis, 16 and 169$^\circ$ for the orbital inclination, and 68 and 284 $M_{Jup}$ for the true mass of the companion (Table \ref{joint_planet_C}). Nevertheless, these results suggest that the outer massive companion is likely a brown dwarf or a low-mass star.

As previously mentioned, there is a GAIA DR3 source situated 1.4 arcseconds west of HD88986. Considering HD88986's parallax value for this source, its semi-major axis deviates by approximately 3.2$\sigma$ from the resulting semi-major axis of the massive companion (see Table \ref{joint_planet_C}). Given the compatibility of the results with a wide range of solutions, it remains uncertain whether this source is the cause of the observed acceleration. Notably, this GAIA source lacks parallax information, raising the possibility that it might be a projected neighbor, unrelated to HD88986.

\subsection{Other constraints from Gaia astrometric excess noise}
\label{astrometry}
We used the Gaia data simulator from the \verb+gaston+ code first developed for the Gaia DR1~\citep{Kiefer2019a,Kiefer2019b,Kiefer2021} to test whether astrometric excess noises (AEN, hereafter) from the Gaia DR3 could lead to complementary mass constraints on the outer companion of HD\,88986 at the orbital period found by coupling to the proper motions of Hipparcos and Gaia. The AEN is a measurement of supplementary motion, beyond proper motion and parallax, in the astrometric data of a source. The AEN is obtained from the RMS of residuals after fitting out the ra-dec position, proper motion, and parallax to the simulated astrometric Gaia measurements by the approximate formula (see also~\citet{Kiefer2019a} and references therein):

\begin{equation}
    \sigma_\text{AL}^2 + \sigma_\text{attitude}^2 + {\rm AEN}_\text{DR3}^2 = \frac{\sum_j R_j^2}{N-5}
\end{equation}

where $R_j$ are the $N$ along-scan (AL) angle residuals of the astrometric fit; $\sigma_\text{AL}$ is the typical AL angle measurement noise; $\sigma_\text{attitude}$ is the spacecraft attitude excess noise, and AEN$_\text{DR3}$ is the AEN. The AL angle measurement noise has a value of $\sigma_\text{AL}$=0.05 mas for targets with a G-magnitude of 6.3 (Fig. A.1 from~\citealt{Lindegren2021}), and the typical attitude noise in the DR3 is $\sigma_\text{attitude}$=0.076\,mas~\citep{Lindegren2021}.

\begin{figure}
 \centering

\includegraphics[width=0.45\textwidth]{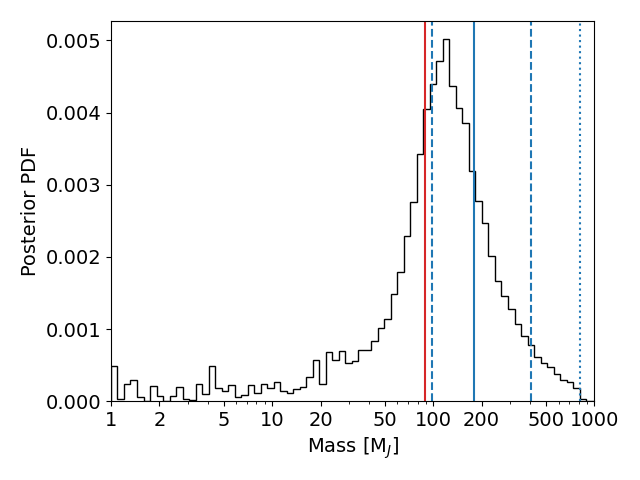}
\caption{Illustration of HD88986 outer companion analysis using GAIA DR3 analysis 
showing the companion mass posterior distribution running \texttt{gaston}. The dotted line shows the 3$\sigma$ upper-limit, the dashed lines show the 1$\sigma$ confidence interval, the solid blue line is the median mass, and the solid red line shows the RV $m\sin i$.}
\label{fig:resgaia}
\end{figure}

\begin{table}
\centering
\caption{Resulting constraints on the orbital inclination and mass of companion "c," and on the predicted photocenter semi-major axis of HD88986, using the AEN from Gaia DR3.}
\begin{tabular}{lccc}
\hline
Parameter & 1$\sigma$ &3$\sigma$ limits \\ \hline
I$_{c}$ (degree) & 33$^{+33}_{-18}$ & $>$2.7 \\
M$_{c}$ (M$_{jup}$)& 180$^{+220}_{-80}$ & $<$810 \\
a$_{\rm phot}$ (mas) & 91$^{+107}_{-47}$ & $<$440 \\
\hline
\end{tabular}
\label{tab:resGaia}
\end{table}

HD88986 has a magnitude of G$_{mag}\sim$6.3 and a color Gb$-$Gr of $\sim$0.8. In the Gaia-DR3 catalog, the typical AEN of single stars at that magnitude and color for sources fitted with five parameters, as HD88986, is 0.14\, mas~(private communication). This nonzero AEN is present for nearly all single stars and is due to a systematic jitter, including instrumental and global modelization noises, that is accounted for in the formal errors used to calculate the $\chi^2$~\citep{Lindegren2021}. The AEN of HD88986 is $\epsilon_\text{DR3}$=0.135 mas. The Gaia DR3 astrometry of this target is thus compatible with a single star without a companion, but it also allows deriving an upper-limit constraint on the mass of the RV-detected companion, given a range of possible orbital periods.

We follow the method from~\citet{Kiefer2019a,Kiefer2021}, using the code \verb+gaston+ adapted to the (E)DR3. The general principle of the method is the same as with the DR1. Fixing P, $m\sin i$, e, $\omega$ \& $T_0$ within their priors derived from combined RVs in Table~\ref{joint_planet_C}, we run several simulations of Gaia measurements of the target along a model of the orbital motion of the system due to the outer massive companion and derive simulated values of AEN that we compare with the actual ${\rm AEN}_\text{DR3}$.

We sample orbital inclination uniformly between 0 and 90$^\circ$ by an MCMC routine based on the \verb+emcee+ code \citep{Foreman_Mackey_2013} and thoroughly explained in~\citet{Kiefer2019a,Kiefer2021}. The orbital inclination changes the amplitude of the astrometric motion due to a different mass of the companion determined from $M$=$m\sin i$/$\sin i$ and thus changes the value of the AEN allowing us to match a range of orbital inclinations to the observed AEN.

Noises, epochs, scan angles, and the number of measurements used in the simulations are updated with respect to the new data reduction of DR3~\citep{Gaia2021}. An epoch is a date when the star is transiting the Gaia field of view; several measurements, typically 9, are performed during a single transit. Those epochs can be found for any target in the Gaia Observation Forecast Tool (or GOST\footnote{https://gaia.esac.esa.int/gost/index.jsp}). We add in our simulated model a jitter of 0.16\,mas, allowing us to reproduce a median AEN of 0.14 mas for single sources at G=6.3 and Gb$-$Gr=0.8. It is modeled as a Gaussian noise changing every epoch of observation. The spacecraft attitude noise is also added to the model as a systematic Gaussian dispersion that changes every observation epoch with a standard deviation of 0.076\, mas. A Gaussian measurement noise of $\sigma_\text{AL}$=0.05\,mas is added to each of the $N_\text{AL}$ astrometric measurements performed at a given epoch.

Table \ref{tab:resGaia} summarises the results of AEN fitting for this star. Fig.~\ref{fig:resgaia} shows the relation between AEN and inclination in the simulations and plots the posterior distribution of companion mass. The posterior distribution on mass gives an upper limit on the mass of the companion below 810\, M$_{jup}$ at 3$\sigma$. This result agrees with the one
presented in Sect. \ref{RV_astrometry}. Finally, we adopted the results from Sect. \ref{RV_astrometry} as it provides a higher level of precision in the mass and orbital inclination of the outer massive companion.

\section{Discussion and conclusion}
\label{Discution}
 
\begin{figure*}
\centering
      \begin{tabular}{cc}
          \includegraphics[width=.45\textwidth]{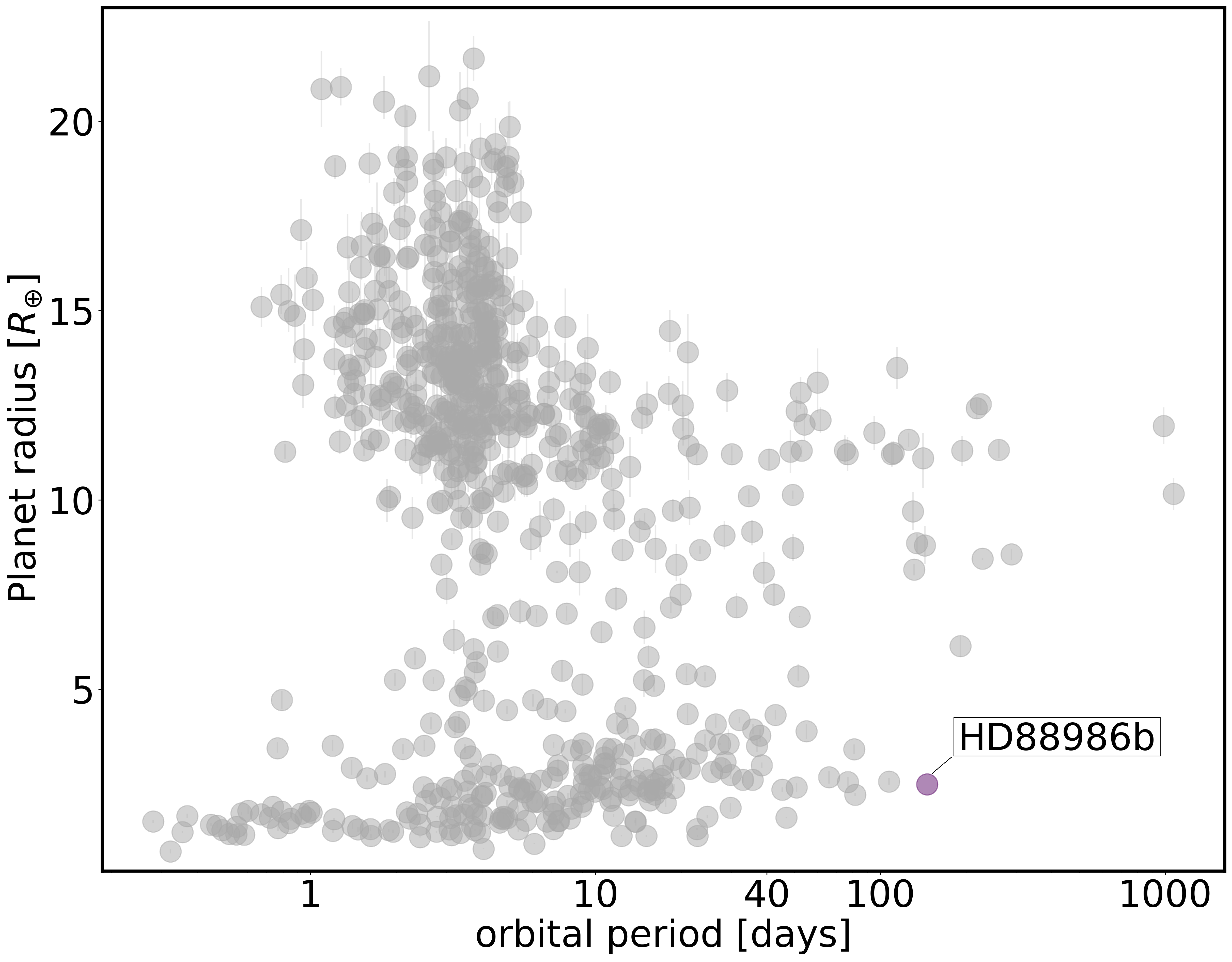} 
          \includegraphics[width=.45\textwidth]{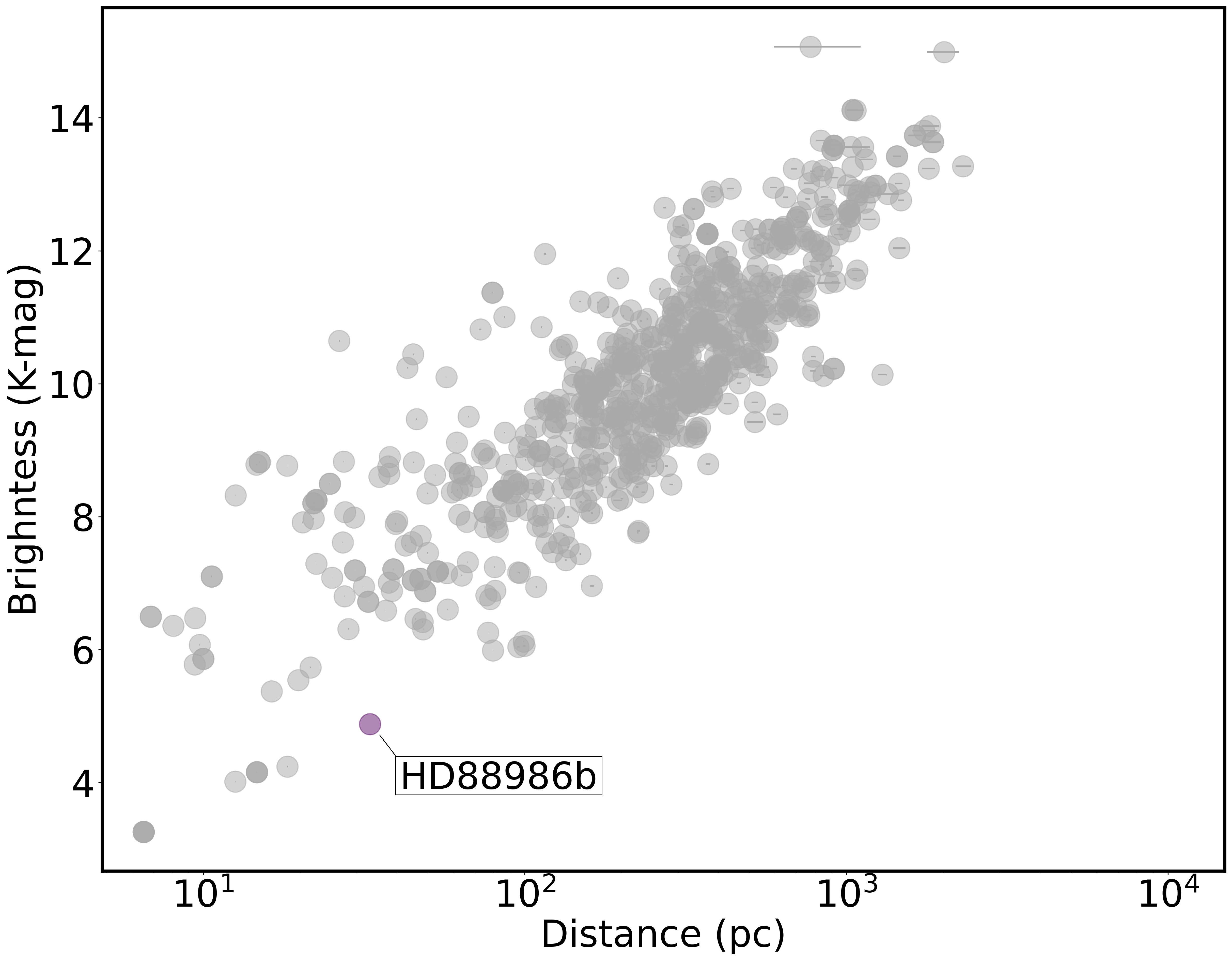}\\
          \includegraphics[width=.45\textwidth]{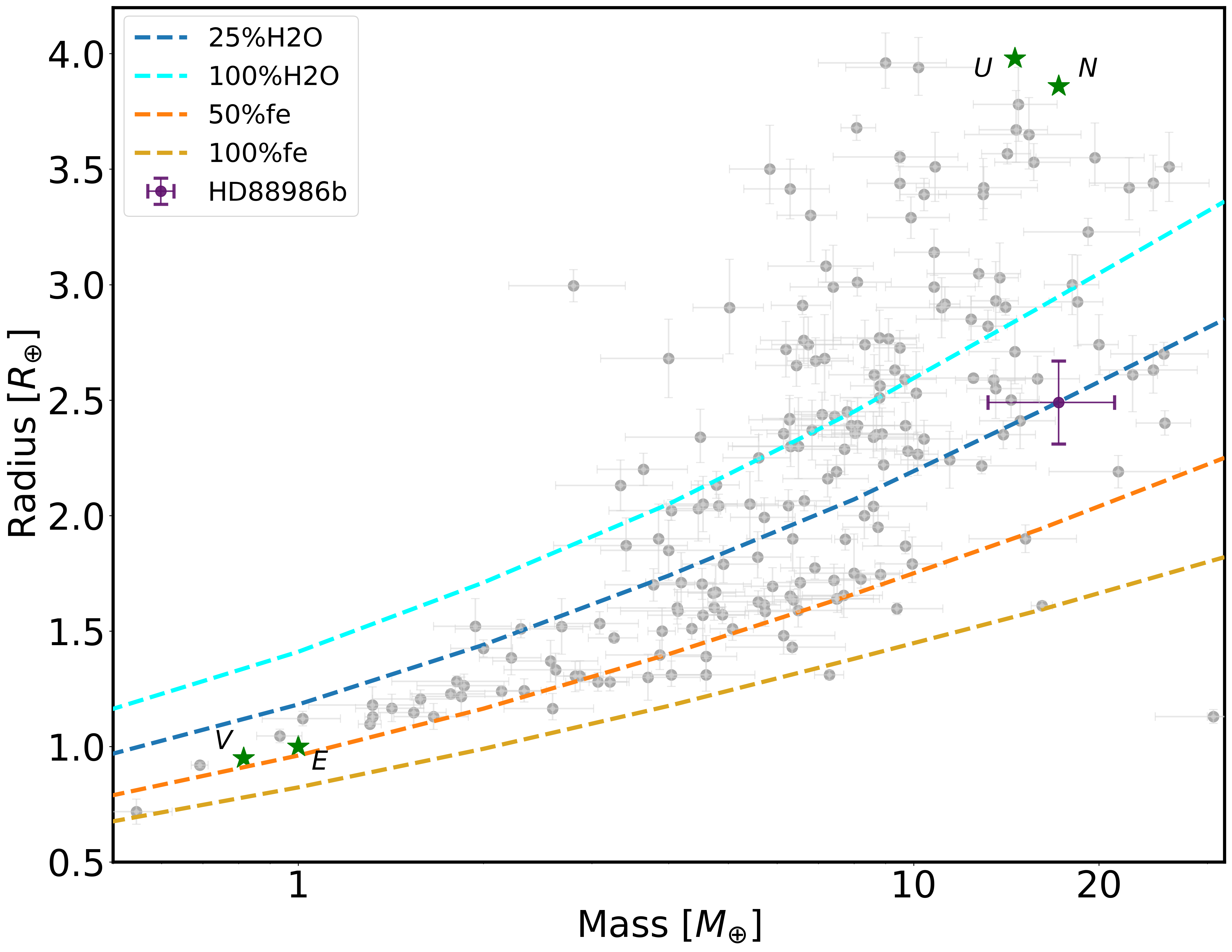}
          \includegraphics[width=.45\textwidth]{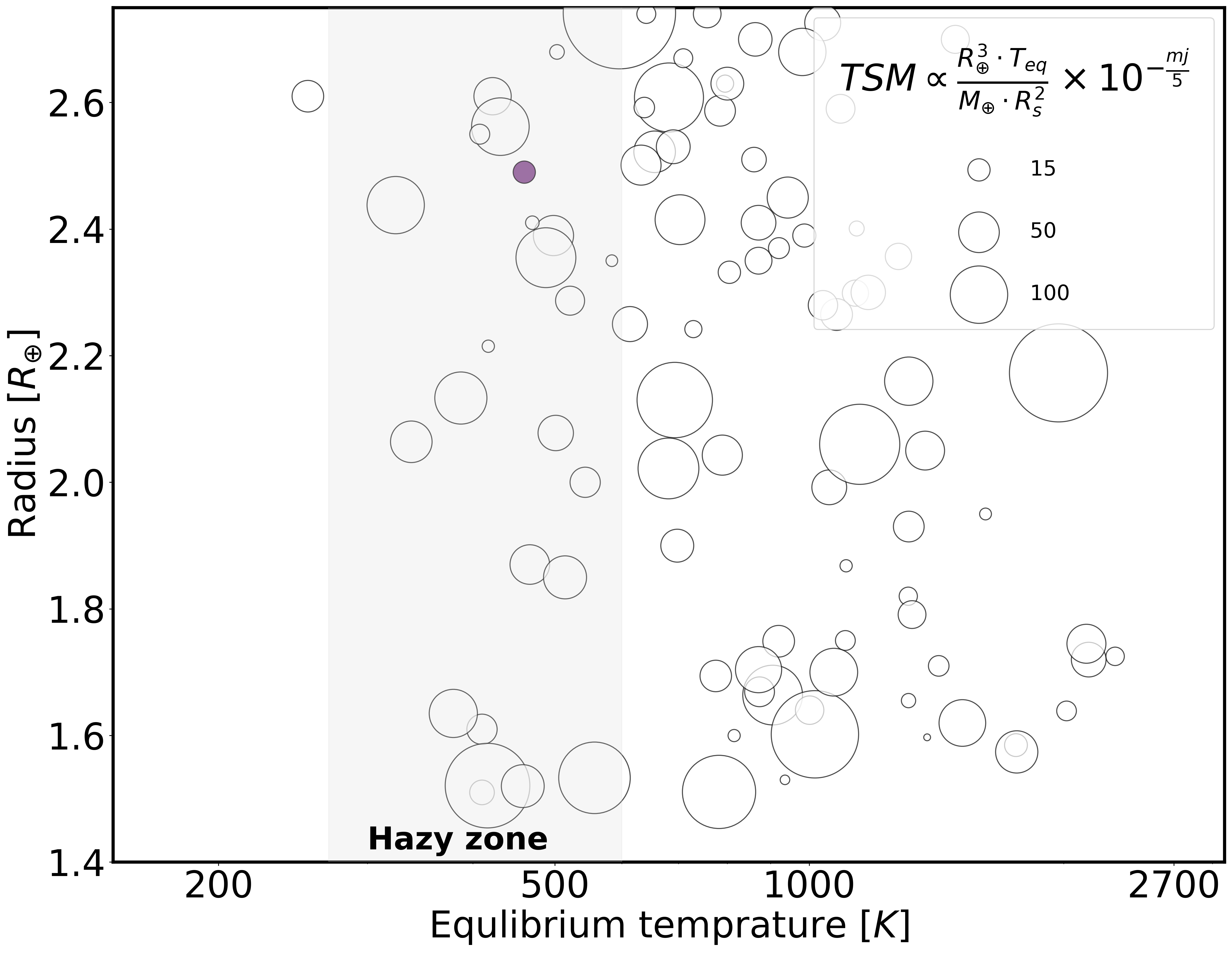}
    \end{tabular}
\caption{Position of HD88986\,b and its host star among known transiting exoplanetary systems, with precise planet mass and radius measurements \citep[$\sigma$M/M = 25\% and $\sigma$R/R = 8\%,][]{otegi2020revisited} from the NASA Exoplanet Data Archive (June 7, 2023). \emph{Top left}: Radius-period diagram of exoplanets. \emph{Top right:} Brightness in the K band versus distance of the exoplanet host star for the same planets. \emph{Bottom left}: Mass-radius diagrams of small planets (R$_{p}<$ 4 R$_{\oplus}$). The colored curves are the two-layer theoretical composition models of \cite{zeng2016mass}. \emph{Bottom right}: Equilibrium temperature-radius diagram of planets within the radius range of 1.5 $R_{\oplus}$ to 2.75 $R_{\oplus}$ with each planet's size scaled according to the propositional TSM introduced by \cite{kempton2018framework}. The gray area is the proposed hazy atmosphere zone by \cite{yu2021haze}. The purple mark indicates HD88986\,b. These planets possess either a known equilibrium temperature or information about the radius and temperature of their host star to estimate their temperature ($T_{eq} \sim T_{*} \sqrt{(R_{*}/2a)}$, \cite{mendez2017equilibrium}). These figures highlight the unique position of the HD88986 system.}    
\label{excitting}
\end{figure*}

We discovered and have characterized HD88986\,b, a sub-Neptune in orbit around a subgiant star, which stands as one of the nearest and brightest (V = 6.47 mag) exoplanet host stars (see Fig. \ref{excitting} top right). Our analysis indicates that this planet is transiting, based on two potential single transit detections in TESS sectors 21 and 48, both of which are consistent with the anticipated transit time from the RV model. By combining data from SOPHIE+ RV measurements and TESS sector 21 photometric data, we determined the following parameters for HD88986\,b: a period of P$_{p}$ = 146.05$^{+0.43}_{-0.40}$~d, a mass of M$_{p}$ =17.2$^{+4.0}_{-3.8}$ M$_{\oplus}$, and a radius of R$_{p}$ = 2.49$\pm$0.18 R$_{\oplus}$, resulting in a high mean density of $\rho_{p}$= 6.1$^{+3.3}_{-2.3}$~ g cm$^{-3}$. The two-layer theoretical composition model developed by \cite{zeng2016mass} indicates that the planet is composed predominantly of rock, accounting for approximately 75\% of its mass, while water makes up the remaining 25\% (Fig. \ref{excitting} bottom left). Additional photometric observations of the system targeting another transit event of HD88986\,b are needed. Such observations would provide a strong confirmation of the planet's transiting nature and yield better estimates for its period and radius.

Additionally, we identified a clear long-term curvature in the RV caused by the presence of a massive companion in the system. The nature of this companion has yet to be confirmed. A joint analysis of RV, Hipparcos, and Gaia astrometric data shows that with a 3$\sigma$ confidence interval, its semi-major axis is between 16.7 and 38.8 au and its mass is between 68 and 284 M$_{Jup}$. SOPHIE+ observations are being conducted to disclose the nature of this massive companion, and in particular better constrain its period and eccentricity. Furthermore, given its extensive semi-major axis, this outer massive companion presents an opportunity for being directly imaged, aiming to provide a more precise characterization of its orbit and mass. This study can be facilitated using the current generation of high-contrast imaging instruments, such as SPHERE.

The top-left panel of Figure \ref{excitting} highlights the unique position of HD88986\,b in the radius-period diagram among other known planets from the NASA Exoplanet Data Archive\footnote{\url{https://exoplanetarchive.ipac.caltech.edu/}} (as of June 7, 2023) with precise mass and radius measurements \citep[$\sigma$M/M = 25\% and $\sigma$R/R = 8\%,][]{otegi2020revisited}. Notably, HD88986\,b has the longest orbital period among the discovered small transiting planets (R$_{p}$< 4 R$_{\oplus}$). This wide orbit suggests that the planet did not undergo significant mass loss due to extreme-ultraviolet radiation, and hence it probably retains its primordial composition \citep{kubyshkina2022mass}. Consequently, HD88986\,b is an excellent candidate for investigating the planet's internal structure and formation conditions.

The bottom-right panel of Figure \ref{excitting} compares the equilibrium temperatures of HD88986\,b with those of other known small planets possessing a precise mass and radius \citep{otegi2020revisited}. HD88986\,b, thanks to its long orbital period, is a cold planet (T$_{eq}$ = 460 $\pm$ 8 K). The study of atmospheric characteristics of cold planets ($\leq$ 500 K) transiting bright hosts is extremely limited by the lack of such planets. In terms of atmospheric chemistry, colder atmospheres may contain disequilibrium chemistry that is useful for understanding atmospheric physics \citep{fortney2020beyond}. Remarkably, HD88986\,b's equilibrium temperature places it within the proposed hazy atmosphere zone, ranging between 270 K and 600 K, as suggested by \cite{yu2021haze}. This intriguing positioning opens up exciting opportunities for studying the haze layer and atmospheric composition above it \citep{kawashima2019detectable}. Hence, for the bottom-right panel of Fig. \ref{excitting}, we examined the proportional transmission spectroscopy metric ( $TSM \propto (\frac{T_{eq} \times R_p^3} { M_p \times R_s^2}) \times 10^{-mJ/5}$) introduced by \cite{kempton2018framework} and scaled the size of each planet accordingly. This metric considers the planet's equilibrium temperature T$_{eq}$, radius $R_p$, mass $M_p$, host star radius $R_s$, and host star magnitude in the J band m$_{J}$. We selected a radius bin of 1.5 $R_{\oplus}$ to 2.75 $R_{\oplus}$ based on \cite{kempton2018framework}'s assumption that planets within the same size bin share similar atmospheric compositions. We exclusively employed the proportional TSM because the constant scale factor in \cite{kempton2018framework}'s TSM is intended for stars with m$_{J}$ > 9, while HD88986 has a magnitude of m$_{J}$ = 5.2 mag. HD88986\,b ranked 19$^{th}$ in comparison with proportional TSM S/N of only 24 planets detected in the hazy atmosphere zone (see Fig. \ref{excitting} bottom-right panel). This relatively low rank is to some extent compensated for by the exceptionally long duration of the transit (16 h). Therefore, HD88986\,b's unique characteristics, such as being a cold exoplanet orbiting a bright star at a distance of 33 pc, make it a good target for atmospheric characterization studies of cold planets residing in the hazy zone.

Furthermore, with a mass of M$_{p}$ = 17.2$^{+4.0}_{-3.8}$ M$_{\oplus}$, HD88986\,b surpasses the critical mass threshold of $\sim$ 10 M$_{\oplus}$ required for envelope accretion \citep{johnson2010giant}. This indicates that it likely formed similarly to the cores of giant planets in our Solar System. However, HD88986\,b failed to accumulate much gas during its formation process. One possible scenario is that HD88986\,b formed at a late stage in the protoplanetary disk when there was little gas present during core assembly \citep{lee2016breeding}. Moreover, according to the minimum mass solar nebular model, it is unlikely for such a massive planet to form in situ at its current location, situated 0.6 AU from its host star \citep{schlichting2014formation}. Instead, it likely formed farther away and subsequently migrated inward over time, potentially influenced by interactions with the detected massive companion in the system. However, to gain a comprehensive understanding of the HD88986 planetary system, additional photometric and spectroscopic observations are required.

\begin{acknowledgements}
      We warmly thank the OHP staff for their support on the observations. We received funding from the French Programme National de Physique Stellaire (PNPS) and the Programme National de Planétologie (PNP) of CNRS (INSU). 
      \\
      N.H. acknowledges CNES postdoctoral funding fellowship. N. H. also acknowledges the financial support of the French embassy in Tehran as well as the Iran Ministry of Science Research and Technology. 
      JSJ acknowledges support by FONDECYT grant 1201371 and from the ANID BASAL project FB210003.
      This paper made use of data collected by the TESS mission which is publicly available
       from the Mikulski Archive for Space Telescopes (MAST) operated by the Space Telescope Science Institute (STScI). Funding for the TESS mission is provided by NASA’s Science Mission Directorate. We acknowledge the use of public TESS data from pipelines at the TESS Science Office and at the TESS Science Processing Operations Center. Resources supporting this work were provided by the NASA High-End Computing (HEC) Program through the NASA Advanced Supercomputing (NAS) Division at Ames Research Center for the production of the SPOC data products. NASA supported DR under award number NNA16BD14C for NASA Academic Mission Services. 
       \\
       A.C., X.D., and T.F. acknowledge support by the French National Research Agency in the framework of the Investissement d’Avenir program (ANR-15-IDEX-02), through the funding of the « Origin of Life » project of the Grenoble-Alpes University. We acknowledge funding from the French ANR under contract number ANR18CE310019 (SPlaSH). This work was supported by FCT - Fundação para a Ciência e a Tecnologia 
through national funds and by FEDER through COMPETE2020 - Programa 
Operacional Competitividade e Internacionalizacão by these grants: 
UID/FIS/04434/2019, UIDB/04434/2020, UIDP/04434/2020, 
PTDC/FIS-AST/32113/2017 \& POCI-01-0145-FEDER- 032113, 
PTDC/FIS-AST/28953/2017 \& POCI-01-0145-FEDER-028953, 
PTDC/FIS-AST/28987/2017 \& POCI-01-0145-FEDER-028987. NCS further 
acknowledges funding by the European Union (ERC, FIERCE, 101052347). 
Views and opinions expressed are however those of the author(s) only and 
do not necessarily reflect those of the European Union or the European 
Research Council. Neither the European Union nor the granting authority 
can be held responsible for them. O.V. acknowledges funding from the ANR project "EXACT" (ANR-21-CE49-0008-01), from
the Centre National d'\'{E}tudes Spatiales (CNES), and from the CNRS/INSU Programme
National de Plan\'etologie (PNP).
\\
 M.H. acknowledges support from ANID-Millennium Science Initiative-ICN12$\_$009. S.D. is funded by the UK Science and Technology Facilities Council (grant number ST/V004735/1). This work has been carried out within the framework of the NCCR PlanetS supported by
the Swiss National Science Foundation under grants 51NF40$\_$182901 and
51NF40$\_$205606. This project has received funding from the European Research
Council (ERC) under the European Union's Horizon 2020 research and innovation
program (project \textsc{Spice Dune}, grant agreement No 947634).
\end{acknowledgements}

\bibliographystyle{aa}
\bibliography{mwd}

\onecolumn

\begin{appendix}
\section{Atmospheric dispersion correction}
\label{Atmospheric}

Atmospheric dispersion can introduce a slope in the spectral continuum, leading to a shift in the mean RV values of the observed targets \citep{pepe2008harps,wehbe2020impact}. In order to achieve a higher RV precision necessary for detecting low-mass planets, it is imperative to consider the impact of atmospheric dispersion. To address this effect in SOPHIE, we implemented a correction method based on the HARPS\footnote{\url{http://www.eso.org/sci/facilities/lasilla/instruments/harps/doc/index.html}} and ESPRESSO DRS \citep{modigliani2019espresso}. This correction involves scaling the target spectrum to match its flux distribution with that of a template. Specifically, we multiplied the target spectrum by the flux ratio between the target spectrum and the template. The template is constructed from a high S/N spectrum of a standard star with the same spectral type, acquired at low air mass conditions. Through the application of this method, the flux distributions of star spectra always have the same distribution, thereby minimizing the influence of atmospheric conditions on the computation of the CCF. 

The method described herein yielded an enhancement in the precision of SOPHIE RV measurements by 8 cm s$^{-1}$ when applied to the high-precision RV measurements from the SOPHIE SP1 star catalog (which comprises 96 stars that have more than 10 observations). This improvement equates to 7\% of the mean error bar of 1.2 m s$^{-1}$ associated with these stars. With the application of this method to the full width at half maximum (FWHM) of the same set of stars, we observed a substantial precision enhancement of 15 m s$^{-1}$ on the FWHM. Notably, this enhancement is nearly five times greater than the average error bars of 2.8 m s$^{-1}$ for the FWHM of these stars. This correction has been incorporated into the SOPHIE DRS and will be used in future planet detections conducted by the SOPHIE instrument. Moreover, the correction method is applied to all SOPHIE RV constant stars (4 super constant stars in addition to $\sim$ 26 other stars with low RV dispersion), resulting in the creation of a more robust RV master constant time series (see Appendix \ref{master_cons}).

\section{Moon contaminated spectra}
\label{moon}

To effectively detect and characterize planets through the RV method, it is imperative to meticulously manage systematic noise sources and eliminate any outliers. One such source of outliers in RV data sets is the spectra contaminated by Moonlight \citep[e.g.,][]{hebrard2008misaligned, santerne2011sophie}. The Moon's reflected light can cause spectral contamination, leading to potential masking of the planet's signals or presenting systematic errors in the properties of the detected planets. Therefore, it is crucial to carefully identify and exclude any contamination by the lunar light from our data sets. 

To achieve this, we developed a recipe that can identify Moon-polluted spectra for star observations made through the simultaneous calibration lamp, where no sky observation is available. This is achieved by taking into account the phase and position of the Moon at the time of observation, using two empirical criteria. 

The first criterion considers whether moonlight contributes significantly to the target spectrum. This can be assumed when either: 1. the Moon phase is more than 68\% at the time of observation and the sky-level is above the mean of the sky-level of all observations, 2. or when the separation between the target and Moon is less than 30 degrees.  We note that the sky-level is a criterion for estimating sky background light (see \cite{bijaoui1980sky,ji2018estimating}). In SOPHIE, it is calculated using SOPHIE DRS and is available in each FITS file spectrum header.

The second criterion we utilized in our study takes into account the proximity of the targets' RV to the Moon's RV. The Barycentric Earth radial velocity (BERV) at the time of observation and in the direction of the target is within approximately 1 km/s of Moon RVs \citep{diaz2012sophie}. Therefore, it is reasonable to consider it as a Moon RV. If the BERV is close to the target radial velocity, with $|RV_{target}-BERV|<$ 2*FWHM, then a spectrum can be considered moonlight polluted.

To test these empirical criteria, we used observations of three stars with simultaneously recorded sky observations where recorded sky spectra were available. Over a total of 59 spectra, 13 data were Moon contaminated. Our criteria allowed us to successfully detect 8 of these contaminated spectra but also flagged three uncontaminated data. So we conclude our criteria are conservative.

Overall, our recipe provides a useful tool for identifying suspected Moon-polluted spectra in RV observations made through simultaneous calibration lamps. This can help improve data quality and ensure accurate scientific results. In the case of HD88986, following these criteria, we identified 31 spectra that were suspected to be contaminated by moonlight. These data conservatively were discarded from our analysis. We note that including or excluding these data had no significant effects
on our final results within the uncertainties, showing our criteria indeed are conservative.

\section{Update on constructing RV master constant timeseries}
\label{master_cons}

\begin{figure}
\centering
\includegraphics[width=0.61\columnwidth]{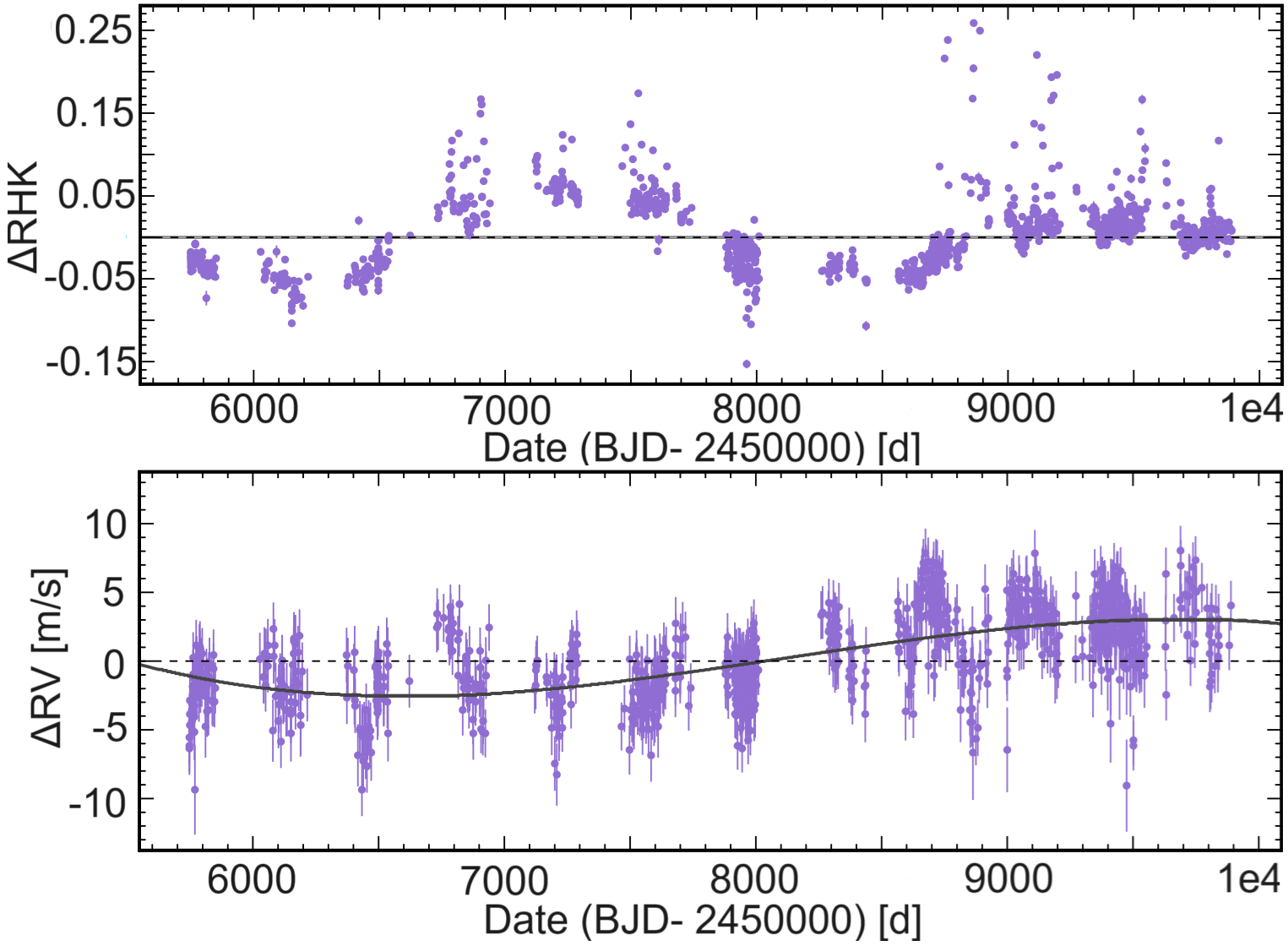}
\caption{$\log R'_{\rm HK}$ (\emph{top}) and RVs (\emph{bottom}) of constant star HD185144. The best fit of a third-order polynomial model (black line \emph{bottom}) is overplotted to the stellar long-period activity. }
\label{rv-activity}
\end{figure}

To monitor long-term instrumental variations, we have been conducting nightly observations of a few constant stars using SOPHIE since 2012. These constant stars include four super constant stars and approximately 25 additional stars with low RV dispersion ($\sigma$ < 3.5 m/s). Then, following the method outlined in \cite{courcol2015sophie}, we create a master constant time series and subtract it from the RVs of each star. However, our latest analysis suggests that one of our constant stars, HD185144, exhibits activity over an extended period (see Fig. \ref{rv-activity}). This activity could potentially affect our master constant time series and consequently impact the RVs of other stars.

We note that this star's activity cycle was already known thanks to HIRES data \citep{isaacson2010chromospheric}. SOPHIE later confirmed this when we built the first master constant correction in 2015 \citep{courcol2015sophie}. At that time, we estimated the semi-amplitude of the signal to be less than 1.5 m/s, which was negligible given the spectrograph's precision. With more data points in 2023, the effect is estimated to have a semi-amplitude of 2.7 $\pm$ 0.1 m/s. Given SOPHIE's improved RV precision, it has become necessary to correct the impact of HD185144's stellar activity on its RVs. Since this star is one of the most frequently observed constant stars by SOPHIE with more than 1100 data points, removing the RVs of this star from our master constant time series is not an option. 

To correct this stellar activity, we utilized a four-step approach. Firstly, we corrected the HD185144 RVs for SOPHIE instrumental variations. To accomplish this, we subtracted the master constant time series, derived from SOPHIE constant stars excluding HD185144, from the RV of HD185144. Next, we fit a third-order polynomial model on the HD185144 RV time series to determine its activity phase (see Fig. \ref{rv-activity} bottom). Subsequently, we subtracted the same polynomial model from the raw RVs of HD185144 (prior to master constant correction). Finally, we used these corrected RVs to construct the final master constant time series along with other stars.

\section{RVs}

\begin{longtable}{llllllll}
\caption{\label{tab:rvs sophie}SOPHIE RVs for HD88986. We note that a value of 999.0 indicates invalid data on the corresponding date. }\\
\hline
BJD (-2400000 d) & RV (km s$^{-1}$ ) & $\sigma_{RV}$ (km s$^{-1}$ ) & BIS (km s$^{-1}$ ) & S-index & $\sigma_{S-index}$&  Na&$\sigma_{Na}$\\ \hline 
\endfirsthead
\multicolumn{4}{c}%
{{\bfseries \tablename\ \thetable{} -- continued from previous page}}\\
\hline 
BJD (-2400000 d) & RV (km s$^{-1}$) & $\sigma_{RV}$ (km s$^{-1}$ ) & BIS (km s$^{-1}$ )  & S-index  & $\sigma_{S-index}$&  Na&$\sigma_{Na}$ \\ \hline 
\endhead
\hline \multicolumn{8}{|r|}{{Continued on next page}} \\ \hline
\endfoot
\hline \hline
\endlastfoot
\hline \hline
SOPHIE:&&&&&\\
54185.43116&    29.0727&        0.0013&&&\\     
54189.45774&    29.0698&        0.0016&&&\\     
54190.42612&    29.0813&        0.0014&&&\\
54190.43296&    29.0717&        0.0035&&&\\
54192.48075&    29.0680&        0.0013&&&\\
54192.48729&    29.0638&        0.0039&&&\\
54236.34361&    29.0812&        0.0014&&&\\     
54249.39304&    29.0765&        0.0015&&&\\
54260.37481&    29.0721&        0.0016&&&\\
54262.37490&    29.0707&        0.0017&&&\\     
54263.36789&    29.0847&        0.0025&&&\\
54264.36478&    29.0817&        0.0017&&&\\
\hline
SOPHIE+:&&&&&\\
56298.72054 & 29.07  & 0.001 & -0.0041 & 0.1759 & 0.0007 & 0.521 & 0.007 \\
56313.59596 & 29.072 & 0.001 & -0.0042 & 0.1709 & 0.0007 & 0.518 & 0.007 \\
56316.56759 & 29.073 & 0.001 & -0.0023 & 0.1676 & 0.0007 & 0.528 & 0.008 \\
56317.68241 & 29.067 & 0.001 & -0.004  & 0.1769 & 0.0012 & 0.525 & 0.01  \\
56319.61938 & 29.069 & 0.001 & -0.0072 & 0.1703 & 0.0011 & 0.53  & 0.01  \\
56322.59358 & 29.069 & 0.001 & -0.0053 & 0.1739 & 0.0008 & 0.516 & 0.008 \\
56327.48507 & 29.07  & 0.002 & -0.0071 & 999.0  & 999.0  & 0.522 & 0.018 \\
56344.58229 & 29.073 & 0.001 & -0.0013 & 0.1737 & 0.0008 & 0.521 & 0.008 \\
56345.50874 & 29.073 & 0.001 & -0.0026 & 0.1739 & 0.001  & 0.528 & 0.009 \\
56346.50675 & 29.072 & 0.001 & -0.0    & 0.1731 & 0.0007 & 0.525 & 0.007 \\
56349.46062 & 29.068 & 0.001 & -0.0043 & 0.1693 & 0.0007 & 0.525 & 0.008 \\
56350.49421 & 29.072 & 0.001 & -0.0017 & 0.1667 & 0.0005 & 0.526 & 0.006 \\
56351.59704 & 29.068 & 0.001 & 0.0018  & 0.1712 & 0.0007 & 0.521 & 0.007 \\
56354.44695 & 29.07  & 0.001 & -0.0012 & 0.1647 & 0.0004 & 0.514 & 0.005 \\
56355.52704 & 29.076 & 0.001 & 0.0006  & 0.171  & 0.0007 & 0.519 & 0.007 \\
56360.42506 & 29.076 & 0.001 & -0.0001 & 0.1657 & 0.0004 & 0.52  & 0.005 \\
56360.51722 & 29.074 & 0.001 & -0.0029 & 0.1751 & 0.0006 & 0.519 & 0.007 \\
56370.50959 & 29.079 & 0.001 & -0.0061 & 0.1709 & 0.0008 & 0.526 & 0.008 \\
56372.49076 & 29.078 & 0.001 & -0.0019 & 0.1677 & 0.0008 & 0.527 & 0.009 \\
56382.36446 & 29.071 & 0.001 & -0.0054 & 0.1694 & 0.0009 & 0.518 & 0.009 \\
56405.33846 & 29.071 & 0.001 & -0.0023 & 0.1722 & 0.0006 & 999.0 & 999.0 \\
56415.36788 & 29.072 & 0.002 & -0.0067 & 0.1747 & 0.0015 & 999.0 & 999.0 \\
56419.4127  & 29.073 & 0.001 & -0.0017 & 0.1707 & 0.0007 & 999.0 & 999.0 \\
56430.35736 & 29.07  & 0.002 & 0.005   & 0.1739 & 0.0017 & 999.0 & 999.0 \\
56435.37259 & 29.069 & 0.002 & 0.0059  & 999.0  & 999.0  & 999.0 & 999.0 \\
56627.71815 & 29.072 & 0.001 & -0.0004 & 0.1761 & 0.0007 & 0.533 & 0.007 \\
56629.7066  & 29.073 & 0.001 & 0.0034  & 0.1658 & 0.0006 & 0.525 & 0.007 \\
56630.6626  & 29.072 & 0.001 & 0.0001  & 0.1669 & 0.0006 & 0.525 & 0.006 \\
56631.68114 & 29.074 & 0.001 & -0.0025 & 0.1707 & 0.0007 & 0.526 & 0.008 \\
56640.58001 & 29.071 & 0.001 & -0.0006 & 0.17   & 0.0009 & 0.523 & 0.008 \\
56653.67292 & 29.073 & 0.001 & 0.0025  & 0.1769 & 0.001  & 0.523 & 0.009 \\
56655.67739 & 29.076 & 0.001 & -0.0039 & 0.1681 & 0.0009 & 0.518 & 0.008 \\
56656.65089 & 29.075 & 0.001 & -0.0055 & 0.1694 & 0.001  & 0.525 & 0.009 \\
56657.63896 & 29.077 & 0.001 & -0.0024 & 0.1717 & 0.001  & 0.516 & 0.01  \\
56663.55861 & 29.074 & 0.001 & -0.0012 & 0.1792 & 0.0008 & 0.515 & 0.008 \\
56665.69581 & 29.075 & 0.001 & 0.0001  & 0.1684 & 0.0008 & 0.514 & 0.008 \\
56672.62532 & 29.074 & 0.001 & -0.0033 & 0.1706 & 0.0011 & 0.527 & 0.01  \\
56680.49332 & 29.078 & 0.001 & -0.0049 & 0.1673 & 0.0006 & 0.522 & 0.007 \\
56681.54471 & 29.081 & 0.001 & -0.0042 & 0.1749 & 0.001  & 0.518 & 0.009 \\
56682.61116 & 29.078 & 0.001 & -0.0012 & 0.1736 & 0.0007 & 0.529 & 0.007 \\
56683.56202 & 29.078 & 0.001 & -0.0058 & 0.1958 & 0.0014 & 0.522 & 0.011 \\
56685.45394 & 29.074 & 0.001 & -0.0031 & 0.1815 & 0.0016 & 0.528 & 0.013 \\
56696.41786 & 29.075 & 0.001 & -0.0011 & 0.1747 & 0.0007 & 0.523 & 0.007 \\
56699.61383 & 29.07  & 0.001 & -0.0019 & 0.1781 & 0.0012 & 0.521 & 0.01  \\
57350.70123 & 29.08  & 0.001 & -0.0014 & 999.0  & 999.0  & 0.524 & 0.009 \\
57390.64252 & 29.078 & 0.001 & -0.0034 & 0.1661 & 0.001  & 0.52  & 0.01  \\
57402.57042 & 29.079 & 0.001 & 0.0022  & 0.1683 & 0.001  & 0.521 & 0.009 \\
57411.45305 & 29.084 & 0.001 & -0.0022 & 0.1812 & 0.0011 & 0.522 & 0.009 \\
57413.56732 & 29.086 & 0.001 & -0.0049 & 0.1688 & 0.001  & 0.516 & 0.009 \\
57436.56638 & 29.078 & 0.001 & -0.0    & 0.1652 & 0.001  & 0.522 & 0.009 \\
57437.64439 & 29.078 & 0.001 & -0.0026 & 0.1912 & 0.0017 & 0.52  & 0.011 \\
57468.4337  & 29.078 & 0.001 & -0.0003 & 999.0  & 999.0  & 0.505 & 0.013 \\
57470.53201 & 29.078 & 0.001 & 0.0011  & 0.1812 & 0.0016 & 0.521 & 0.01  \\
57476.412   & 29.082 & 0.001 & -0.003  & 0.1676 & 0.0006 & 0.524 & 0.007 \\
57498.40682 & 29.082 & 0.001 & -0.0006 & 0.1677 & 0.0006 & 999.0 & 999.0 \\
57503.40866 & 29.08  & 0.001 & -0.005  & 0.1687 & 0.001  & 999.0 & 999.0 \\
57504.43547 & 29.083 & 0.001 & -0.0055 & 0.1802 & 0.0014 & 999.0 & 999.0 \\
57507.38017 & 29.085 & 0.001 & -0.0005 & 0.1688 & 0.0009 & 999.0 & 999.0 \\
57512.4085  & 29.083 & 0.001 & 0.0056  & 0.1609 & 0.0016 & 999.0 & 999.0 \\
57528.33925 & 29.082 & 0.001 & -0.002  & 0.1675 & 0.0014 & 999.0 & 999.0 \\
57699.67587 & 29.082 & 0.001 & -0.0013 & 0.1793 & 0.0011 & 0.523 & 0.01  \\
57701.63228 & 29.084 & 0.002 & -0.0009 & 999.0  & 999.0  & 0.522 & 0.015 \\
57728.6449  & 29.08  & 0.001 & -0.0002 & 0.1704 & 0.0004 & 0.521 & 0.006 \\
57730.6722  & 29.079 & 0.001 & -0.0019 & 0.1725 & 0.0005 & 0.526 & 0.007 \\
57757.69077 & 29.079 & 0.001 & -0.0017 & 0.1758 & 0.001  & 0.522 & 0.01  \\
57758.5679  & 29.079 & 0.001 & -0.0029 & 0.1795 & 0.0012 & 0.529 & 0.011 \\
57759.52851 & 29.082 & 0.001 & -0.0001 & 0.1793 & 0.0013 & 0.524 & 0.012 \\
57766.64339 & 29.082 & 0.001 & 0.0009  & 0.1733 & 0.0007 & 0.519 & 0.008 \\
57770.67824 & 29.08  & 0.001 & -0.0025 & 0.1797 & 0.001  & 0.526 & 0.01  \\
57771.63633 & 29.082 & 0.001 & -0.0023 & 0.1758 & 0.0007 & 0.526 & 0.008 \\
57772.53428 & 29.082 & 0.001 & -0.0012 & 0.1738 & 0.0005 & 0.527 & 0.006 \\
57800.52831 & 29.085 & 0.001 & 0.0007  & 0.1718 & 0.0005 & 0.516 & 0.006 \\
57801.52514 & 29.083 & 0.001 & 0.0021  & 0.1733 & 0.0006 & 0.526 & 0.007 \\
57820.5185  & 29.084 & 0.001 & 0.004   & 0.1773 & 0.0013 & 0.527 & 0.012 \\
57821.46905 & 29.082 & 0.001 & -0.0001 & 0.172  & 0.0008 & 0.516 & 0.009 \\
57823.50327 & 29.083 & 0.001 & -0.0002 & 0.1706 & 0.0006 & 0.522 & 0.007 \\
57826.55084 & 29.084 & 0.001 & 0.0014  & 0.1831 & 0.0009 & 0.514 & 0.008 \\
57827.4408  & 29.086 & 0.001 & -0.0014 & 0.1709 & 0.0005 & 0.508 & 0.006 \\
57829.50094 & 29.086 & 0.001 & 0.0009  & 0.1731 & 0.0005 & 0.524 & 0.006 \\
57830.47692 & 29.082 & 0.001 & -0.0018 & 0.1725 & 0.0007 & 0.523 & 0.008 \\
57831.48271 & 29.084 & 0.002 & 0.0045  & 999.0  & 999.0  & 0.512 & 0.014 \\
57832.46256 & 29.081 & 0.001 & 0.0009  & 0.1662 & 0.0006 & 0.519 & 0.007 \\
57833.47721 & 29.08  & 0.001 & 0.0045  & 0.1682 & 0.0012 & 0.515 & 0.011 \\
57848.41243 & 29.089 & 0.001 & -0.0026 & 0.1752 & 0.0009 & 999.0 & 999.0 \\
57849.36053 & 29.089 & 0.001 & 0.0019  & 0.1889 & 0.001  & 999.0 & 999.0 \\
57850.416   & 29.082 & 0.001 & 0.0018  & 0.1717 & 0.0005 & 999.0 & 999.0 \\
57851.41507 & 29.084 & 0.001 & 0.0014  & 0.1703 & 0.0005 & 999.0 & 999.0 \\
57852.42093 & 29.082 & 0.001 & 0.0004  & 0.1734 & 0.0006 & 999.0 & 999.0 \\
57853.46839 & 29.083 & 0.001 & -0.0017 & 0.1746 & 0.0006 & 999.0 & 999.0 \\
57854.45925 & 29.081 & 0.001 & 0.0001  & 0.1723 & 0.0008 & 999.0 & 999.0 \\
57855.44274 & 29.082 & 0.001 & -0.0026 & 0.1731 & 0.0005 & 999.0 & 999.0 \\
57856.36935 & 29.082 & 0.001 & -0.0019 & 0.1722 & 0.0008 & 999.0 & 999.0 \\
57858.36212 & 29.08  & 0.001 & -0.0016 & 0.1696 & 0.0007 & 999.0 & 999.0 \\
57859.3784  & 29.08  & 0.001 & -0.0013 & 0.1724 & 0.0009 & 999.0 & 999.0 \\
57860.40603 & 29.077 & 0.001 & -0.0005 & 0.1727 & 0.0012 & 999.0 & 999.0 \\
58072.69214 & 29.081 & 0.001 & -0.0007 & 999.0  & 999.0  & 0.525 & 0.008 \\
58073.61653 & 29.08  & 0.001 & -0.0027 & 999.0  & 999.0  & 0.524 & 0.007 \\
58074.69324 & 29.08  & 0.001 & 0.0014  & 999.0  & 999.0  & 0.524 & 0.009 \\
58075.67984 & 29.082 & 0.001 & -0.0027 & 999.0  & 999.0  & 0.525 & 0.009 \\
58076.67888 & 29.078 & 0.002 & -0.0096 & 999.0  & 999.0  & 0.508 & 0.015 \\
58077.70279 & 29.078 & 0.001 & -0.0023 & 999.0  & 999.0  & 0.525 & 0.01  \\
58078.70137 & 29.075 & 0.001 & -0.0051 & 999.0  & 999.0  & 0.526 & 0.01  \\
58090.72611 & 29.084 & 0.001 & 0.0002  & 999.0  & 999.0  & 0.525 & 0.01  \\
58091.68905 & 29.084 & 0.001 & -0.0011 & 999.0  & 999.0  & 0.524 & 0.009 \\
58093.62615 & 29.083 & 0.001 & 0.0016  & 999.0  & 999.0  & 0.524 & 0.007 \\
58100.68476 & 29.085 & 0.001 & -0.0007 & 999.0  & 999.0  & 0.528 & 0.009 \\
58102.70122 & 29.083 & 0.002 & -0.0034 & 999.0  & 999.0  & 0.486 & 0.016 \\
58104.64993 & 29.082 & 0.001 & 0.0011  & 999.0  & 999.0  & 0.525 & 0.011 \\
58108.69417 & 29.088 & 0.001 & -0.0026 & 999.0  & 999.0  & 0.521 & 0.009 \\
58110.70972 & 29.085 & 0.001 & 0.0028  & 999.0  & 999.0  & 0.525 & 0.008 \\
58111.62003 & 29.088 & 0.001 & 0.0024  & 999.0  & 999.0  & 0.523 & 0.007 \\
58130.60025 & 29.092 & 0.001 & -0.0036 & 0.1649 & 0.0008 & 0.525 & 0.008 \\
58131.66504 & 29.09  & 0.001 & -0.0011 & 0.161  & 0.0011 & 0.503 & 0.009 \\
58141.61493 & 29.086 & 0.001 & 0.0048  & 0.1638 & 0.0009 & 0.513 & 0.009 \\
58142.5731  & 29.087 & 0.001 & 0.0003  & 0.1675 & 0.0005 & 0.522 & 0.006 \\
58146.57834 & 29.083 & 0.001 & 0.0     & 0.1651 & 0.0008 & 0.525 & 0.008 \\
58147.54992 & 29.087 & 0.001 & 0.0005  & 0.1634 & 0.0009 & 0.495 & 0.008 \\
58148.5671  & 29.087 & 0.001 & 0.0021  & 0.1636 & 0.001  & 0.522 & 0.01  \\
58151.67525 & 29.088 & 0.001 & -0.0058 & 0.1557 & 0.0018 & 0.526 & 0.013 \\
58152.626   & 29.081 & 0.002 & -0.0048 & 0.1493 & 0.002  & 0.525 & 0.015 \\
58153.56616 & 29.088 & 0.001 & 0.0016  & 0.1621 & 0.0014 & 0.525 & 0.012 \\
58185.49837 & 29.085 & 0.001 & 0.0025  & 0.1661 & 0.0007 & 0.529 & 0.008 \\
58186.47613 & 29.086 & 0.001 & 0.0018  & 0.1672 & 0.0008 & 0.519 & 0.008 \\
58209.54029 & 29.089 & 0.002 & -0.0035 & 999.0  & 999.0  & 999.0 & 999.0 \\
58210.44872 & 29.086 & 0.001 & -0.0005 & 0.17   & 0.0006 & 999.0 & 999.0 \\
58213.54147 & 29.086 & 0.001 & 0.0009  & 999.0  & 999.0  & 999.0 & 999.0 \\
58214.39055 & 29.088 & 0.001 & 0.0004  & 0.1676 & 0.0006 & 999.0 & 999.0 \\
58215.45449 & 29.085 & 0.001 & -0.003  & 0.1674 & 0.0011 & 999.0 & 999.0 \\
58216.45408 & 29.082 & 0.001 & -0.0002 & 0.165  & 0.0011 & 999.0 & 999.0 \\
58218.47574 & 29.084 & 0.002 & -0.0085 & 999.0  & 999.0  & 999.0 & 999.0 \\
58234.37191 & 29.084 & 0.001 & -0.0024 & 0.1684 & 0.0007 & 999.0 & 999.0 \\
58235.44114 & 29.091 & 0.001 & -0.0005 & 0.1707 & 0.0015 & 999.0 & 999.0 \\
58236.38366 & 29.084 & 0.001 & -0.0035 & 0.1733 & 0.0008 & 999.0 & 999.0 \\
58242.46149 & 29.087 & 0.002 & 0.0037  & 999.0  & 999.0  & 999.0 & 999.0 \\
58243.4582  & 29.083 & 0.001 & 0.0004  & 999.0  & 999.0  & 999.0 & 999.0 \\
58245.36182 & 29.091 & 0.001 & -0.0011 & 0.1736 & 0.0011 & 999.0 & 999.0 \\
58255.40949 & 29.09  & 0.001 & 0.0     & 0.1766 & 0.0013 & 999.0 & 999.0 \\
58257.39281 & 29.089 & 0.001 & 0.0039  & 0.1773 & 0.0008 & 999.0 & 999.0 \\
58258.37469 & 29.088 & 0.001 & -0.0016 & 0.1732 & 0.0008 & 999.0 & 999.0 \\
58262.35981 & 29.091 & 0.001 & -0.0016 & 0.1706 & 0.0007 & 999.0 & 999.0 \\
58263.35275 & 29.089 & 0.001 & -0.002  & 0.1729 & 0.0008 & 999.0 & 999.0 \\
58264.35148 & 29.089 & 0.001 & 0.0003  & 0.1723 & 0.0018 & 999.0 & 999.0 \\
58270.34924 & 29.088 & 0.001 & -0.0018 & 0.1716 & 0.0006 & 999.0 & 999.0 \\
58271.35578 & 29.092 & 0.001 & -0.0042 & 0.1714 & 0.001  & 999.0 & 999.0 \\
58272.34719 & 29.084 & 0.002 & -0.0026 & 999.0  & 999.0  & 999.0 & 999.0 \\
58274.38435 & 29.085 & 0.003 & -0.0218 & 999.0  & 999.0  & 999.0 & 999.0 \\
58440.66644 & 29.094 & 0.001 & 0.0009  & 0.1704 & 0.0008 & 0.526 & 0.008 \\
58441.63117 & 29.086 & 0.001 & -0.0025 & 0.1672 & 0.001  & 0.523 & 0.009 \\
58456.70082 & 29.094 & 0.001 & 0.004   & 0.1691 & 0.0008 & 0.515 & 0.009 \\
58457.72054 & 29.091 & 0.001 & 0.0012  & 0.1673 & 0.0005 & 0.52  & 0.006 \\
58459.65022 & 29.091 & 0.001 & -0.0035 & 0.1502 & 0.0013 & 0.513 & 0.012 \\
58460.67423 & 29.095 & 0.001 & -0.0035 & 0.1588 & 0.0012 & 0.51  & 0.01  \\
58467.67996 & 29.092 & 0.001 & -0.0031 & 0.1644 & 0.0011 & 0.53  & 0.01  \\
58475.70316 & 29.085 & 0.001 & -0.0029 & 0.1599 & 0.0013 & 0.519 & 0.011 \\
58486.64971 & 29.091 & 0.002 & 0.001   & 999.0  & 999.0  & 0.521 & 0.017 \\
58487.69909 & 29.089 & 0.003 & 0.0024  & 999.0  & 999.0  & 999.0 & 999.0 \\
58488.69742 & 29.095 & 0.001 & 0.0051  & 0.1672 & 0.001  & 0.529 & 0.01  \\
58489.70608 & 29.09  & 0.002 & 0.0028  & 0.1606 & 0.0019 & 0.527 & 0.014 \\
58490.67879 & 29.094 & 0.001 & -0.002  & 0.1692 & 0.0006 & 0.527 & 0.007 \\
58496.63178 & 29.094 & 0.002 & -0.0059 & 999.0  & 999.0  & 999.0 & 999.0 \\
58497.61321 & 29.09  & 0.001 & 0.0023  & 0.1681 & 0.0014 & 0.516 & 0.012 \\
58498.66066 & 29.094 & 0.001 & 0.0002  & 0.1667 & 0.0013 & 0.528 & 0.011 \\
58499.59632 & 29.091 & 0.001 & 0.0017  & 0.1701 & 0.0006 & 0.526 & 0.007 \\
58526.57706 & 29.098 & 0.001 & 0.0008  & 0.1725 & 0.0011 & 0.533 & 0.011 \\
58527.54347 & 29.09  & 0.001 & 0.0006  & 0.1721 & 0.0012 & 0.527 & 0.011 \\
58528.50643 & 29.091 & 0.001 & 0.0025  & 0.1712 & 0.0005 & 0.527 & 0.006 \\
58529.5612  & 29.09  & 0.001 & 0.001   & 0.1712 & 0.0005 & 0.522 & 0.006 \\
58530.5159  & 29.089 & 0.001 & 0.0004  & 0.1718 & 0.0006 & 0.526 & 0.007 \\
58531.59703 & 29.09  & 0.001 & -0.0016 & 0.1696 & 0.0005 & 0.53  & 0.006 \\
58532.53022 & 29.088 & 0.001 & -0.0036 & 0.17   & 0.0008 & 0.521 & 0.008 \\
58533.4996  & 29.089 & 0.001 & -0.003  & 0.1696 & 0.0007 & 0.526 & 0.007 \\
58534.50645 & 29.091 & 0.001 & -0.0008 & 0.1702 & 0.0007 & 0.527 & 0.008 \\
58535.59314 & 29.091 & 0.001 & -0.0023 & 0.171  & 0.001  & 0.521 & 0.009 \\
58543.57619 & 29.096 & 0.001 & 0.0017  & 0.1692 & 0.001  & 0.525 & 0.009 \\
58544.51305 & 29.097 & 0.001 & -0.0024 & 0.1706 & 0.0011 & 0.528 & 0.01  \\
58545.4534  & 29.094 & 0.001 & -0.0008 & 0.1707 & 0.0011 & 0.522 & 0.01  \\
58546.37522 & 29.094 & 0.001 & 0.0014  & 0.1728 & 0.0006 & 0.526 & 0.006 \\
58548.37912 & 29.093 & 0.001 & -0.0014 & 0.1662 & 0.0009 & 0.52  & 0.009 \\
58550.39724 & 29.093 & 0.001 & 0.0003  & 0.1691 & 0.0017 & 0.522 & 0.013 \\
58551.42881 & 29.092 & 0.001 & 0.0005  & 0.1702 & 0.0007 & 0.524 & 0.008 \\
58552.40384 & 29.099 & 0.001 & -0.0004 & 0.1689 & 0.0009 & 0.519 & 0.009 \\
58553.41709 & 29.094 & 0.001 & -0.0061 & 0.1676 & 0.0008 & 0.525 & 0.009 \\
58554.49277 & 29.09  & 0.002 & -0.001  & 999.0  & 999.0  & 0.524 & 0.014 \\
58556.44137 & 29.095 & 0.001 & 0.0004  & 0.1689 & 0.0009 & 0.526 & 0.009 \\
58557.48776 & 29.095 & 0.001 & 0.0006  & 0.1676 & 0.0007 & 0.518 & 0.008 \\
58559.44023 & 29.095 & 0.001 & 0.0006  & 0.1693 & 0.0007 & 0.527 & 0.008 \\
58560.51233 & 29.092 & 0.001 & -0.0027 & 0.1706 & 0.0012 & 0.529 & 0.011 \\
58561.43576 & 29.096 & 0.001 & 0.0005  & 0.1728 & 0.0009 & 0.522 & 0.009 \\
58563.51862 & 29.098 & 0.001 & -0.0005 & 0.1745 & 0.0008 & 0.529 & 0.008 \\
58586.3874  & 29.094 & 0.001 & 0.0043  & 0.1703 & 0.0009 & 999.0 & 999.0 \\
58587.40305 & 29.096 & 0.001 & 0.0017  & 0.1676 & 0.0013 & 999.0 & 999.0 \\
58588.37817 & 29.094 & 0.001 & -0.0019 & 0.1749 & 0.0017 & 999.0 & 999.0 \\
58590.4154  & 29.095 & 0.001 & -0.0001 & 0.169  & 0.0004 & 999.0 & 999.0 \\
58591.34231 & 29.095 & 0.001 & 0.0027  & 0.1631 & 0.0012 & 999.0 & 999.0 \\
58592.45683 & 29.096 & 0.002 & -0.0032 & 999.0  & 999.0  & 999.0 & 999.0 \\
58593.42747 & 29.092 & 0.001 & 0.0006  & 0.1745 & 0.0008 & 999.0 & 999.0 \\
58598.49199 & 29.095 & 0.003 & -0.002  & 999.0  & 999.0  & 999.0 & 999.0 \\
58600.48275 & 29.091 & 0.001 & -0.0043 & 999.0  & 999.0  & 999.0 & 999.0 \\
58601.42492 & 29.093 & 0.001 & -0.0005 & 0.1736 & 0.0012 & 999.0 & 999.0 \\
58602.32023 & 29.091 & 0.001 & -0.0008 & 0.1715 & 0.001  & 999.0 & 999.0 \\
58603.34279 & 29.093 & 0.001 & -0.0003 & 0.1727 & 0.0009 & 999.0 & 999.0 \\
58604.36121 & 29.094 & 0.001 & -0.0023 & 0.1736 & 0.0006 & 999.0 & 999.0 \\
58605.37671 & 29.097 & 0.001 & -0.0013 & 0.1786 & 0.0009 & 999.0 & 999.0 \\
58606.38135 & 29.093 & 0.001 & -0.0018 & 0.1775 & 0.001  & 999.0 & 999.0 \\
58607.37926 & 29.089 & 0.001 & 0.0054  & 0.176  & 0.0016 & 999.0 & 999.0 \\
58608.39479 & 29.093 & 0.001 & -0.0009 & 0.1739 & 0.0012 & 999.0 & 999.0 \\
58610.38085 & 29.098 & 0.001 & 0.0005  & 0.1761 & 0.0007 & 999.0 & 999.0 \\
58613.41864 & 29.091 & 0.002 & -0.006  & 999.0  & 999.0  & 999.0 & 999.0 \\
58616.33982 & 29.096 & 0.001 & -0.0001 & 0.1745 & 0.0013 & 999.0 & 999.0 \\
58617.33881 & 29.092 & 0.001 & -0.0012 & 0.1719 & 0.0006 & 999.0 & 999.0 \\
58619.4107  & 29.086 & 0.001 & -0.0036 & 0.1778 & 0.0012 & 999.0 & 999.0 \\
58620.38094 & 29.089 & 0.001 & 0.0032  & 0.1739 & 0.0009 & 999.0 & 999.0 \\
58624.41277 & 29.093 & 0.001 & -0.0002 & 999.0  & 999.0  & 999.0 & 999.0 \\
58625.39701 & 29.085 & 0.002 & 0.0102  & 999.0  & 999.0  & 999.0 & 999.0 \\
58626.36614 & 29.09  & 0.001 & 0.0019  & 0.1777 & 0.0007 & 999.0 & 999.0 \\
58627.37623 & 29.089 & 0.001 & 0.0022  & 0.18   & 0.0011 & 999.0 & 999.0 \\
58630.37143 & 29.092 & 0.001 & 0.0002  & 0.1747 & 0.0008 & 999.0 & 999.0 \\
58631.35476 & 29.092 & 0.001 & 0.0002  & 0.172  & 0.0013 & 999.0 & 999.0 \\
58632.35378 & 29.085 & 0.001 & -0.0019 & 0.1656 & 0.0013 & 999.0 & 999.0 \\
58633.35156 & 29.088 & 0.001 & -0.0037 & 0.169  & 0.001  & 999.0 & 999.0 \\
58634.34584 & 29.09  & 0.001 & 0.0031  & 0.1734 & 0.0007 & 999.0 & 999.0 \\
58635.34845 & 29.089 & 0.001 & -0.0007 & 0.1727 & 0.0008 & 999.0 & 999.0 \\
58636.36255 & 29.091 & 0.001 & -0.0017 & 0.1777 & 0.001  & 999.0 & 999.0 \\
58637.35109 & 29.09  & 0.001 & -0.0007 & 0.174  & 0.0008 & 999.0 & 999.0 \\
58820.63063 & 29.091 & 0.001 & -0.003  & 0.1801 & 0.0011 & 0.516 & 0.009 \\
58821.60302 & 29.094 & 0.001 & -0.0038 & 0.1796 & 0.0013 & 0.512 & 0.01  \\
58824.67421 & 29.095 & 0.001 & 0.0003  & 0.1764 & 0.0007 & 0.52  & 0.007 \\
58828.61484 & 29.094 & 0.001 & 0.0015  & 0.1735 & 0.0007 & 0.53  & 0.008 \\
58852.6337  & 29.097 & 0.002 & 0.0044  & 999.0  & 999.0  & 0.527 & 0.017 \\
58853.55459 & 29.105 & 0.001 & 0.0028  & 0.189  & 0.0016 & 0.524 & 0.013 \\
58854.58145 & 29.096 & 0.001 & -0.0021 & 0.1818 & 0.0007 & 0.525 & 0.008 \\
58855.57747 & 29.094 & 0.001 & -0.0005 & 0.1844 & 0.0009 & 0.53  & 0.009 \\
58856.6402  & 29.099 & 0.001 & 0.0013  & 0.185  & 0.0011 & 0.514 & 0.01  \\
58856.73025 & 29.096 & 0.001 & 0.0001  & 0.1882 & 0.0014 & 0.518 & 0.011 \\
58875.6166  & 29.099 & 0.001 & -0.0043 & 0.1808 & 0.0006 & 0.525 & 0.006 \\
58877.61726 & 29.099 & 0.001 & -0.0014 & 0.1804 & 0.0008 & 0.515 & 0.008 \\
58881.59325 & 29.099 & 0.001 & 0.0011  & 0.1845 & 0.0009 & 0.512 & 0.008 \\
58882.53817 & 29.095 & 0.001 & -0.0039 & 0.193  & 0.0016 & 0.512 & 0.012 \\
58883.54223 & 29.096 & 0.001 & -0.0011 & 0.1874 & 0.0011 & 0.521 & 0.01  \\
58884.46423 & 29.1   & 0.001 & -0.0048 & 0.1967 & 0.0018 & 0.527 & 0.014 \\
58885.38578 & 29.101 & 0.002 & -0.0014 & 999.0  & 999.0  & 0.517 & 0.019 \\
58886.50785 & 29.102 & 0.001 & 0.0004  & 0.1805 & 0.0006 & 0.535 & 0.007 \\
58887.45579 & 29.097 & 0.001 & -0.0026 & 0.1833 & 0.0011 & 0.524 & 0.011 \\
58888.50882 & 29.096 & 0.001 & -0.0002 & 0.1791 & 0.0005 & 0.524 & 0.005 \\
58891.56505 & 29.097 & 0.001 & 0.0004  & 0.1891 & 0.0013 & 0.52  & 0.011 \\
58892.55219 & 29.1   & 0.001 & -0.0005 & 0.1843 & 0.0007 & 0.532 & 0.008 \\
58893.5583  & 29.098 & 0.001 & 0.0028  & 0.191  & 0.0013 & 0.524 & 0.011 \\
58894.48852 & 29.099 & 0.001 & -0.0009 & 0.18   & 0.0007 & 0.524 & 0.008 \\
58897.58099 & 29.094 & 0.003 & -0.0036 & 999.0  & 999.0  & 999.0 & 999.0 \\
58898.50343 & 29.096 & 0.001 & 0.0014  & 0.185  & 0.001  & 999.0 & 999.0 \\
58906.53712 & 29.095 & 0.001 & 0.001   & 0.1838 & 0.0013 & 0.518 & 0.011 \\
58907.50513 & 29.096 & 0.001 & 0.0035  & 0.1838 & 0.0013 & 0.526 & 0.011 \\
58911.46747 & 29.098 & 0.001 & 0.0016  & 0.1893 & 0.0015 & 0.527 & 0.012 \\
58912.49592 & 29.094 & 0.001 & 0.0019  & 0.187  & 0.0017 & 0.526 & 0.014 \\
58913.4537  & 29.097 & 0.001 & 0.0052  & 0.1752 & 0.001  & 0.52  & 0.01  \\
58914.41822 & 29.1   & 0.001 & 0.0015  & 0.1894 & 0.0014 & 0.519 & 0.012 \\
58916.42498 & 29.099 & 0.001 & -0.0002 & 0.1903 & 0.0013 & 0.535 & 0.012 \\
58918.44042 & 29.101 & 0.001 & 0.0024  & 0.1933 & 0.0016 & 0.527 & 0.013 \\
58919.48435 & 29.099 & 0.002 & 0.0003  & 999.0  & 999.0  & 0.474 & 0.012 \\
58920.43465 & 29.096 & 0.001 & 0.0004  & 0.18   & 0.0007 & 0.522 & 0.008 \\
58924.46581 & 29.101 & 0.001 & 0.0001  & 0.1776 & 0.0009 & 0.528 & 0.009 \\
59170.59242 & 29.108 & 0.001 & -0.0077 & 0.2018 & 0.0013 & 0.518 & 0.009 \\
59171.67277 & 29.104 & 0.001 & -0.0012 & 0.1707 & 0.0005 & 0.524 & 0.006 \\
59172.59053 & 29.105 & 0.001 & -0.0036 & 0.2086 & 0.0011 & 0.515 & 0.006 \\
59175.67902 & 29.1   & 0.001 & -0.0027 & 0.1753 & 0.0007 & 0.519 & 0.008 \\
59182.69165 & 29.103 & 0.001 & 0.0006  & 0.172  & 0.0005 & 0.516 & 0.006 \\
59183.6644  & 29.104 & 0.001 & -0.0018 & 0.1798 & 0.0006 & 0.523 & 0.007 \\
59184.66932 & 29.098 & 0.002 & 0.001   & 0.1895 & 0.0018 & 0.513 & 0.014 \\
59186.62898 & 29.105 & 0.001 & -0.0007 & 0.1748 & 0.0007 & 0.526 & 0.008 \\
59197.73091 & 29.102 & 0.001 & -0.001  & 999.0  & 999.0  & 0.524 & 0.007 \\
59203.65678 & 29.109 & 0.002 & -0.0044 & 0.1674 & 0.0019 & 0.514 & 0.015 \\
59205.62881 & 29.108 & 0.001 & 0.0043  & 0.1709 & 0.0006 & 0.518 & 0.007 \\
59206.6322  & 29.098 & 0.001 & -0.0025 & 0.1701 & 0.0011 & 0.516 & 0.011 \\
59247.50061 & 29.103 & 0.001 & -0.001  & 0.1749 & 0.0014 & 999.0 & 999.0 \\
59248.63037 & 29.108 & 0.001 & -0.001  & 0.174  & 0.0006 & 0.516 & 0.007 \\
59249.56861 & 29.105 & 0.001 & -0.003  & 999.0  & 999.0  & 0.518 & 0.01  \\
59263.47458 & 29.106 & 0.001 & -0.002  & 0.1724 & 0.0006 & 0.526 & 0.007 \\
59265.54794 & 29.105 & 0.001 & -0.0031 & 0.176  & 0.0009 & 0.525 & 0.009 \\
59266.41158 & 29.104 & 0.001 & 0.0002  & 0.1763 & 0.0008 & 0.526 & 0.008 \\
59267.37249 & 29.106 & 0.002 & -0.0034 & 999.0  & 999.0  & 0.52  & 0.02  \\
59269.38117 & 29.105 & 0.001 & 0.0014  & 0.1739 & 0.0007 & 0.525 & 0.008 \\
59271.4168  & 29.106 & 0.001 & 0.0011  & 0.1725 & 0.0005 & 0.53  & 0.007 \\
59272.37643 & 29.105 & 0.001 & 0.0042  & 0.1777 & 0.0009 & 0.525 & 0.009 \\
59273.50018 & 29.103 & 0.001 & 0.0036  & 0.1775 & 0.0009 & 0.522 & 0.008 \\
59274.52433 & 29.107 & 0.001 & -0.0007 & 0.1729 & 0.0005 & 0.527 & 0.007 \\
59275.53869 & 29.109 & 0.001 & -0.0002 & 0.1728 & 0.0006 & 0.526 & 0.007 \\
59277.48064 & 29.109 & 0.001 & -0.0001 & 0.1775 & 0.0009 & 0.522 & 0.009 \\
59278.45468 & 29.109 & 0.001 & -0.0007 & 0.1737 & 0.0008 & 0.523 & 0.008 \\
59279.43298 & 29.108 & 0.001 & 0.0007  & 0.175  & 0.0006 & 0.524 & 0.007 \\
59280.4677  & 29.104 & 0.001 & -0.0049 & 0.1808 & 0.0012 & 0.517 & 0.01  \\
59327.42527 & 29.104 & 0.001 & -0.0025 & 0.1785 & 0.0009 & 999.0 & 999.0 \\
59328.43132 & 29.106 & 0.001 & -0.0037 & 0.1742 & 0.0007 & 999.0 & 999.0 \\
59329.37507 & 29.106 & 0.001 & -0.0012 & 0.1727 & 0.0006 & 999.0 & 999.0 \\
59330.42242 & 29.105 & 0.002 & -0.0119 & 999.0  & 999.0  & 999.0 & 999.0 \\
59336.36732 & 29.108 & 0.001 & 0.0007  & 0.1743 & 0.0011 & 999.0 & 999.0 \\
59337.39674 & 29.11  & 0.001 & 0.0027  & 0.1788 & 0.0009 & 999.0 & 999.0 \\
59339.38104 & 29.109 & 0.001 & 0.0026  & 0.1717 & 0.0008 & 999.0 & 999.0 \\
59340.36174 & 29.11  & 0.001 & 0.0012  & 0.1725 & 0.0007 & 999.0 & 999.0 \\
59342.33408 & 29.11  & 0.001 & 0.0017  & 0.1732 & 0.0007 & 999.0 & 999.0 \\
59343.35287 & 29.103 & 0.001 & -0.0023 & 0.1718 & 0.0006 & 999.0 & 999.0 \\
59346.36233 & 29.101 & 0.001 & -0.0025 & 0.1721 & 0.0012 & 999.0 & 999.0 \\
59347.33788 & 29.11  & 0.001 & -0.0001 & 0.1705 & 0.0006 & 999.0 & 999.0 \\
59349.34036 & 29.106 & 0.001 & -0.0041 & 0.1816 & 0.0013 & 999.0 & 999.0 \\
59351.4423  & 29.107 & 0.002 & -0.0066 & 999.0  & 999.0  & 999.0 & 999.0 \\
59351.45289 & 29.107 & 0.002 & 0.0035  & 999.0  & 999.0  & 999.0 & 999.0 \\
59352.34585 & 29.108 & 0.001 & -0.0014 & 0.1771 & 0.0009 & 999.0 & 999.0 \\
59354.35096 & 29.106 & 0.001 & -0.001  & 0.1764 & 0.0014 & 999.0 & 999.0 \\
59355.35012 & 29.107 & 0.002 & -0.0035 & 999.0  & 999.0  & 999.0 & 999.0 \\
59357.37102 & 29.106 & 0.001 & -0.0012 & 999.0  & 999.0  & 999.0 & 999.0 \\
59363.3513  & 29.107 & 0.001 & 0.0004  & 0.1755 & 0.001  & 999.0 & 999.0 \\
59586.50556 & 29.115 & 0.001 & -0.0067 & 0.186  & 0.0008 & 0.53  & 0.008 \\
59590.52159 & 29.117 & 0.001 & -0.0015 & 0.19   & 0.0013 & 0.504 & 0.011 \\
59593.4655  & 29.11  & 0.001 & -0.0032 & 0.1857 & 0.0007 & 0.529 & 0.007 \\
59594.5332  & 29.112 & 0.001 & -0.0003 & 0.1765 & 0.0006 & 0.53  & 0.007 \\
59602.6255  & 29.12  & 0.001 & -0.0001 & 0.1818 & 0.0006 & 0.526 & 0.007 \\
59605.44167 & 29.12  & 0.001 & 0.0011  & 0.1895 & 0.0008 & 0.527 & 0.008 \\
59606.47317 & 29.119 & 0.001 & -0.0017 & 0.185  & 0.0009 & 0.521 & 0.008 \\
59607.57129 & 29.118 & 0.001 & 0.0046  & 0.1907 & 0.0009 & 0.528 & 0.009 \\
59608.67797 & 29.116 & 0.001 & -0.0013 & 0.1938 & 0.0012 & 0.521 & 0.01  \\
59609.60292 & 29.117 & 0.001 & 0.0011  & 0.1878 & 0.001  & 0.525 & 0.009 \\
59610.51254 & 29.115 & 0.001 & 0.0021  & 0.1853 & 0.0008 & 0.523 & 0.009 \\
59620.52934 & 29.111 & 0.001 & -0.0019 & 0.1789 & 0.0007 & 0.523 & 0.008 \\
59621.56096 & 29.112 & 0.001 & -0.0011 & 0.1847 & 0.001  & 0.513 & 0.009 \\
59622.47061 & 29.112 & 0.001 & -0.0006 & 0.1833 & 0.0008 & 0.522 & 0.008 \\
59623.53538 & 29.116 & 0.001 & -0.0011 & 0.1872 & 0.001  & 0.525 & 0.01  \\
59625.55919 & 29.112 & 0.001 & -0.0029 & 0.1812 & 0.0008 & 0.52  & 0.008 \\
59628.46108 & 29.111 & 0.001 & -0.0024 & 0.1871 & 0.0012 & 0.519 & 0.011 \\
59630.49989 & 29.115 & 0.001 & -0.0012 & 0.1978 & 0.0015 & 0.526 & 0.012 \\
59633.41679 & 29.118 & 0.001 & -0.0023 & 0.1834 & 0.0008 & 0.529 & 0.008 \\
59634.56414 & 29.115 & 0.001 & 0.0021  & 0.178  & 0.0008 & 0.525 & 0.008 \\
59635.48433 & 29.12  & 0.001 & -0.0003 & 0.1852 & 0.001  & 0.523 & 0.01  \\
59636.55304 & 29.121 & 0.001 & 0.0003  & 0.1847 & 0.0009 & 0.528 & 0.009 \\
59637.57302 & 29.111 & 0.001 & 0.0009  & 0.181  & 0.0008 & 0.526 & 0.008 \\
59638.51495 & 29.115 & 0.001 & 0.002   & 0.1796 & 0.0009 & 0.506 & 0.008 \\
59639.54117 & 29.113 & 0.001 & 0.0006  & 0.1851 & 0.001  & 0.525 & 0.01  \\
59640.4762  & 29.112 & 0.001 & -0.0016 & 0.1914 & 0.0013 & 0.525 & 0.011 \\
59641.56528 & 29.118 & 0.001 & 0.0018  & 0.1905 & 0.0009 & 0.524 & 0.008 \\
59642.51285 & 29.113 & 0.001 & -0.0019 & 0.1843 & 0.0008 & 0.522 & 0.008 \\
59644.46157 & 29.122 & 0.001 & 0.0049  & 0.1885 & 0.0013 & 0.523 & 0.01  \\
59645.56396 & 29.122 & 0.001 & 0.0024  & 0.2027 & 0.0015 & 0.509 & 0.01  \\
59646.45423 & 29.121 & 0.001 & 0.0013  & 0.1883 & 0.0009 & 0.527 & 0.009 \\
59648.51781 & 29.12  & 0.001 & 0.0021  & 0.1851 & 0.0006 & 0.525 & 0.007 \\
59658.43432 & 29.114 & 0.001 & -0.0043 & 0.1906 & 0.001  & 0.526 & 0.01  \\
59660.5041  & 29.113 & 0.001 & 0.0004  & 0.1848 & 0.001  & 0.528 & 0.009 \\
59661.43835 & 29.112 & 0.001 & 0.0012  & 0.1801 & 0.0008 & 0.532 & 0.008 \\
59662.40771 & 29.118 & 0.001 & 0.0017  & 0.1786 & 0.0006 & 0.525 & 0.007 \\
59663.4597  & 29.113 & 0.001 & -0.0003 & 0.1785 & 0.0006 & 0.528 & 0.007 \\
59678.33034 & 29.113 & 0.001 & -0.0036 & 0.1757 & 0.0005 & 999.0 & 999.0 \\
59683.40842 & 29.116 & 0.001 & 0.0003  & 0.1769 & 0.0006 & 999.0 & 999.0 \\
59686.49836 & 29.118 & 0.001 & 0.0032  & 0.2257 & 0.0017 & 999.0 & 999.0 \\
59889.7015  & 29.123 & 0.001 & 0.0008  & 0.1778 & 0.0006 & 0.518 & 0.007 \\
59890.67476 & 29.126 & 0.001 & 0.0012  & 999.0  & 999.0  & 0.521 & 0.014 \\
59902.69906 & 29.118 & 0.001 & -0.0017 & 0.1811 & 0.0011 & 0.524 & 0.01  \\
59903.66273 & 29.122 & 0.001 & 0.0032  & 0.185  & 0.0011 & 0.523 & 0.01  \\
59919.63625 & 29.128 & 0.001 & -0.0002 & 0.1802 & 0.0008 & 0.527 & 0.009 \\
59920.64785 & 29.132 & 0.001 & -0.0011 & 0.1769 & 0.0007 & 0.52  & 0.008 \\
59921.57429 & 29.126 & 0.001 & 0.0026  & 0.181  & 0.0009 & 0.526 & 0.008 \\
59937.67352 & 29.121 & 0.001 & -0.0072 & 0.1894 & 0.0015 & 0.51  & 0.012 \\
59958.60953 & 29.123 & 0.001 & -0.0013 & 0.1789 & 0.0009 & 0.528 & 0.009 \\
59959.57843 & 29.118 & 0.002 & -0.0036 & 999.0  & 999.0  & 0.506 & 0.014 \\
59972.57361 & 29.124 & 0.001 & 0.0028  & 0.1877 & 0.0009 & 0.527 & 0.009 \\
59973.54181 & 29.125 & 0.001 & -0.001  & 0.1816 & 0.0007 & 0.521 & 0.008 \\
59974.51399 & 29.127 & 0.001 & -0.0011 & 0.178  & 0.0006 & 0.529 & 0.007 \\
59975.61481 & 29.133 & 0.001 & 0.0006  & 0.1797 & 0.0012 & 0.524 & 0.011 \\
59976.53289 & 29.128 & 0.001 & -0.0018 & 0.1851 & 0.0011 & 0.515 & 0.01  \\
59978.45779 & 29.121 & 0.001 & -0.0021 & 0.1819 & 0.0008 & 0.527 & 0.008 \\
59980.4728  & 29.119 & 0.001 & 0.0011  & 0.179  & 0.0014 & 0.529 & 0.012 \\
59983.48087 & 29.127 & 0.001 & -0.0004 & 0.1876 & 0.0012 & 0.529 & 0.011 \\
59984.62393 & 29.12  & 0.001 & -0.0009 & 0.1786 & 0.0008 & 0.527 & 0.008 \\
59988.46075 & 29.123 & 0.001 & -0.0013 & 999.0  & 999.0  & 0.524 & 0.006 \\
59990.5449  & 29.122 & 0.001 & -0.0003 & 999.0  & 999.0  & 0.527 & 0.006 \\
60006.45527 & 29.125 & 0.001 & 0.0014  & 0.1775 & 0.0006 & 0.526 & 0.007 \\
60007.43307 & 29.129 & 0.001 & 0.0012  & 0.178  & 0.0006 & 0.529 & 0.007 \\
60008.49984 & 29.129 & 0.001 & -0.0003 & 0.1798 & 0.0007 & 0.534 & 0.008 \\
60009.54012 & 29.129 & 0.001 & 0.0006  & 0.1808 & 0.0007 & 0.53  & 0.007 \\
60010.53674 & 29.131 & 0.001 & 0.0025  & 0.1809 & 0.0007 & 0.528 & 0.008 \\
60013.52425 & 29.129 & 0.001 & -0.0012 & 0.1891 & 0.0013 & 0.508 & 0.011 \\
60016.39097 & 29.134 & 0.001 & -0.002  & 0.186  & 0.001  & 0.524 & 0.009\\
\end{longtable}

\twocolumn

\onecolumn

\begin{table}\centering
\caption{HIRES RVs for HD88986}
\label{tab:rvs hires}
\begin{tabular}{llll}
\hline
BJD (-2400000 d)&       RV (m s$^{-1}$ )  & $\sigma_{RV}$ (m s$^{-1}$ )& S-index \\
\hline
HIRES:&&&\\
50420.10946 & 22.494 & 1.643 & 0.1382 \\
50463.00185 & 20.158 & 1.406 & 0.1397 \\
50545.86786 & 20.750 & 1.560 & 0.1422 \\
50787.06992 & 26.429 & 3.740 & 0.1542 \\
50787.07411 & 28.104 & 1.608 & 0.1521 \\
50838.01219 & 19.714 & 1.467 & 0.1527 \\
50954.80494 & 29.031 & 1.550 & 0.1540 \\
50955.75722 & 25.449 & 1.522 & 0.1510 \\
51171.99838 & 20.624 & 1.395 & 0.1508 \\
51229.00435 & 10.522 & 1.389 & 0.1529 \\
51341.82219 & 21.790 & 1.538 & 0.1604 \\
51551.06319 & 20.628 & 1.536 & 0.1382 \\
51982.00707 & 9.603  & 1.384 & 0.1272 \\
52308.02544 & 10.857 & 1.870 & 0.1273 \\
52601.15939 & 1.742  & 1.715 & 0.1373 \\
52712.96215 & 0.661  & 1.464 & 0.1309 \\
52988.06928 & 2.112  & 1.521 & 0.1224\\
\hline
HIRES+:&&&\\
53370.00523 & 8.157   & 1.533 & 0.1336 \\
53841.88252 & 5.906   & 1.406 & 0.1361 \\
55289.71865 & -1.037  & 1.585 & 0.1270 \\
55339.84111 & 3.448   & 1.336 & 0.1425 \\
55669.86880 & -0.145  & 1.349 & 0.1300 \\
55707.73738 & -3.202  & 1.436 & 0.1382 \\
55719.83472 & 1.936   & 1.365 & 0.1346 \\
55719.83589 & -1.386  & 1.374 & 0.1379 \\
55720.79264 & 1.047   & 1.365 & 0.1410 \\
55720.79490 & -3.926  & 1.315 & 0.1405 \\
55721.81729 & -8.243  & 1.434 & 0.1409 \\
55721.81885 & -5.055  & 1.434 & 0.1436 \\
55721.82031 & -4.837  & 1.285 & 0.1427 \\
55749.77312 & -6.456  & 1.447 & 0.1434 \\
55749.77472 & -4.918  & 1.348 & 0.1487 \\
55750.76293 & 0.377   & 1.447 & 0.1403 \\
55750.76616 & -2.757  & 1.536 & 0.1417 \\
55912.09590 & 7.933   & 1.321 & 0.1332 \\
55912.09693 & 3.431   & 1.350 & 0.1327 \\
55912.09794 & 3.180   & 1.400 & 0.1307 \\
55971.91094 & -2.303  & 1.531 & 0.1245 \\
55971.91211 & -3.924  & 1.473 & 0.1274 \\
55971.91332 & -1.016  & 1.531 & 0.1264 \\
55997.01689 & -10.852 & 1.618 & 0.1317 \\
56024.88359 & -8.907  & 1.289 & 0.1367 \\
56024.88929 & -14.024 & 1.346 & 0.1373 \\
56024.89104 & -10.796 & 1.289 & 0.1362 \\
56024.89298 & -6.688  & 1.221 & 0.1371 \\
56329.97966 & -5.863  & 1.541 & 0.1314 \\
56329.98069 & -6.265  & 1.472 & 0.1343 \\
56329.98178 & -5.186  & 1.561 & 0.1368 \\
56676.89512 & -3.038  & 1.427 & 0.1370 \\
56676.89610 & -3.379  & 1.465 & 0.1361\\
\hline
\hline
\end{tabular}
    
\end{table}     

\onecolumn

\begin{table}[h]\centering
\caption{ELODIE RVs for HD88986}
\label{tab:rvs elodie}
\begin{tabular}{lll}
\hline
BJD (-2400000 d)&       RV (m s$^{-1}$ )  & $\sigma_{RV}$ (m s$^{-1}$ )\\
\hline
50508.4203 & 29009.240546 & 7.201594  \\
50509.5491 & 28987.240514 & 7.943309  \\
50509.56   & 28998.240514 & 7.467904  \\
50510.5337 & 28995.240487 & 7.111546  \\
50533.4545 & 29000.739841 & 7.351913  \\
50554.435  & 29012.139250 & 8.299547  \\
50554.4436 & 29012.139250 & 8.843668  \\
50584.3377 & 29016.838409 & 8.021465  \\
50770.7146 & 29026.833161 & 7.937892  \\
50821.6156 & 29032.831728 & 7.838441  \\
50858.5533 & 29017.030688 & 7.324449  \\
50885.4545 & 28999.829930 & 7.506361  \\
50886.4527 & 29013.829902 & 8.826301  \\
50887.4928 & 29010.829873 & 8.564956  \\
50939.3686 & 29016.328412 & 7.989163  \\
50939.3776 & 29010.328412 & 8.438047  \\
50973.345  & 29005.327455 & 7.360800  \\
51153.7066 & 29009.222377 & 9.974239  \\
51235.5637 & 29004.820072 & 7.097413  \\
51505.6993 & 29002.412466 & 9.181249  \\
51562.5653 & 29016.910865 & 9.396184  \\
51725.365  & 29003.006281 & 9.713716  \\
51901.6319 & 28986.201318 & 9.321223  \\
51982.4489 & 29000.399042 & 8.947926  \\
52723.4565 & 28984.278178 & 7.554854  \\
52989.6915 & 28981.170682 & 10.367050 \\
52995.6586 & 28991.170514 & 7.902043  \\
53033.6212 & 28987.869445 & 11.367766\\
\hline
\hline
\end{tabular}
    
\end{table}     

\section{Background contamination of the calibration lamp from the SOPHIE spectra}
\label{activity_back}

The combined analysis of RVs and activity indicators plays a crucial role in determining the origin of a signal. To do so, having accurate activity indicators is essential for effectively interpreting the data. In CCD spectrograph images, the recorded light from fibers A and B in a spectral order are located close to each other (e.g., within approximately 17 pixels for SOPHIE). This proximity introduces a small but non-negligible amount of light diffusion, primarily caused by the calibration lamp's light from fiber B to fiber A \citep{lovis2011harps}. Consequently, before deriving the activity indexes such as $\log R'_{\rm HK}$ and $H \alpha$, it becomes imperative to subtract the background light originating from the diffuse light emitted by either the Th-Ar or FP calibration lamp present in the star spectrum.

To address this contamination issue, the SOPHIE DRS employs various methods, depending on the calibration lamps used. In the case of the Th-Ar lamp, a background is estimated from the flux of fiber B in the same spectral order by fitting a polynomial function on local minima \citep{boisse2010sophie}. On the other hand, for the FP lamp, which has been installed since semester 2017B, the background is directly measured using a Dark-FP frame, that is, illumination of fiber B with the FP calibration lamp while keeping fiber A completely dark \citep{hobson2019exoplanet,lovis2011harps}. However, as more years of observations were conducted, it became evident that there was a noticeable discrepancy in the data obtained from the two calibration lamps. This discrepancy can be attributed to the utilization of different background correction methods and likely over-estimation of background contamination in the method used in Th-Ar data. In order to rectify this issue, we employed a direct measuring method \citep{hobson2019exoplanet,lovis2011harps} for both calibration lamps. By implementing this approach, we successfully corrected this discrepancy and significantly improved the consistency between data sets obtained using different calibration lamps. This method has been implemented into the SOPHIE DRS and will be utilized in forthcoming SOPHIE planet publications. We applied this method to the SOPHIE spectra prior to deriving the H$\alpha$ and $\log R'_{\rm HK}$ (or S-index) activity indices for HD88986. We note that we did not use the H$\alpha$ activity indexes in our final analysis due to its high contamination by the telluric.

\section{Priors on HD88986\,b for RV-only model}

\begin{table*}[h!]
\centering
\caption{Priors and description of parameters used within \texttt{juliet} to model RVs of HD88986 in Sect. \ref{rv_several}.}
\resizebox{0.7\columnwidth}{!}{%
\begin{tabular}{lccccc}
\hline
Parameter & prior & description \\
\hline
Planet parameters & & \\
P (d) & $\mathcal{U}(135,155)$ & Period of HD88986\,b \\
T$_{c}$-2400000 (d) & $\mathcal{U} (58850,58850+146)$ & Center of transit time for HD88986\,b \\
K (m/s) & $\mathcal{U}(0,10)$ & RV semi-amplitude for HD88986\,b \\
e & $\mathcal{U}(0,1)$ or 0 (fixed) & Eccentricity \\
$\omega$ ($^\circ$) & $\mathcal{U}(0,360)$ or 90 (fixed) & Argument of periastron \\
\\
Telescope Parameters & & \\
$\sigma_{SOPHIE+}$ (m/s) & $\mathcal{U} (1e-3, 100.)$ & RV jitter \\
mu$_{SOPHIE+}$ (m/s) & $\log \mathcal{U} (28995,29196)$ & Instrumental offset \\
\\
Drift on SHOPHIE+ & & \\
A (m/s) & $\mathcal{U} (-5,5)$ & Linear RV drift \\
Q (m/s) & $\mathcal{U} (-5,5)$ & Quadratic RV drift \\
\\
QP-GP on SOPHIE+ & & \\
B$_{GP}$ (m/s) & $\mathcal{J} (10^{-5},100)$ & Amplitude of the GP kernel \\
C$_{GP}$ (m/s) & $10^{-20}$ (fixed) & Constant scaling term of the GP kernel \\
P$_{rot}$ (d) & $\mathcal{N} (29,3)$ & Rotation period of the GP kernel \\
L$_{GP}$ (d) & $\mathcal{J} (10^{-20},300)$ & Correlation time-scale of the GP kernel \\
\\
EXP-GP on SOPHIE+ & & \\
T$_{GP}$ & $\log \mathcal{U} (1e-13,100.)$ & Length-scale of the GP kernel \\
$\sigma_{GP}$ (m/s) & $\log \mathcal{U} (1e-13,100.0)$ & Amplitude of the GP kernel \\
\\
Sinusoidal on SOPHIE+ & & \\
P (d) & $\mathcal{U}(25,35)$ & Period for the additional sinusoid \\
K (m/s) & $\mathcal{U}(58870,58870+30)$ & RV semi-amplitude for the additional sinusoid \\
T$_{c}$-2400000 (d) & $\mathcal{U} (0,10)$ & Center of transit time for additional sinusoid \\
\hline
\end{tabular}%
}
\tablefoot{The prior labels of $\mathcal{N}$, $\mathcal{U}$, $\log \mathcal{U}$ indicate normal, uniform, and uniform logarithms of distributions.}
\label{prior_rv-only}
\end{table*}

\section{False positive tests}
\begin{figure*}[h!]
\centering
\includegraphics[width=0.48\columnwidth]{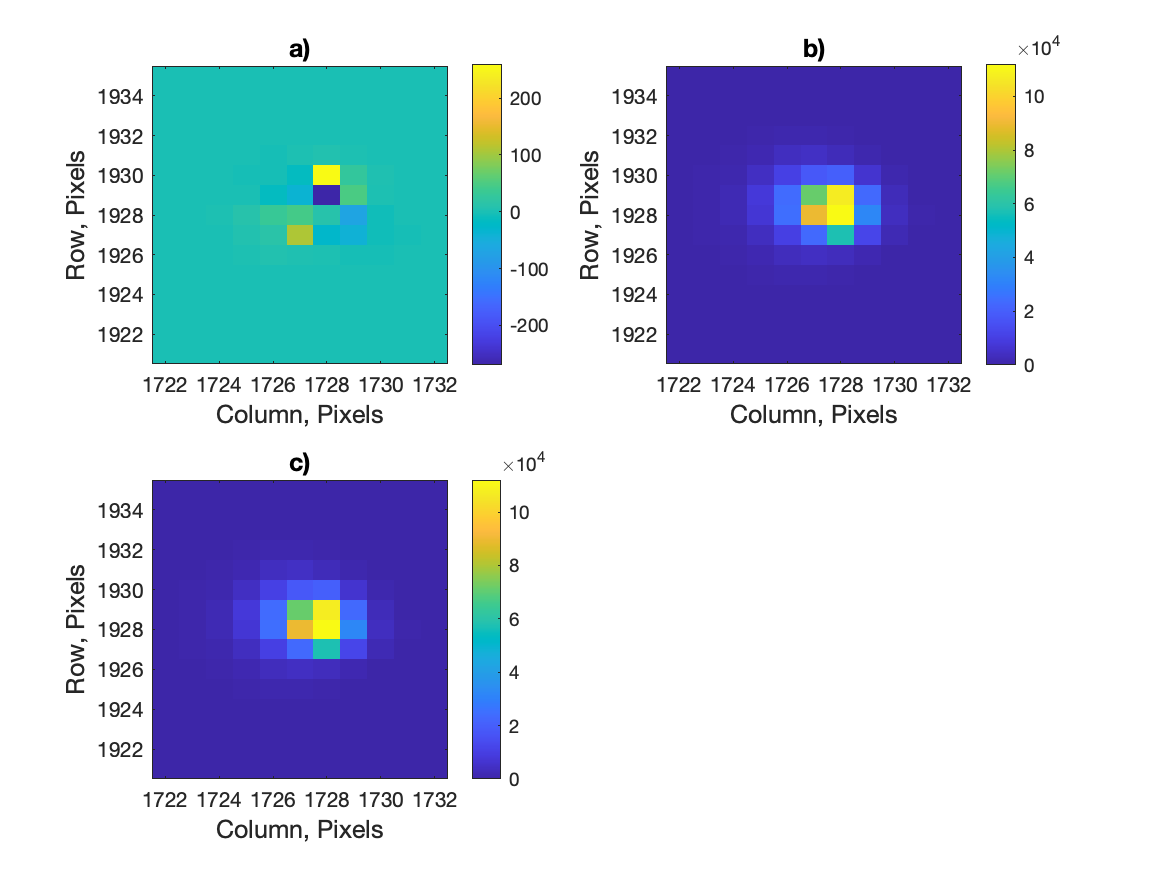}
\includegraphics[width=0.48\columnwidth]{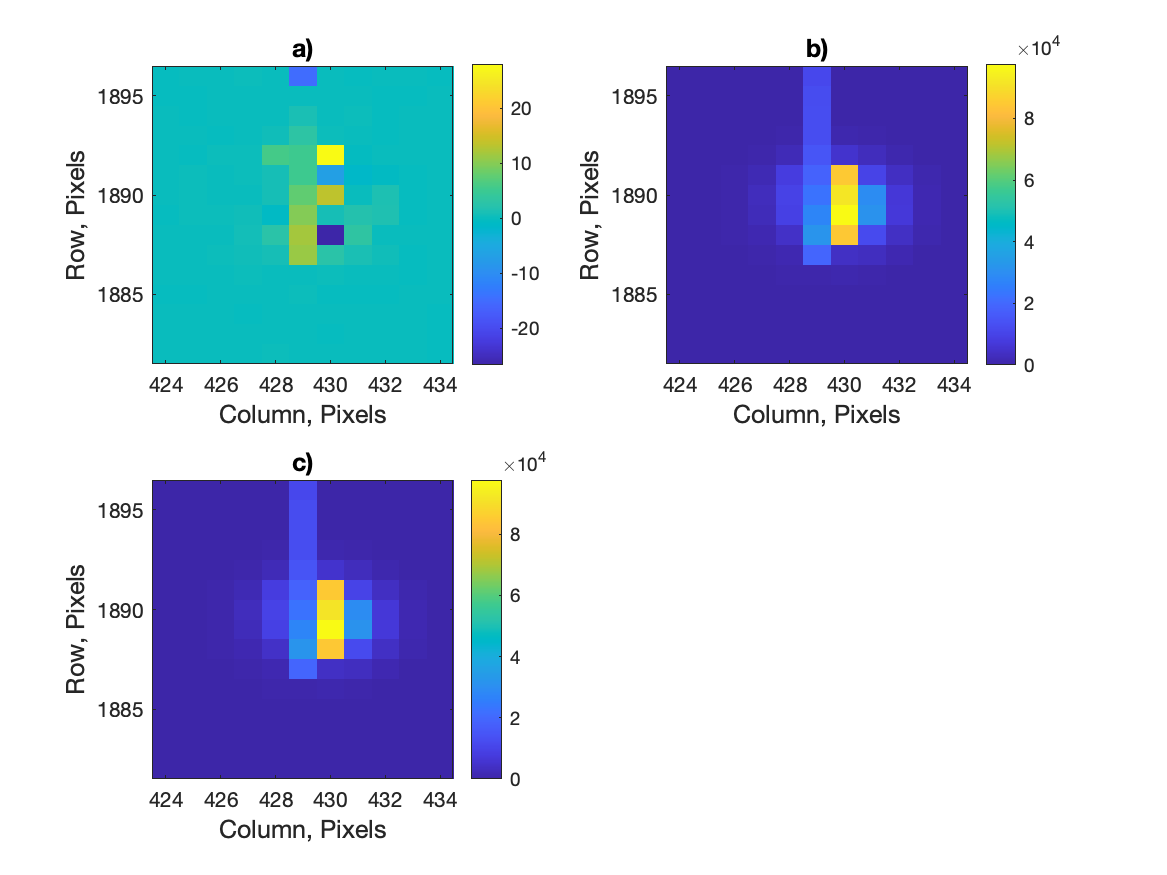}
\caption{Centroid analysis for HD88986. Difference images (panels a) for the potential transits in sector 21 (\emph{left}) and sector 48 (\emph{right}), along with the mean out-of-transit image (panel b) and the mean in-transit image (panel c). The difference images are obtained by subtracting the mean in-transit image from the mean out-of-transit image and ideally appear as an isolated stellar image of the host star. We note that in sector 48, the host star is located just beside a bad column. While interpreting the difference images from saturated stars such as HD88986 is particularly challenging, we observed that most of the energy in the transit feature is associated with the upper end of the bleed of the saturated pixels in the core of the stellar image. Thus, the transit features are likely associated with the host star in both sectors.}
\label{test_spoc}
\end{figure*}

\begin{figure*}
\centering
\includegraphics[width=0.8\columnwidth]{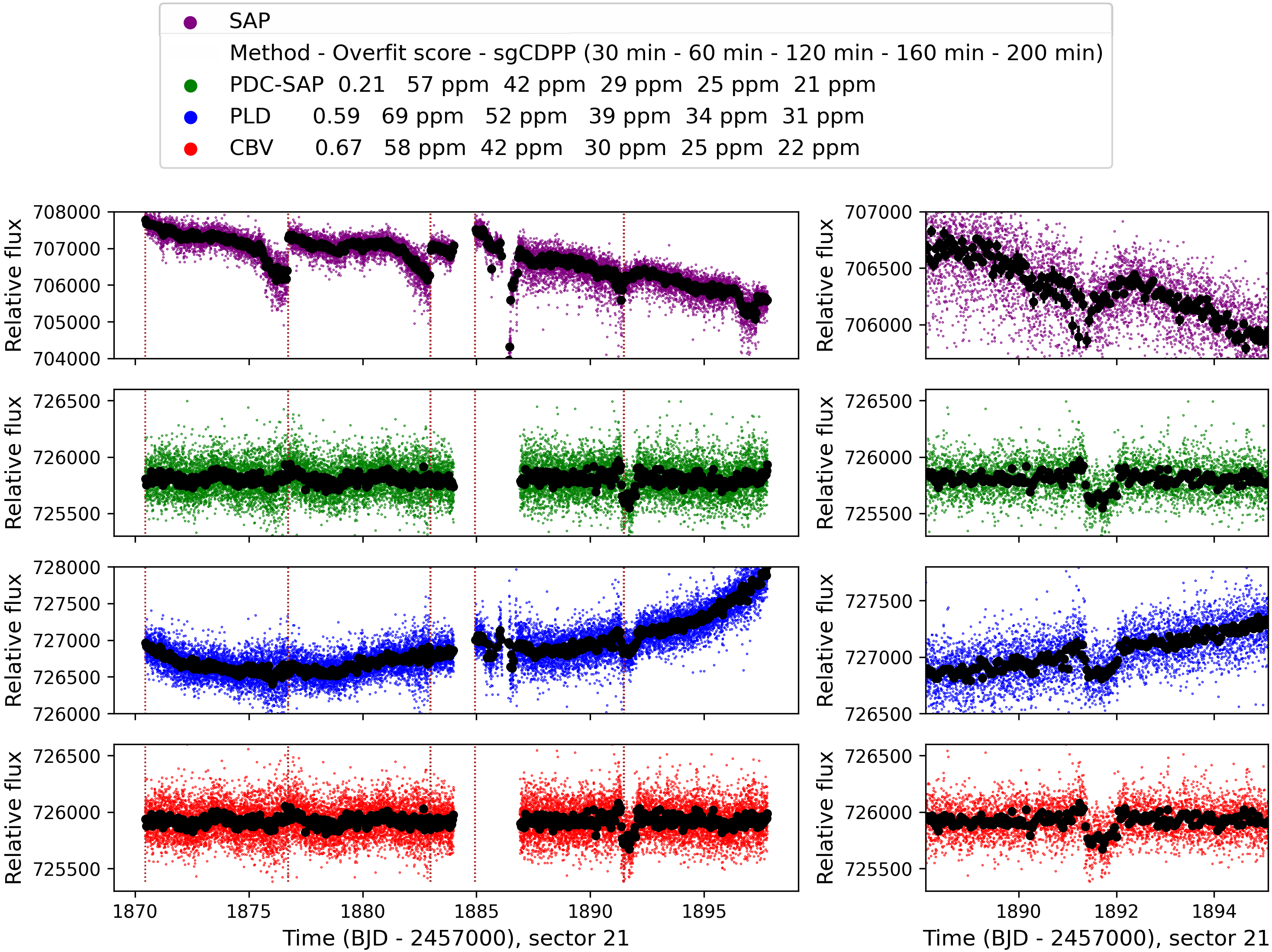}

\includegraphics[width=0.8\columnwidth]{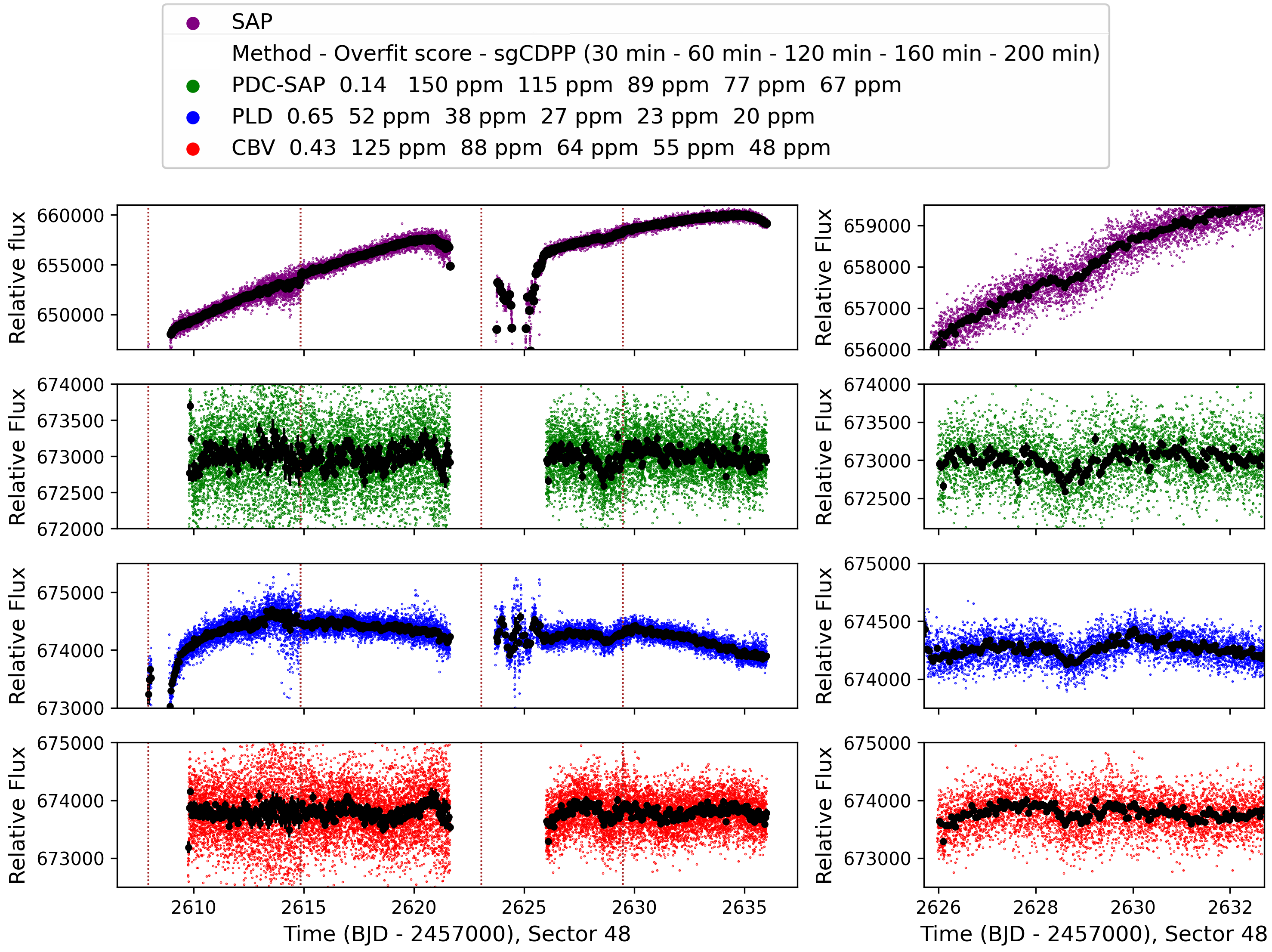}
\caption{TESS light curves reproduced using PLD and CBV approaches (\emph{left panels}) and zoomed in the potential single transits (\emph{right panels}) for sector 21 (\emph{top}) and sector 48 (\emph{bottom}). The SAP and PDC-SAP data are also plotted to provide a reference for comparison. The data are binned (black points) in 1 hour. The legend includes an overfitting score and the sgCDPP metric to facilitate an assessment of the different light curves. The brown vertical lines are the telescope momentum dumps.}
\label{pld_rcq}
\end{figure*}

\begin{figure*}
\centering
\includegraphics[width=0.8\columnwidth]{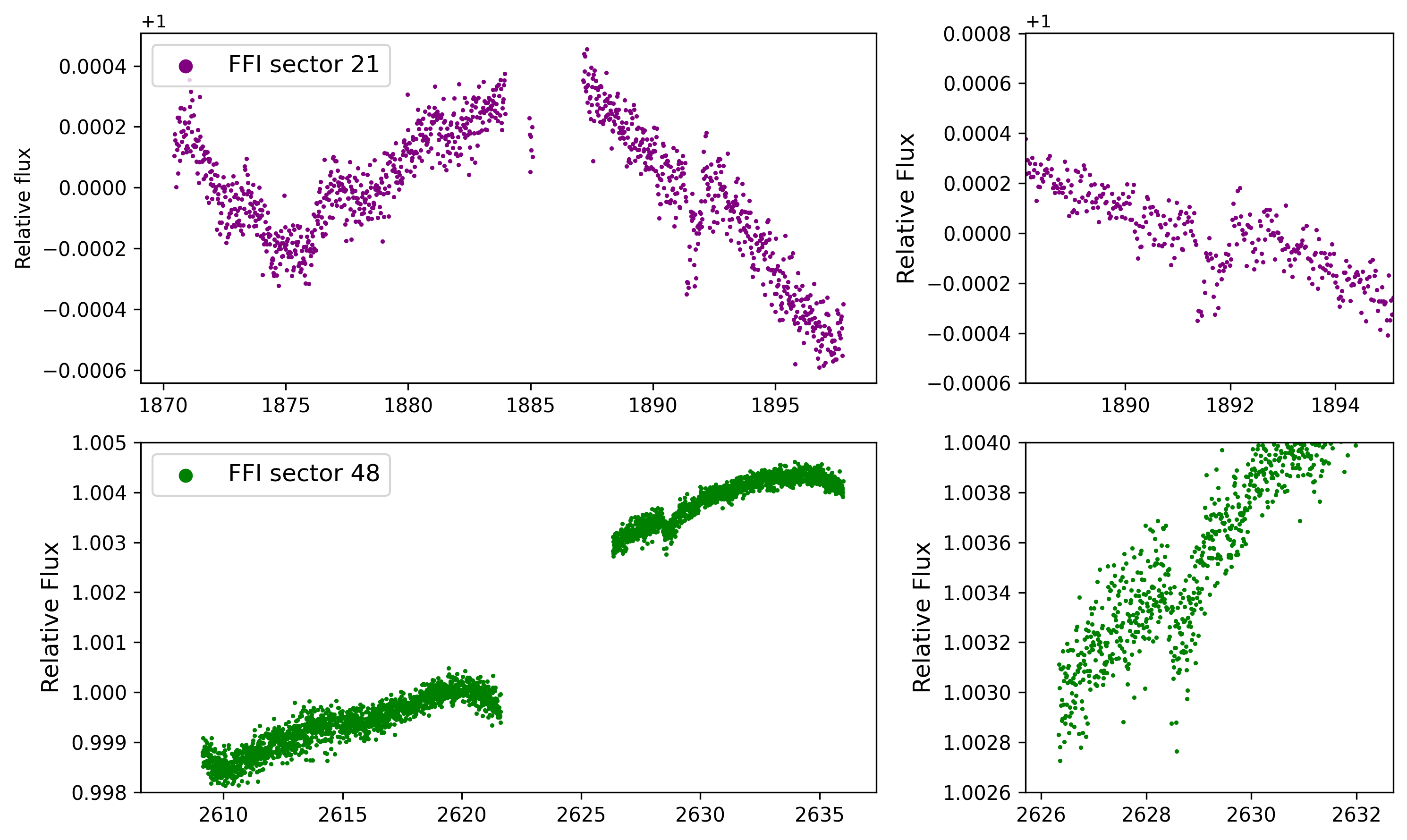}
\caption{TESS data extracted from FFI for sector 21 (\emph{top}) and sector 48 (\emph{bottom}). The data are zoomed in on the potential single transits in the right panels.}
\label{ffi}
\end{figure*}

\begin{figure}
\centering
\includegraphics[width=0.48\columnwidth]{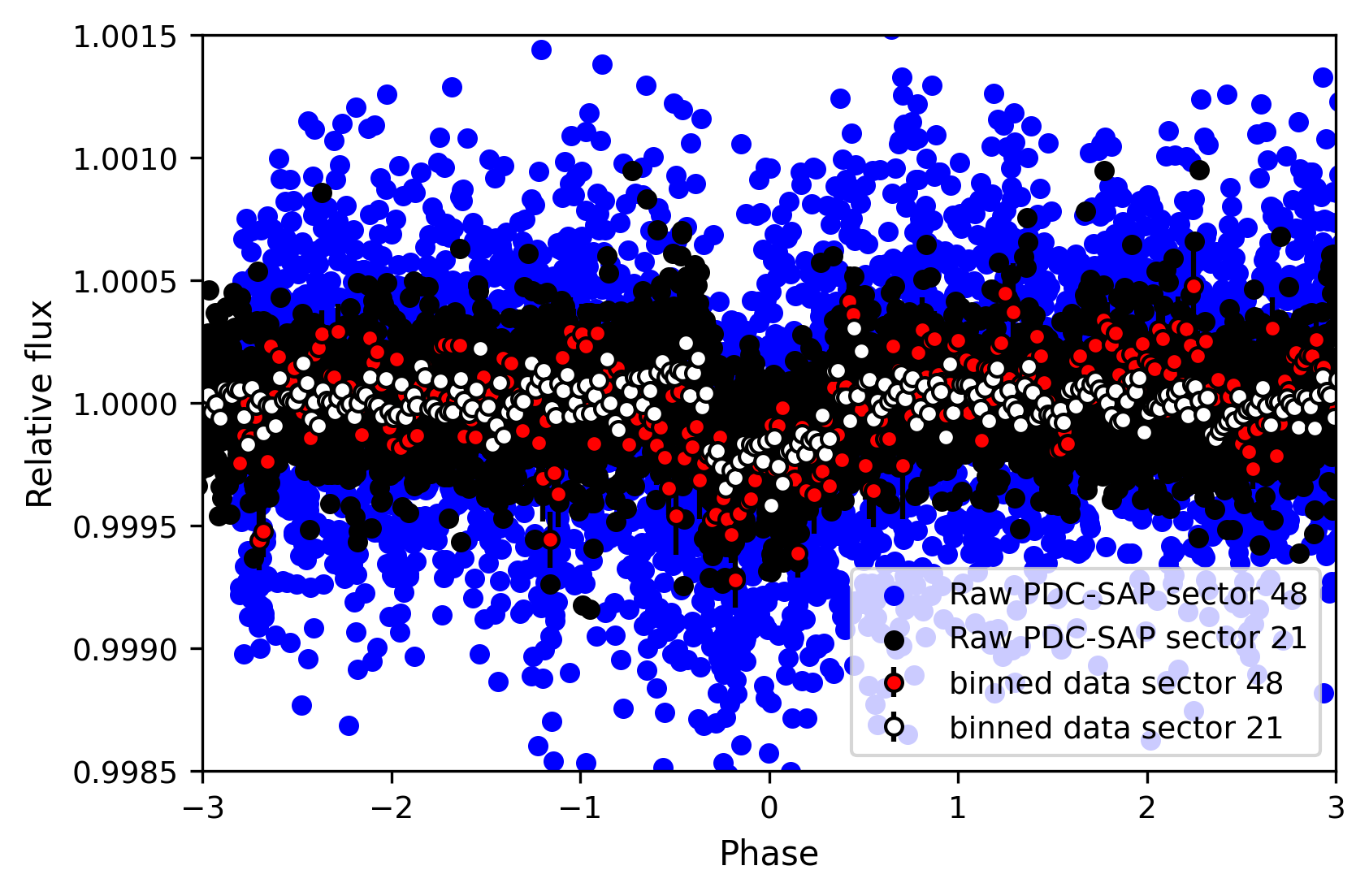}
\includegraphics[width=0.48\columnwidth]{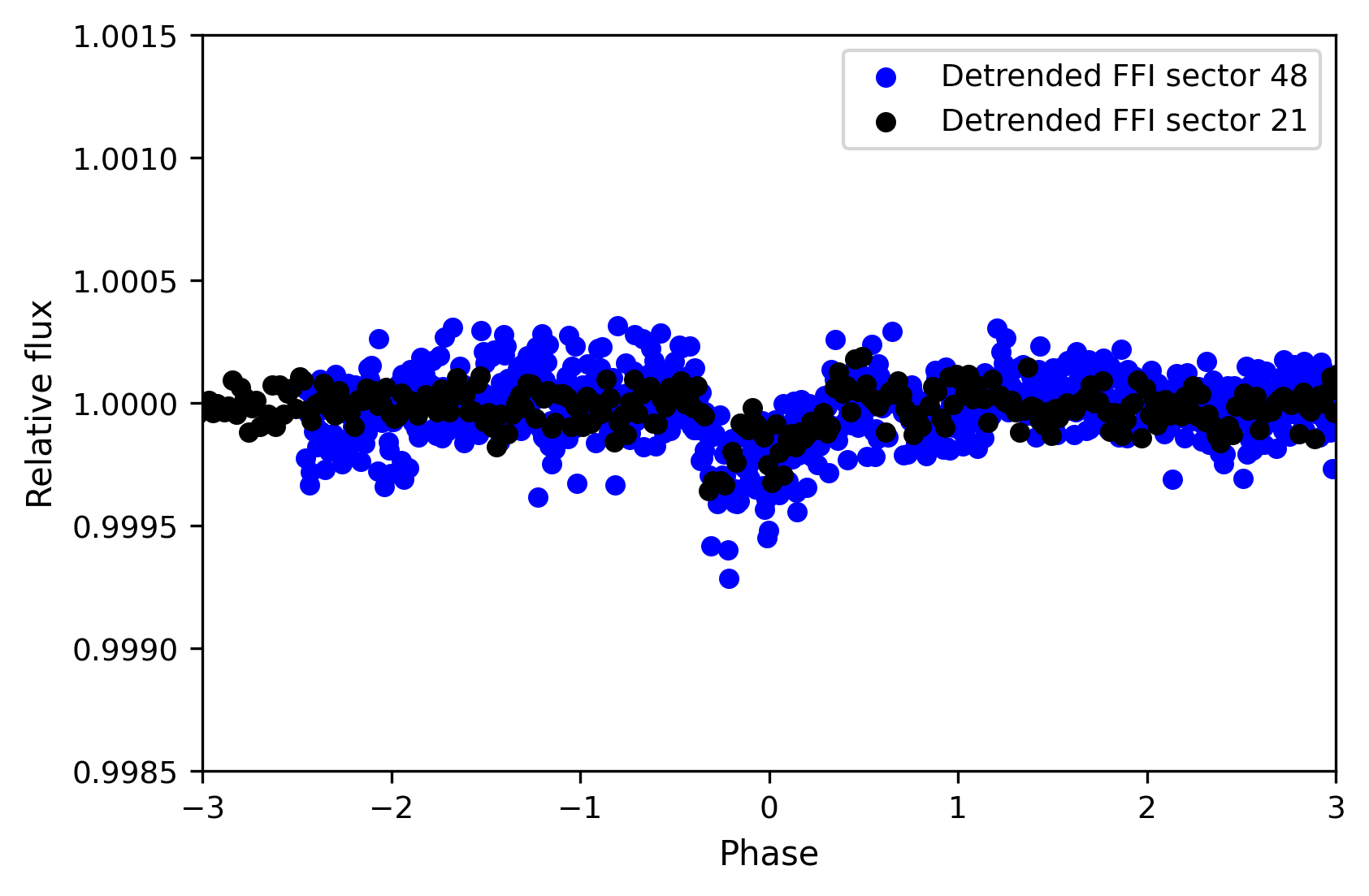}
\caption{Raw PDC-SAP (\emph{left}) and detrended FFI (\emph{right}) phase-folded data from sector 21 and sector 48 with a period of 147.4 d. The PDC-SAP data were binned in 30 minutes to ensure compatibility with the FFI data.}
\label{2_min_compatib}
\end{figure}

\section{Priors on joint modeling of RV and photometric data}

\begin{table*}[h!]
\centering
\caption{Priors for the joint modeling of RV and photometric data with \texttt{juliet} in Sect. \ref{joint2}}
\resizebox{0.7\columnwidth}{!}{%
\begin{tabular}{lccccc}
\hline
Parameter & prior & description \\
\hline
Stellar parameters & & \\
$\rho_{*}$ (kg/m$^3$) & $\mathcal{N}(479.768,40)$ & Stellar density \\
\\
Planet parameters & & \\
P (d) & $\mathcal{U}(135,155)$ & Period of HD88986\,b \\
T$_{c}$-2400000 (d) & $\mathcal{N} (2458891.6,5)$ & Center of the transit time for HD88986\,b \\
K (m/s) & $\mathcal{U}(0,10)$ & RV semi-amplitude for HD88986\,b \\
e & $\mathcal{U}(0,0.9)$ & Eccentricity \\
$\omega$ ($^\circ$) & $\mathcal{U}(0,360)$ & Argument of periastron \\
b & $\mathcal{U}(0,1)$ & Transit impact parameter \\
R$_{P}$/R$_{*}$ & $\mathcal{U}(0,1)$ & Planet-to-star radius ratio \\
\\
Drift on SHOPHIE+ & & \\
A (m/s) & $\mathcal{U} (-5,5)$ & Linear RV drift \\
Q (m/s) & $\mathcal{U} (-5,5)$ & Quadratic RV drift \\
\\
QP-GP on SOPHIE+ & & \\
B$_{GP}$ (m/s) & $\mathcal{J} (10^{-5},100)$ & Amplitude of the GP kernel \\
C$_{GP}$ (m/s) & $10^{-20}$ (fixed) & Constant scaling term of the GP kernel \\
P$_{rot}$ (d) & $\mathcal{N} (29,3)$ & Rotation period of the GP kernel \\
L$_{GP}$ (d) & $\mathcal{J} (10^{-20},300)$ & Correlation time-scale of the GP kernel \\
\\
SOPHIE instrumental Parameters & & \\
$\sigma_{SOPHIE+}$ (m/s) & $\log \mathcal{U} (1e-3, 100.)$ & RV jitter \\
mu$_{SOPHIE+}$ (m/s) & $\mathcal{U} (28995,29196)$ & Instrumental offset \\
\\
TESS instrumental parameters & & \\
D$_{TESS}$ & 1.0 (fixed) & Dilution factor \\
M$_{TESS}$ & $\mathcal{N} (0.,1)$ & Relative flux offset \\
$\sigma_{TESS}$ & $\log \mathcal{U} (0.1,1000.)$ & Extra jitter term \\
q & $\mathcal{U}(0,1)$ & Linear limb-darkening \\
\hline
\end{tabular}%
}
\tablefoot{The prior labels of $\mathcal{N}$, $\mathcal{U}$, $\log \mathcal{U}$ indicate normal, uniform, and uniform logarithms of distributions.}
\label{prior_joint}
\end{table*}

\section{Priors on joint modeling of RV and Hipparcos/Gaia astrometric data}

\begin{table*}[h!]
\centering
\caption{Priors for the joint modeling of RV and Hipparcos/Gaia astrometric data.}
\resizebox{0.7\columnwidth}{!}{%
\begin{tabular}{lccccc}
\hline
Parameter & prior & description \\
\hline
Stellar parameters & & \\
M$_{*}$ (M$_{\odot}$) & $\mathcal{N}(1.2,0.06)$ & Stellar mass \\
Parallax (mas) & $\mathcal{N}(30.025,0.023)$ & Stellar parallax \\
$\sigma_{*}$ (m/s) & $\mathcal{U}(0,10)$ & Stellar jitter \\
\\

Planet parameters & & \\
$\textit{a}$ (au) & $\mathcal{U}(1,100)$ & Semi-major axis of HD88986 c \\
e & $\mathcal{U}(0,0.99)$ & Eccentricity \\
$\omega$ ($^{\circ}$) & $\mathcal{U}(0,360)$ & Argument of periastron \\
$\textit{I}$ ($^{\circ}$) & sin(0,180) & Orbital inclination \\
$\Omega$ ($^{\circ}$) & $\mathcal{U}(0,360)$ & Longitude of ascending node \\
M$_{c}$ ($M_{Jup}$) & $\mathcal{U}(1,500)$ & Mass of HD88986 c \\
\\
SOPHIE instrumental Parameters & & \\
mu$_{ELODIE}$ (m/s) & $\mathcal{U} (28000,30000)$ & ELODIE offset \\
mu$_{SOPHIE}$ (m/s) & $\mathcal{U} (28000,30000)$ & SOPHIE offset \\
mu$_{SOPHIE+}$ (m/s) & $\mathcal{U} (28000,30000)$ & SOPHIE+ offset \\
mu$_{HIRES}$ (m/s) & $\mathcal{U} (-1000,1000)$ & HIRES offset \\
mu$_{HIRES+}$ (m/s) & $\mathcal{U} (-1000,1000)$ & HIRES+ offset \\
\hline
\end{tabular}%
}
\tablefoot{The prior labels of $\mathcal{N}$, $\mathcal{U}$, sin indicate normal, uniform, and sinusoidal distributions.}
\label{prior_joint_RV_Hip_Gaia}
\end{table*}

\section{Corner plot of the joint modeling SOPHIE+ RVs and TESS sector 21 photometry}

\begin{figure}[h]
\includegraphics[width=\textwidth]{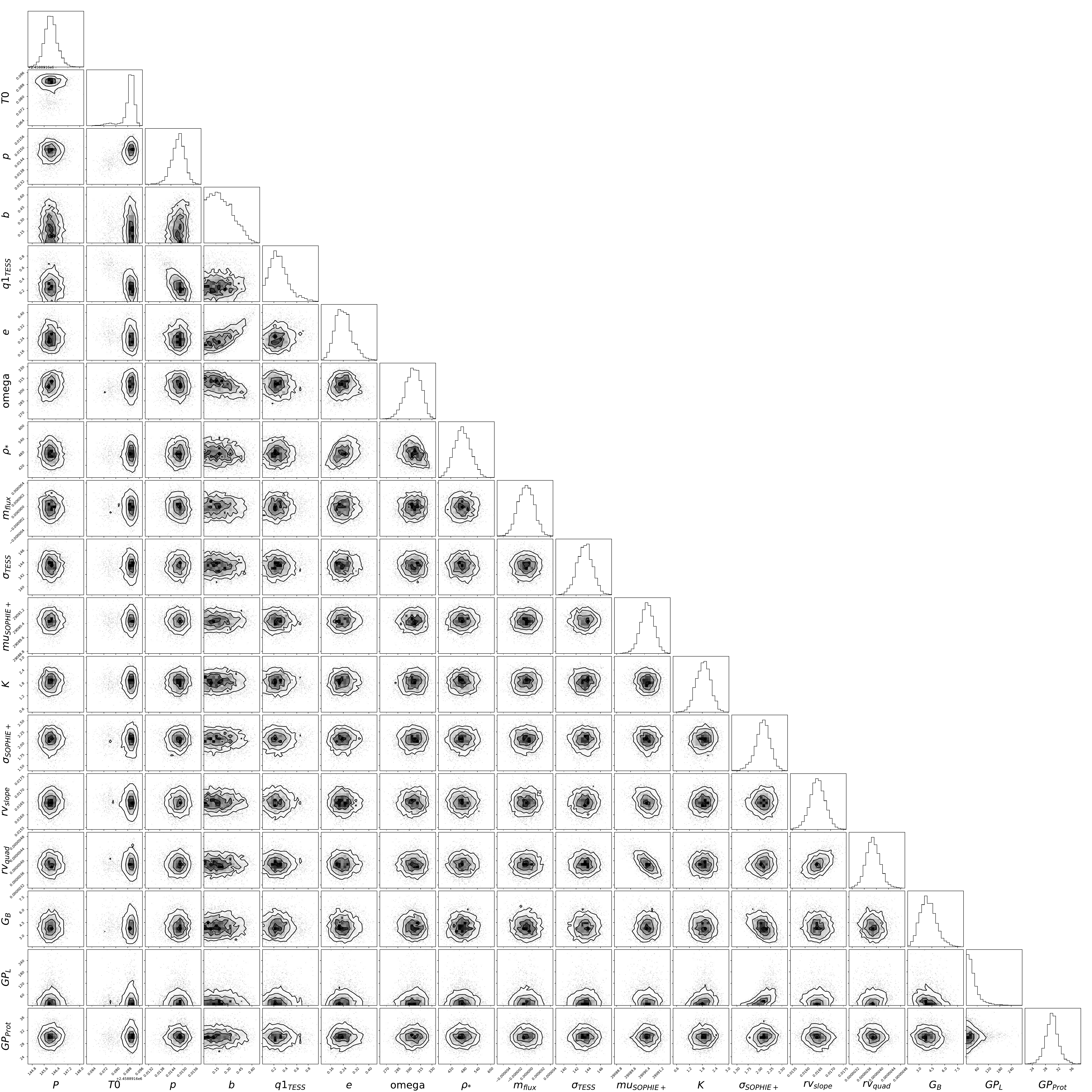}
\caption{Nested samples distribution of fitted parameters of HD88986\,b on joint modeling of SOPHIE+ RVs and TESS sector 21 light curve. The 1, 2, and 3$\sigma$ confidence levels of the posterior samples are shown by the contours.}
\label{fig:corner}
\end{figure}

\section{Corner plot of the joint modeling of combined HD88986 RVs, Hipparcos and Gaia astrometry}
\begin{figure}[h]
\includegraphics[width=\textwidth]{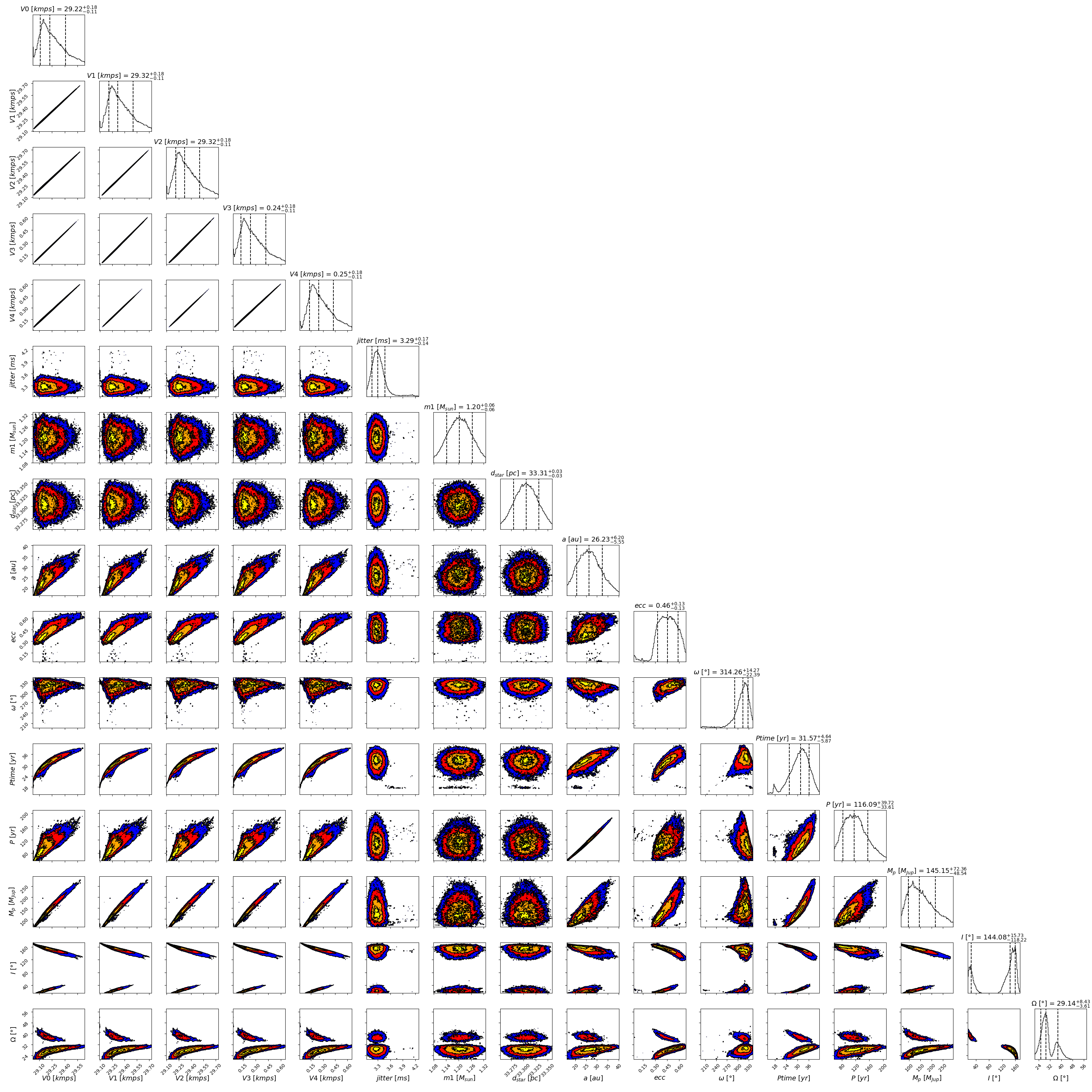}
\caption{MCMC samples distribution of fitted parameters of HD88986 outer massive companion on joint modeling of HD88986 RVs and Hipparcos/Gaia astrometry. V0, V1, V2, V3, and V4 correspond to the ELODIE, SOPHIE, SOPHIE+, HIRES, and HIRES+ RV dataset, respectively.}
\label{fig:corner_MCMC}
\end{figure}

\section{APT photometric data}

\centering
\begin{table*}[h!]
\centering
\caption{Summary of APT photometric observation for HD88986}
\begin{tabular}{lcccc}
\hline
Observing & $N_{obs}$ &Date Range & $\sigma$ & Seasonal Mean \\
Season & & (HJD $-$ 2,400,000)& (mag) & (mag) \\
\hline
   1995--96   &  18 & 50192--50235 & 0.00129 & $-$1.16507(30) \\
   1996--97   &  46 & 50395--50597 & 0.00083 & $-$1.16476(12) \\
   1997--98   &  68 & 50755--50978 & 0.00140 & $-$1.16513(17)\\
   1998--99   &  72 & 51115--51339 & 0.00146 & $-$1.16529(17) \\
   1999--00   &  60 & 51480--51706 & 0.00116 & $-$1.16500(15) \\
   2000--01   &  43 & 51862--52051 & 0.00104 & $-$1.16486(16)\\
   2001--02   &  60 & 52202--52445 & 0.00114 & $-$1.16421(15) \\
   2002--03   &  75 & 52572--52805 & 0.00108 & $-$1.16476(13) \\
   2003--04   &  78 & 52930--53171 & 0.00099 & $-$1.16492(11) \\
   2004--05   & 104 & 53308--53529 & 0.00132 & $-$1.16491(13)  \\
   2005--06   & 109 & 53663--53904 & 0.00138 & $-$1.16509(13) \\
   2006--07   &  99 & 54032--54259 & 0.00119 & $-$1.16453(12)\\
   2007--08   &  62 & 54391--54630 & 0.00101 & $-$1.16479(13) \\
   2008--09   &  44 & 54774--54983 & 0.00086 & $-$1.16511(13)\\
   2009--10   &  75 & 55139--55351 & 0.00116 & $-$1.16494(13) \\
   2010--11   &  71 & 55513--55709 & 0.00105 & $-$1.16474(12)\\
   2011--12   &  61 & 55893--56069 & 0.00109 & $-$1.16472(14)\\
   2012--13   &  38 & 56236--56447 & 0.00068 & $-$1.16502(11)\\
   2013--14   &  51 & 56607--56809 & 0.00097 & $-$1.16539(14)\\
   2014--15   &  67 & 56967--57176 & 0.00118 & $-$1.16532(14)\\
   2019--20   &  34 & 58856--58975 & 0.00117 & $-$1.16518(20)\\
   \hline
\end{tabular}
\label{aptdata}
\end{table*}

\end{appendix}

\end{document}